\documentstyle[12pt]{article}
\newcommand{\bff}[1]{{\mbox{\boldmath $#1$}}}
\begin{document}
\setlength{\baselineskip}{0.5cm}

%%%%%%%%%%%%%%%%%%%%%%%%%%%%%%%%%%%%%%%%%%%%%%%%%%%%%%%%%%%%
%%  Title page 
%%%%%%%%%%%%%%%%%%%%%%%%%%%%%%%%%%%%%%%%%%%%%%%%%%%%%%%%%%%%
\title{ Relativistic Mean Field Theory in Rotating Frame:
 Single-particle Properties at Superdeformation}
\author{A.\ V.\ Afanasjev$^{1,2,3,4}$, G.A. Lalazissis$^{1,5}$ 
and P.\ Ring$^{1}$ \\
$^1$ Physik-Department, Technische 
Universit{\"a}t M{\"u}nchen\\
 D-85747, Garching, Germany\\
$^2$Department of Mathematical Physics\\
Lund Institute of Technology\\
 PO Box 118, S-22100, Lund, Sweden\\
$^3$NORDITA-Nordisk Institut for Teoretisk Fysik\\
 Blegdamsvej 17, DK-2100, 
Copenhagen 0, Denmark \\
$^4$Nuclear Research Center, Latvian Academy of Sciences\\
 LV-2169, Salaspils, Miera str.\ 31, Latvia\\
$^5$Department of Theoretical Physics\\
Aristotle University of
Thessaloniki\\
GR-54006, Thessaloniki, Greece}

\date{}
\maketitle
\begin{abstract}

Single-particle properties at superdeformation are
investigated within the Cranked Relativistic Mean Field
(CRMF) theory on the example of superdeformed rotational
bands observed in the $A\sim 140-150$ mass region.
Applying the  effective alignment approach it is shown that
CRMF theory provides a reasonable description of the
alignment properties of the single-particle orbitals. The
agreement with experiment is good in most of the cases.
This  suggests that many features of the observed
superdeformed bands can be well understood in terms of an
almost undisturbed single-particle motion. The stability of
the results with respect to the parameterizations used in
Relativistic Mean Field (RMF) theory is also investigated,
employing some frequently used non-linear effective forces.
It turns out, with the exception of the single-particle
ordering in superdeformed minimum, that the dependence of
the calculated observables on the parameterization is
rather small.
\end{abstract}

%%%%%%%%%%%%%%%%%%%%%%%%%%%%%%%%%%%%%%%%%%%%%%%%%%%%%%%%%%
%%%%  Introduction  %%%%%%%%%%%%%%%%%%%%%%%%%%%%%%%%%%%%%%
%%%%%%%%%%%%%%%%%%%%%%%%%%%%%%%%%%%%%%%%%%%%%%%%%%%%%%%%%%

\section{Introduction}

Superdeformation at high spin is one of the most
interesting topics of the low energy nuclear physics.  The
large amount of experimental information, which has become
available in recent years for superdeformed (SD) rotational
bands at various mass regions \cite{BFS.96,Zn62},  has also
attracted much theoretical interest for detailed studies of
their structure and properties.  However, there are still
very important quantities not yet measured.  Linking
transitions from superdeformed states have been identified
only for  $^{194}$Hg \cite{Hg194} and $^{194}$Pb
\cite{Pb194} nuclei. Therefore, spins, parities as well as
excitation energies of SD bands are not yet known in most
of the cases. A direct test of the structure of the wave
functions of the single-nucleonic orbitals (e.g. via
magnetic moments) is considered extremely difficult due to
the very short life-times of the SD high-spin states. All
these make a straightforward comparison between experiment
and theory almost impossible in most of the cases.
Therefore, the relative properties of different SD bands
play an important role in our understanding of their
structure (see for example Refs.
\cite{BRA.88,NWJ.89,Rag91,Rag93,Gd46-50,AKR.96}).  One of
the main goals is to understand which changes of the
single-particle orbitals are involved on going from one SD
band to another and how they affect physical observables,
like dynamic moments of inertia $J^{(2)}$, charge
quadrupole moments $Q_0$ etc. The theoretical analysis of
the underlying structure of two SD bands in two neighboring
nuclei, differing by the occupation of one single-nucleonic
orbital, provides a less model dependent comparison
compared with the one among bands which differ in
occupation by several nucleonic orbitals. This is because
of possible inadequacies in the theoretical description of
the properties of the single-particle orbitals. There are
different ways to identify the underlying orbital, however,
all of them are not free of disadvantages.  The most
reliable experimental data is obtained from precise
measurements of transition energies within the SD bands.
Therefore the analysis can be based either on the relative
properties of dynamic moments of inertia $J^{(2)}$, or on
the effective alignment $i_{eff}$ of two different SD
bands.  In the present article, these two  approaches are
discussed and compared in detail. Moreover, the
applications of these approaches to the analyses of SD
bands within the framework of different models are also
overviewed.

Relativistic mean field (RMF) theory has recently gained
considerable success in describing various facets of
nuclear structure properties \cite{R.96}. The cranked
relativistic mean field (CRMF) theory
\cite{KR.89,KR.90,KR.93} provides the extension for the
description of rotating nuclei.  The CRMF theory has been
successfully applied for the study of twin bands in
rotating SD nuclei \cite{KR.93}.  Recently, it has been
also used for a  detailed description of the SD bands
observed in the $A \sim 80$ \cite{AKR.96pl} and $A \sim
140-150$ \cite{AKR.96} regions, where RMF theory provides
also an excellent description of the ground state
properties \cite{lala1,lala2}. It is noted in passing, that
RMF theory excellently reproduces the recently reported
excitation energies of the SD bands relative to the ground
state of $^{194}$Hg and $^{192}$Pb nuclei \cite{LR.97}.
    
It has been shown \cite{AKR.96,AKR.96pl} that CRMF theory
provides a good agreement with experiment.  It should be
noted, however, that this analysis has been based on the
properties of dynamic moment of inertia $J^{(2)}$ only
\cite{AKR.96,AKR.96pl}.  So far, the additional information
that can be obtained by means of the effective alignment
approach for the absolute alignments of the single-particle
orbitals and for their changes with increasing rotational
frequency, has not been exploited.  Such an analysis seems
also to be necessary in order to test to which extent CRMF
theory is able to describe not only collective but also
single-particle properties of rotating nuclei.

The present work has two parts. In the first part we study
the alignment properties of single-particle orbitals in the
superdeformed minimum.  This study is based on SD bands
observed in the $A\sim 140-150$ mass region. There are two
reasons that motivate this selection.  The first reason is
connected with the fact that, as it is rather commonly
accepted, many features of the SD bands in this mass region
can be well understood in terms of an almost undisturbed
single-particle motion
\cite{Rag91,Rag93,Gd46-50,AKR.96,Satula}.  This implies
that a study without pairing correlations is a reasonable
approximation. Therefore, as in our previous investigations
for this mass region \cite{AKR.96}, pairing correlations
are not taken into account. The motivation as well as the
validity  of such an approximation have been discussed in
detail in Ref. \cite{AKR.96}.  Secondly, the effective
alignment approach based on the cranked Nilsson-Strutinsky
(CN) model has been extensively used in this mass region
\cite{Rag91,Rag93,Gd46-50,Tb152a,Gd147c,Eu142a}.  This provides the
opportunity to make a comparison with the CN approach, in
which bulk and single-particle properties are treated
separately.  We have carried out a detailed analysis of the
alignment properties of some single-particle orbitals
involved in the SD bands observed in nuclei in the
neighborhood of $^{152}$Dy and $^{143}$Eu. We have limited
ourselves to SD bands which do not undergo unpaired band
crossing. One should note, however, that in some cases band
crossings may play an important role to identify some
orbitals via the characteristic shape of $J^{(2)}$ in
crossing region and the characteristic crossing
frequencies. This has been illustrated, for example, in
Refs. \cite{Rag91,Gd46-50,Gd147c}.

In the second part of the article we examine the dependence
of the calculated observables on the Lagrangian
parameterizations of the RMF theory. So far, only the
frequently used NL1 force \cite{RRM.86} has been employed
in our investigations \cite{AKR.96,AKR.96pl} and
satisfactory results have been obtained.  Difficulties have
been observed, however, in $^{146-148}$Gd isotopes,
for a consistent determination of the yrast and first
excited SD bands. NL1 predicts a rather large asymmetry
energy in nuclear matter calculations and therefore does
not seem to be appropriate for nuclei with large neutron
excess \cite{NL3}.  We have carried out calculations using
the NLSH \cite{NLSH} parameterization as well as the
recently proposed  parameter set NL3 \cite{NL3} for
several rare-earth nuclei. The results are compared with
experiment and those obtained with NL1.

The article is organized as follows.  In section 2 the main
features of the CRMF theory are briefly outlined.  Details
of the calculations are also given. In section 3 the
effective alignment approach is discussed and compared with
the one based on dynamic moment of inertia.
Applications of the two approaches within the framework of
different models are also overviewed.  In sections 4 and 5
the results of the CRMF theory for nuclei around $^{152}$Dy
and $^{143}$Eu, respectively, are analyzed using the
effective alignment approach.  In section 6 the dependence
of collective and single-particle properties on the
parameterizations of the RMF Lagrangian is discussed.
Finally, section 7 summarizes our main conclusions.

%%%%%%%%%%%%%%%%%%%%%%%%%%%%%%%%%%%%%%%%%%%%%%%%%%%
%
\section{Cranked relativistic mean field theory}
%
%%%%%%%%%%%%%%%%%%%%%%%%%%%%%%%%%%%%%%%%%%%%%%%%%%%

The starting point of the RMF theory is the well known
local Lagrangian density \cite{SW.86} where nucleons are
described as Dirac spinors which interact via the exchange
of several mesons such as the $\sigma$-meson responsible
for the large scalar attraction at intermediate distances,
the $\omega$-meson for the vector repulsion at short
distances and the $\rho$-meson for the asymmetry properties
of nuclei with large neutron or proton excess.

The local Lagrangian density can be written in the
following form: 
\begin{eqnarray}
{\cal L}&=&\bar{\psi}(i\gamma ^\mu \partial _\mu -m)\psi
+\frac{1}{2}(\partial_\mu \sigma \partial ^\mu 
\sigma -m_\sigma ^2\sigma ^2)
-\frac{1}{3}g_2\sigma^3-\frac{1}{4}g_3\sigma^4
\nonumber\label{eq1}\\ 
&&-\frac{1}{4}{\sl\Omega}_{\mu\nu}{\sl\Omega}^{\mu\nu}
  +\frac{1}{2}m_\omega^2\omega_\mu\omega^\mu 
  ~-~\frac{1}{4}{\vec R}_{\mu\nu}{\vec R}^{\mu\nu}
  +\frac{1}{2}m_\rho^2\vec{\rho}_\mu\vec{\rho\,}^\mu  
  ~-~\frac{1}{4}F_{\mu\nu}F^{\mu\nu}
\\
&&-g_\sigma\bar{\psi}\sigma\psi~-~
   g_\omega \bar{\psi}\gamma^\mu\omega_\mu\psi~-~
   g_\rho\bar{\psi}\gamma^\mu\vec{\tau}\vec{\rho}_\mu\psi~-~ 
   e\,\bar{\psi}\gamma^\mu\,\frac{1-\tau_3}{2}A_\mu\psi, 
\nonumber
\end{eqnarray}
where the non-linear self-coupling of the $\sigma$-field,
which is important for an adequate description of nuclear
surface properties and the deformations of finite nuclei,
is taken into account according to Ref. \cite{BB.77}.  The
field tensors for the vector mesons and the photon field
are:
\begin{equation}
{\sl\Omega}_{\mu\nu}=\partial_\mu\omega_\nu-\partial_\nu\omega_\mu,\qquad 
{\vec
R}_{\mu\nu}=\partial_\mu\vec\rho_\nu-\partial_\nu\vec\rho_\mu,
\qquad F_{\mu\nu }=\partial_\mu A_\nu-\partial_\nu A_\mu.  
\label{eq2}
\end{equation}
In the present state of the art of RMF theory, the meson
and photon fields are treated as classical fields.

The Lagrangian (\ref{eq1}) contains as parameters the
masses of the mesons $m_{\sigma}$, $m_{\omega}$ and
$m_{\rho}$, the coupling constants $g_{\sigma}$,
$g_{\omega}$ and $g_{\rho}$ and the non-linear terms $g_2$
and $g_3$.

The mesons are considered as effective particles carrying
the most important quantum numbers and generating the
interaction in the corresponding channels in a Lorentz
invariant manner by a local coupling to the nucleons. In
this sense, the Lagrangian (1) is an effective Lagrangian
constructed for the mean field approximation.

The RMF theory has been extended for the description of
rotating nuclei by employing the cranking model
\cite{KR.89,KR.90}.  The rotation of a nucleus with the
angular velocity (rotational frequency) ${\sl\Omega}_x$ is
considered along the $x$-axis in a one-dimensional cranking
approximation. The rotational frequency is determined by
the condition that the expectation value of the total
angular momentum at spin $I$ has a definite value:
\begin{eqnarray}
J({\sl\Omega}_x)=\langle{\sl\Phi_\Omega}\mid\hat{J}_x
\mid {\sl\Phi_\Omega}\rangle~=~\sqrt{I(I+1)}.  
\label{eq3}
\end{eqnarray}
In this concept the rotation of nuclei can be described by
transforming the effective Lagrangian density (\ref{eq1})
from the laboratory system into a frame which rotates with
constant angular velocity ${\sl\Omega}_x$  along the
$x$-axis. Subsequently, the variational principle leads to
time-independent inhomogeneous Klein-Gordon equations for
the mesonic fields in the rotating frame (for details see
Ref. \cite{KR.89} or \cite{KNM.93})
\begin{eqnarray}
\left\{-\Delta-({\sl\Omega}_x\hat{L}_x)^2+m_\sigma^2\right\}~
\sigma(\bff r)&=&
-g_\sigma\left[\rho_s^p(\bff r)+\rho_s^n(\bff r)\right] 
-g_2\sigma^2(\bff r)-g_3\sigma^3(\bff r),  
\nonumber \\
\left\{-\Delta-({\sl\Omega}_x\hat{L}_x)^2+m_\omega^2\right\}
\omega_0(\bff r)&=&
g_\omega\left[\rho_v^p(\bff r)+\rho_v^n(\bff r)\right],  
\nonumber \\
\left\{-\Delta-({\sl\Omega}_x(\hat{L}_x+\hat{S}_x))^2+
m_\omega^2\right\}~
\bff\omega(\bff r)&=&
g_\omega\left[\bff j^p(\bff r)+\bff j^n(\bff r)\right],  
\nonumber \\
\left\{-\Delta-({\sl\Omega}_x\hat{L}_x)^2+m_\rho^2\right\} 
\rho_0(\bff r)&=&
g_\rho\left[\rho_v^n(\bff r)-\rho_v^p(\bff r)\right],  
\label{eq4}\\
\left\{-\Delta-({\sl\Omega}_x(\hat{L}_x+\hat{S}_x))^2+
m_\rho^2\right\}~
\bff\rho(\bff r)&=&
g_\rho\left[\bff j^n(\bff r)-\bff j^p(\bff r)\right],  
\nonumber \\
-\Delta~A_0(\bff r)&=&e\rho_v^p(\bff r),  
\nonumber \\
-\Delta~\bff A(\bff r)&=&e\bff j^p(\bff r),
\nonumber
\end{eqnarray}
with source terms involving the various nucleonic densities
and currents
\begin{equation}
\rho_s^{n,p}=\sum_{i=1}^{N,Z}\bar{\psi}_i\psi _i,\qquad 
\rho_v^{n,p}=\sum_{i=1}^{N,Z}\psi _i^{+}\psi _i,\qquad 
\bff j^{n,p}=\sum_{i=1}^{N,Z}\bar\psi_i\bff\gamma\psi_i,
\label{eq5}
\end{equation}
where the labels $n$ and $p$ are used for neutrons and
protons, respectively.  In the equations above, the sums
run over the occupied positive-energy shell model states
only ({\it no-sea approximation}) \cite{RRM.86}.  In the
Eq.(\ref{eq4}), the four-vector components of the vector
fields $\omega^\mu $, $\rho^\mu$ and $A^\mu$ are separated
into the time-like ($\omega _0$, $\rho_0$ and $A_0$) and
space-like [$\bff\omega\equiv(\omega^x,\omega^y,\omega^z)$,
$\bff\rho\equiv(\rho^x,\rho^y,\rho^z)$, and $\bff
A\equiv(A^x,A^y,A^z)$] components. The contribution of
space-like components vanish in the systems with
time-reversal symmetry \cite{SW.86}.  Note that in mean
field approximation the contribution of the charged
$\rho$-mesons vanish too. The uncharged components are
defined as $\rho_0$ and $\bff\rho$.
                                                  
In the Hartree approximation, the stationary Dirac equation
for the nucleons in the rotating frame can be written as
\begin{eqnarray}
\left\{\bff\alpha(-i\bff\nabla-\bff V(\bff r))~+~
V_0(\bff r)~+~\beta(m+S(\bff r))-{\sl\Omega}_x\hat{J}_x\right\} 
\psi_i~=~\epsilon_i\psi_i  
\label{eq6}
\end{eqnarray}
where $V_{0}({\bf r})$ represents a repulsive vector 
potential:
\begin{equation}
V_0(\bff r)~=~g_\omega\omega_0(\bff r)+g_\rho\tau_3\rho_0(\bff r)
+e\frac{1-\tau_3}{2}A_0(\bff r)
\end{equation}
and $S({\bf r})$ an attractive scalar potential:
\begin{equation}
S(\bff r)=g_\sigma\sigma(\bff r).
\end{equation}
The latter contributes to the effective mass as:
\begin{equation}
M^{\ast}({\bf r}) = M + S({\bf r}).
\end{equation}
A magnetic potential $\bff V(\bff r)$: 
\begin{equation}
\bff V(\bff r)~=~g_\omega\bff\omega(\bff r)
+g_\rho\tau_3\bff\rho(\bff r)+
e\frac{1-\tau_3}{2}\bff A(\bff r)
\end{equation}
originates from the spatial components of the vector
mesons. It breaks the time reversal symmetry and removes
the degeneracy between nucleonic states related via this
symmetry. This effect is commonly referred to as {\bf
nuclear magnetism} \cite{KR.89}.  The role of the magnetic
potential is quite important for a proper description of
the moments of inertia \cite{KR.93}. Therefore, in the
present work {\it the spatial components of the vector
mesons are taken into account in a fully self-consistent
way}.

Finally the term
\begin{equation}
{\sl\Omega}_x\hat{J}_x~=~
{\sl\Omega}_x(\hat{L}_x+\frac{1}{2}\hat{\Sigma}_x)
\end{equation}
represents the Coriolis field. It is noted that 
time-reversal symmetry is also broken by the Coriolis
field.

%%%%%%%%%%%%%%%%%%%%%%%%%%%%%%%%%%%%%%%%% 
\subsection{Details of the calculations}
%%%%%%%%%%%%%%%%%%%%%%%%%%%%%%%%%%%%%%%%%

The numerical calculations have been performed along the
lines presented in section 2.2 of Ref. \cite{AKR.96}. The
experimental data for SD bands is taken from the references
given in Table 2 shown below. However, in the present investigation we
are mainly interested in the  single-particle properties.
This suggests that a  more accurate truncation scheme
should be used.  In view of this, more mesh points (16)
have been used for the  Gauss-Hermite integration to ensure
better stability in the final results. All bosonic states
below the energy cut-off $E^{cut-off}_B\leq 16.5
\hbar\omega_0^B$ and all fermionic states below the energy
cut-off $E^{cut-off}_F\leq 11.5\hbar \omega _0^F$ have been
used in the diagonalization. The basis deformation
$\beta_0=0.5$ has been used for all nuclei considered in
this study.

The accuracy of the calculations using this truncation
scheme has been tested against the calculations in an
extended basis with fermionic and bosonic states up to
$15.5\hbar \omega_0^F$ and $16.5\hbar \omega_0^B$,
respectively. This extended basis is very stable with
respect to changes of its parameters $\hbar \omega_0$ and
$\beta_0$.  We have carried out test calculations using
different combinations for  $\hbar \omega_0$ and $\beta_0$.
Specifically, the values  $\hbar \omega_0=41$A$^{-1/3}$
MeV, $\hbar \omega_0=44$A$^{-1/3}$ MeV, $\beta_0=0.5$ and
$\beta_0=0.7$ have been used.  For the lowest SD
configuration of $^{152}$Dy at $\Omega_x=0.8$ MeV, we have
found  the following maximal variations for several
quantities: total energy - 0.02\%, total spin - 0.5\%,
kinematic moment of inertia $J^{(1)}$ - 0.5\%, charge
quadrupole moments $Q_0$ - 0.17\%, mass hexadecupole moment
$Q_{40}$ - 0.11\%. In Table 3 given below we show some 
results based on
this truncation scheme.  The results obtained with this
basis are close to those obtained in Ref.  \cite{AKR.96}.
They are typically within the uncertainty range of the
truncation scheme used in the previous calculations.  It
should be noted, however, that such calculations require
enormously large computation time and a systematic study
becomes practically impossible.
  
The comparison with the calculations in an extended basis
shows that our truncated basis provides reasonable accuracy
and the computation time allows a systematic investigation.
The numerical uncertainty in the calculation of the
effective alignments $i_{eff}$ is typically less than
$0.1\hbar$ dependent on the active orbital. For example,
the numerical uncertainty in the calculation of the
effective alignments $i_{eff}$ of the $\nu[402]5/2(r=\pm
i)$ orbitals is less than $0.05\hbar$, while for some
high-$N$ orbitals it is close to the upper limit. The
accuracy in the relative charge quadrupole moments $\Delta
Q_0$ is of about 0.02 $e$b. That is sufficient for drawing
meaningful physical conclusions.

In the present work we have mainly used the non-linear
parameter set NL1 \cite{RRM.86}. The Lagrangian
parameterizations NL3 and NLSH  have been also employed in
the second part of our study. In Table 1 we list all the
parameter sets used in this work. They have been adjusted
in the literature to bulk properties of several spherical
nuclei and no information about single-particle properties
have been taken into account in their determination.  For
the coupling of the Coulomb field we use the experimental
value $e^2/4\pi\equiv\alpha = 1/137$.

%%%%%%%%%%%%%%%%%%%%%%%%%%%%%%%%%%%%%%%%%%%%%%%%%%%%%
%
\section{The effective alignment approach}
%
%%%%%%%%%%%%%%%%%%%%%%%%%%%%%%%%%%%%%%%%%%%%%%%%%%%%%

One way to identify the single-particle orbital by which
two SD bands differ is to compare the difference of their
experimental dynamic moments of inertia $J^{(2)}$ with the
calculated ones \cite{BRA.88}. The advantage of the
analysis in terms of the dynamic moment of inertia
$J^{(2)}$ is that, no knowledge of the spin values is
required.  It should be noted, however, that the $J^{(2)}$
values are calculated either from the second derivative of the
total energy as a function of the spin:
\begin{eqnarray}
J^{(2)}({\sl\Omega}_x)&=&\left\{\frac{d^2E}{dJ^2}\right\}^{-1}
\label{eq18}
\end{eqnarray}
or using a finite difference approximation for  Eq.
(\ref{eq18}):
\begin{eqnarray}
J^{(2)}(I)&=&\frac{4}{E_\gamma((I+2)\rightarrow I)- 
E_\gamma(I\rightarrow(I-2))}~~~~{\rm MeV^{-1}}
\label{eq21}
\end{eqnarray}
where $E_\gamma(I)$  are the energies of the
$\gamma$-transitions within a band.  This implies that such
a comparison is very sensitive to the accuracy with which
$J^{(2)}$  is described within a certain  model. A  more
important disadvantage of the method based on $J^{(2)}$ is
that the orbitals with nearly constant alignment, as a
function of rotational frequency (or spin), should have only
a small impact on $J^{(2)}$. Indeed, the fractional  change
$(FC)$ in $J^{(2)}$ of two bands $A$ and $B$ (the criterion
most frequently used for selection of identical bands) is
\cite{BHN.95}
\begin{eqnarray}
FC_{B,A}=\left.\frac{J^{(2)}_B-J^{(2)}_A}{J^{(2)}_A}\right|_{\Omega_x}=
\left.\frac{d(I_B-I_A)}{d I_A}\right|_{\Omega_x}=
\left.\frac{di^{B,A}}{dI_A}\right|_{\Omega_x}=
\left.\frac{di^{B,A}}{d\Omega_x}/J^{(2)}_A\right|_{\Omega_x}
\label{FC}
\end{eqnarray}
where $i^{B,A}$ is the relative alignment of the two bands,
i.e. the difference between their spins at constant
$\Omega_x$:
\begin{eqnarray}
i^{B,A}(\Omega_x)=I_B(\Omega_x)-I_A(\Omega_x)
\label{alig}
\end{eqnarray}
Equations  (\ref{FC}) and (\ref{alig}) suggest that two
bands with constant (or nearly constant) relative alignment
as a function of the rotational frequency (or spin) should
have the same (or very similar) dynamic moments of inertia.
In SD bands of the $A\sim 140-150$ mass region, many of the
$N=3, 4, 5$ non-intruder orbitals, sitting close to the
Fermi level, have this property. The rather general
uncertainties connected with the single-particle ordering in
the SD minimum prevent a unique definition of the
underlying configuration in terms of non-intruder orbitals,
if the approach is based only on properties of $J^{(2)}$.
A large difference in the dynamic moments of inertia
$J^{(2)}$ of two bands is expected if only the relative
alignment of the two bands has large variations as a
function of the rotational frequency (or spin). This is
usually the case when two bands differ by a particle in the
high-$N$ intruder orbital ($N=6,7$ for $A\sim 140-150$)
\cite{BRA.88,KRA.97a}.  The density of these  orbitals in
the vicinity of the Fermi level is rather small, thus the
identification of the high $N$-orbital, by which two bands
differ, is expected to be rather realistic \cite{BRA.88}.
However, even in this case, such an analysis does not
reveal how well theory is able to describe the relative
alignment of two bands because the difference in
$J^{(2)}$'s of the two bands is not sensitive to the
absolute value of the relative alignment.

An alternative way to analyze the contributions coming from
specific orbitals and thus to identify the configuration,
is the effective alignment approach  suggested by I.\
Ragnarsson in Ref. \cite{Rag91} and widely used within the
CN model for the analysis of the structure of SD bands
observed in the $A\sim 140-150$
\cite{Rag93,Gd46-50,Tb152a,Gd147c,Eu142a,Rag96} and 
$A\sim 135$ mass regions \cite{AR.96}. Recently, 
it has been also used for the description of relative
properties of smooth terminating bands observed in the
$A\sim 110$ mass region \cite{AR.98}.
The effective alignment is
defined as $i^{B,A}_{eff}=i^{B,A}$ and the notation simply
reflects the fact that the spins of the bands under
consideration are not experimentally determined.
Experimentally, it includes both the alignment of the
single-particle orbital and the effects associated with
changes in deformation, pairing etc. between the two bands.
This approach exploits the  fact that the spin is quantized,
integer for even nuclei and half-integer for odd nuclei. In
addition it is constrained by the signature. Its advantage
is that it allows us to study the absolute value of the
relative alignment of two bands, providing additional
information about alignment properties of the
single-particle orbitals by which two band differ.

It should be noted that this method is not free from some
deficiencies. It requires the knowledge of the absolute
spin values of the SD bands, which, however, are not
experimentally known.  Therefore, some spin value is
usually assumed for one band, typically for the yrast band
of a doubly-magic nucleus, based on model arguments (e.g.
comparison between calculated and 'experimental' kinematic
moments of inertia) or on some experimental estimates.
Then the effective alignment approach is used to establish
the relative spins with respect to this "reference" band
for SD bands in neighboring nuclei. The ability to
establish the relative spins is the clear advantage of the
effective alignment approach compared with the method based
on $J^{(2)}$.  One should note that, once the
configurations and specifically the signatures are fixed,
the relative spins can only be changed in steps of $\pm
2\hbar n$ (n is integer number)  providing thus a
reasonable estimate for these quantities, based on a
comparison between theory and experiment. Only in the case
of signature degenerated bands, the relative spins can be
changed in steps of $\pm 1\hbar n$.  There are two
constraints for the absolute values of $i_{eff}$: a) At
large deformation most of the  single-particle orbitals
have similar $\langle j_x \rangle$ values (usually, in the
range $-2\hbar < \langle j_x \rangle <2\hbar$), and the
deviation comes only from the intruder orbitals
\cite{KRA.97a}, b) Most of the signature degenerated
orbitals should have $\langle j_x \rangle$ in the range
$-1\hbar < \langle j_x \rangle <1\hbar$ because in most
cases large $\langle j_x \rangle$ values imply large
signature splitting between signature partner orbitals.  In
such a way it is possible to map the relative spins for a
large number of SD bands in the whole region. This has been
done, for example, in the framework of the CN model
\cite{Rag93,Gd46-50} for the $A \sim 140-150$ mass region.

It should be also taken into consideration that there are
cases where several orbitals close to the Fermi level have
similar effective alignments. There is no unique way to
distinguish them.  This is overcome, in some cases,
exploiting specific experimental features which allow us to
distinguish amongst several possibilities. For example,
accidental degeneracy in the excitation energy of two
states in bands 2 and 4 of $^{150}$Gd indicates that these
bands have the same parity and signature \cite{Gd150}.
Without this observation, two orbitals namely, $\nu
[402]5/2$ and $\nu [514]9/2$, having rather similar
alignment properties, could be considered equally  possible
candidates for the signature degenerated bands with
$i_{eff}\approx 0\hbar$ observed in nuclei around
$^{152}$Dy.  In the existing interpretation band 2 in
$^{150}$Gd has positive parity.  This suggests that the
$\nu [402]5/2(r=\pm i)$ orbitals should be occupied in such
bands \cite{Tb152a,Gd150}.

The effective alignment approach is not free from
uncertainties related to the single-particle energies in
the SD minimum. However, different theoretical approaches
(CN, cranked Wood-Saxon (CWS), cranked Hartree-Fock (CHF)
with Skyrme forces and CRMF) indicate the same group of
specific orbitals lying in the vicinity of the SD shell
gaps. Therefore, one expects to be able to specify the most
probable orbitals which are occupied or emptied in the SD
bands with the help of the effective alignment approach.
Further test may be done using the different deformation
dependence of single-particle orbitals by considering
a large group of SD nuclei with different equilibrium
deformations.

Precise measurements of charge quadrupole moments could
provide additional information for the involved
single-particle orbitals.  However, an  analysis based on
this physical quantity should be considered complementary
because of the relatively large experimental error-bars. In
addition, SD bands with different structure could have
similar charge quadrupole moments.

Recent investigations of the additivity of the quadrupole
moments, performed using the CHF approach with Skyrme
forces \cite{Satula} and the CN model \cite{KRA.97b} for
the $A\sim 150$ mass region, have shown that the relative
quadrupole moments between two different SD bands can be
written as the sum of the independent contributions of the
single-particle/hole states around the doubly magic SD core
of $^{152}$Dy. A similar principle of the additivity for
effective alignments has been discussed long ago in
Ref. \cite{Rag91}. However, only a restricted number of SD
bands in the $^{146-148}$Gd nuclei have been considered,
which indeed show that the experimental effective
alignments follow a simple rule of additivity \cite{Rag91}.
One should note that a detailed theoretical investigation
of the additivity of the effective alignments is lacking
till now. Similarly to the case of the quadrupole moments,
the problem can be formulated in the following way:
whether the effective alignments $i_{eff}^{AB}$ between
configurations A and B, calculated at their self-consistent
deformations, can be written as a sum of 'independent'
effective alignments $i_{eff}$ calculated from the
single-particle/hole orbitals (by which configurations A
and B differ) around the doubly magic SD core of
$^{152}$Dy: 
\begin{eqnarray}
i_{eff}^{AB} \approx \sum_{orbitals} i_{eff}
\label{addit}
\end{eqnarray}
In eq. (\ref{addit}), the sum runs over the orbitals by
which the configurations A and B differ.  We do not aim to
perform a detailed investigation for the additivity of the
effective alignments considering that similarly to the case
of quadrupole moments \cite{Satula} it should involve the
calculations of a considerable number of SD bands. On the
other hand, general features related to the additivity of
the effective alignments can be understood by considering
some extreme cases. In order to see to which extend the
additivity of the effective alignments is fulfilled we
introduce the quantity $\Delta i_{eff}$ defined as 
\begin{eqnarray}
\Delta i_{eff} = i_{eff}^{AB} - \sum_{orbitals} i_{eff}
\label{addit2}
\end{eqnarray}
The results of the calculations for $\Delta i_{eff}$ are
shown in Fig. 1. In the extreme case which involves the
$\nu[402]5/2(r=\pm i)$ orbitals with an effective alignment
close to $0\hbar$, (see Fig. 3 below) the additivity
of the effective alignments is fulfilled with rather high
accuracy. On the other hand, in other two extreme cases
which involve the $\nu[770](r=-i)$ and the
$\pi[651]3/2(r=\pm i)$ orbitals with very pronounced
rotational frequency dependence of the effective alignments
(see Figs. 2, 3 and 10 below) the
additivity of the effective alignments is broken. The
origin of this feature is connected with the fact that the
alignment properties of these single-particle orbitals are
strongly depended on the deformation (see for example Fig.
2 and Table 4).  In comparison, the effective
alignments of the $\nu[402]5/2(r=\pm i)$ orbitals do not
show a considerable dependence on the deformation (see Fig.
3). Note also that in both cases we have compared
configurations with rather large differences in the
self-consistent deformations. This suggests core
polarization effects.

These results indicate that the effective alignments follow
with reasonable accuracy the rules of additivity only in
cases where the difference between the compared configurations
is connected with the occupation of orbitals having
alignments varying only weakly with rotational frequency 
and deformation. In the SD minimum, these are typically
non-intruder orbitals. As a result, considerable deviations
of the experimental effective alignments, connected with
such orbitals, from the additivity rules of Eq.
(\ref{addit}) could be considered as a fingerprint of
residual interactions not accounted in a pure
single-particle picture. Note, that even in this case it is
reasonable to expect that deviations from the additivity
rules will increase with an increasing number of orbitals,
by which the compared configurations differ. On
the other side it is clear that the effective alignments of
high-$N$ intruder orbitals are less additive than the
effective alignments of non-intruder orbitals.  This is
connected with the fact that the angular momentum
alignments of these orbitals have a pronounced rotational
frequency and deformation dependence. In addition, their
occupation typically has considerable impact on the
self-consistent deformation.

The present paper is the first work where CRMF theory is
used to describe SD bands within the effective alignment
approach. Most of the theoretical studies of SD bands have
employed the approach based on $J^{(2)}$, although the
effective alignments of different SD bands could provide an
additional test on the validity of the models. As discussed
above the effective alignment approach has been  widely
used only within the CN model. In the framework of the CWS
approach, there is only one study for the SD bands of
$^{82,83}$Sr nuclei \cite{Sr82}, where shape polarization
effects are taken into account.  Finally, within the
microscopic non-relativistic effective theories there is
also only one study on the market. The effective alignment
associated with the $\pi [301]1/2$ orbital has been studied
within the CHF approach using various Skyrme forces (SIII,
SkM$^{*}$ and SkP) \cite{DD.95}. It was shown that it was
more difficult to reproduce  the relative alignments of the
pairs of bands  than the identity of the dynamic moments of
inertia. The authors suggested that a readjustment of the
parameters of effective forces in the time-odd channel
seems to be necessary for a better description of the
relative alignments of different bands.

%%%%%%%%%%%%%%%%%%%%%%%%%%%%%%%%%%%%%%%%%%%%%%%%%%%%%%%%%%%
%
\section{Effective alignments in nuclei around $^{152}$Dy}
%
%%%%%%%%%%%%%%%%%%%%%%%%%%%%%%%%%%%%%%%%%%%%%%%%%%%%%%%%%%%

Our starting point is the lowest SD configuration, namely
$\pi 6^4 \nu 7^2 (+,+1)$, of the doubly magic nucleus
$^{152}$Dy. Theoretical approaches within different models
agree that this configuration is indeed the lowest one
\cite{BRA.88,NWJ.89,AKR.96,DD.95,RA.86,BFH.96}.  The  magic
character of $^{152}$Dy is  consistent with the measured
intensities of the excited SD bands, which are smaller than
the corresponding ones in all neighboring nuclei
\cite{Twi.95}.  Adding particles or holes to this
configuration, one can study the single-particle orbitals
which are available in the vicinity of the $N=86$ and
$Z=66$ SD shell gaps by means of the effective alignment
approach \cite{Rag93}. We report and discuss our results
separately for each active orbital. We restrict ourselves
mainly to configurations calculated in Ref. \cite{AKR.96}.

Linking transitions from SD to normal-deformed (or
spherical) states of Dy, Tb and Gd nuclei have not yet been
observed.  Therefore, no direct information about the
absolute spin values of the SD bands is available. For some
SD bands, however, there are experimental estimates  of
their values with typical uncertainty  $1-2~\hbar$
\cite{BFS.96}.  In our analysis, we have selected  the spin
of the lowest observed state in band 1 of $^{152}$Dy to be
$I_0=24^+$. It corresponds to the $\gamma$-transition $26^+
\rightarrow 24^+$ with energy 602.4 keV.  The reason for
this choice is that the spins, which have been assigned to
the SD bands in the Dy, Tb and Gd  nuclei using the
effective alignment approach (see Table 2), should be, on
the average, within the uncertainties of the available
experimental estimates \cite{BFS.96}.  One should note,
however, that the comparison with the calculated kinematic
moment of inertia $J^{(1)}$ favors a higher spin value for
the lowest observed state in band 1 of $^{152}$Dy
\cite{AKR.96}.  This might be connected with pairing, which
has been ignored in our study. Similar observations are
made in CWS calculations as well as in CHF calculations
with Skyrme forces \cite{EDD.97}, where the value
$I_0=26~\hbar$ is obtained, if pairing is omitted.  The
inclusion of pairing in the CWS approach lowers the spin
value down to $I_0=24\hbar$ and leads to a better agreement
with experimental estimates.  Further investigation  within
the CRMF theory including pairing is necessary to clarify the
nature of this deviation from the empirical estimate.

%%%%%%%%%%%%%%%%%%%%%%%%%%%%%%%%%%%%%%%%%%%%%%
\subsection{The $\nu[770]1/2(r=-i)$ orbital:} 
%%%%%%%%%%%%%%%%%%%%%%%%%%%%%%%%%%%%%%%%%%%%%%

In  Fig. 2d the calculated effective alignments of the
$^{151}$Dy(1)/$^{152}$Dy(1), $^{150}$Tb(1)/$^{151}$Tb(1)
and $^{149}$Gd(1)/$^{150}$Gd(1) pairs are compared with
those extracted from experiment. They reflect the angular
momentum alignment of the $\nu[770]1/2(r=-i)$ orbital. It
is seen that the observed effective alignments are
overestimated in CRMF calculations. For example, even at
the highest observed rotational frequencies, where the role
of pairing is negligible, the CRMF results show an
overestimate which is $\approx 0.8 \hbar$ for
$^{151}$Dy(1)/$^{152}$Dy(1), $\approx 0.4 \hbar$ for
$^{150}$Tb(1)/$^{151}$Tb(1) and $\approx 0.5 \hbar$ for
$^{149}$Gd(1)/$^{150}$Gd(1).  We recall in passing, that
for this orbital the CN calculations underestimate the
experimental effective alignments (see Fig. 3 in Ref.
\cite{Rag93}). For the $^{150}$Tb(1)/$^{151}$Tb(1) and
$^{149}$Gd(1)/$^{150}$Gd(1) pairs the deviation from
experiment decreases on going to higher rotational
frequencies. The  rather large discrepancy at low frequency
is most likely connected with the influence of pairing
correlations.  The effect is especially pronounced for
$^{149}$Gd(1)/$^{150}$Gd(1) and should reflect the large
variations of $J^{(2)}$ in $^{150}$Gd(1) band.  This
variation was explained within the CWS approach in terms of
consecutive neutron and proton alignments \cite{NWJ.89}.
The discrepancy at low frequencies is smaller for the
$^{150}$Tb(1)/$^{151}$Tb(1) pair, which suggests that the
influence of pairing is weaker in these bands. This is also
consistent with the observed features of the dynamic
moments of inertia.
 
In contrast to the other two pairs, the discrepancy between
theory and experiment, for $^{151}$Dy(1)/$^{152}$Dy(1),
increases with the increase of the rotational frequency.
This feature could be correlated with the behavior of
$J^{(2)}$ in band 1 of  $^{151}$Dy, which also increases
with the frequency. It should be noted that our previous
calculations have not been able to reproduce this increase
in $J^{(2)}$ \cite{AKR.96}.  It is also seen that at the
highest observed point the experimental effective alignment
in $^{151}$Dy(1)/$^{152}$Dy(1) is lower by $\sim 0.5 \hbar$
compared with the two other pairs.

%%%%%%%%%%%%%%%%%%%%%%%%%%%%%%%%%%%%%%%%%%%
\subsection{The $\pi[651]3/2(r=+i)$ orbital:} 
%%%%%%%%%%%%%%%%%%%%%%%%%%%%%%%%%%%%%%%%%%%

Based on CN calculations it was suggested in Ref.
\cite{Rag93}, that the difference in the configurations of
the $^{150}$Tb(1)/$^{151}$Dy(1) and
$^{151}$Tb(1)/$^{152}$Dy(1) pairs is related to the proton
hole in the $\pi [651]3/2(r=+i)$ orbital.  The same
conclusion holds also within CRMF theory \cite{AKR.96}.  
In Fig. 2a the calculated effective alignments are compared
with experiment. It is observed that at high rotational
frequencies $(\Omega_x>0.5$ MeV), where pairing is
essentially gone, data is reasonably reproduced.  For the
$^{150}$Tb(1)/$^{151}$Dy(1) pair the agreement is
excellent, while for $^{151}$Tb(1)/$^{152}$Dy(1), the CRMF
theory systematically overestimates the observed effective
alignment by $\approx 0.3\hbar$ at $\Omega_x>0.5$ MeV.
Compared, however, with the CN results the CRMF
calculations are in much better agreement with experiment.
The CN calculations at the highest observed frequencies
show large discrepancies with experiment ($\approx
0.9~\hbar$ for $^{150}$Tb(1)/$^{151}$Dy(1) and  $\approx
0.5~\hbar$ for $^{151}$Tb(1)/$^{152}$Dy(1) (Fig. 3 in
Ref. \cite{Rag93}). The deviations increase with decreasing
frequencies.  Our calculations suggest that the spins of
the $^{150}$Tb(1) and $^{151}$Tb(1) bands relative to the
$^{152}$Dy(1) band should be increased by $2\hbar$ compared
with the CN results (see Table 2).  This change will affect
the relative spins of the compared SD bands only in  cases
where only one of them has the proton hole in the
$\pi[651]3/2(r=+i)$ orbital.

%%%%%%%%%%%%%%%%%%%%%%%%%%%%%%%%%%%%%%%%%%%
\subsection{The $\pi[651]3/2(r=-i)$ orbital:} 
%%%%%%%%%%%%%%%%%%%%%%%%%%%%%%%%%%%%%%%%%%%

The yrast bands of $^{149,150}$Gd isotopes have two proton
holes in the proton core of the doubly magic $^{152}$Dy
nucleus. The results of the CRMF theory \cite{AKR.96} and
those of the  CN model \cite{Rag93} suggest that these
holes are in the $\pi [651]3/2(r=\pm i)$ orbitals.
Analyzing the $^{150}$Gd(1)/$^{151}$Tb(1) and
$^{149}$Gd(1)/$^{150}$Tb(1) pairs, we can study the
effective alignment of the $\pi[651]3/2(r=-i)$ orbital
because compared bands have the same neutron configuration.
Our results are shown in Fig. 2b. It is seen, that
similarly to the $\pi[651]3/2(r=+i)$ orbital, the
discrepancy between theory and experiment decreases with
the increase of the rotational frequency.  The agreement
with experiment at high frequencies is much better within
CRMF theory than in the CN model (see also Fig. 3 of Ref.
\cite{Rag93}).

Another example is the effective alignment in the  
$^{150}$Tb(2)/$^{151}$Dy(1) pair (see Fig. 3a) 
It was discussed in Refs. \cite{AKR.96,Dy151a} that the
difference in the configurations of these two bands is
connected with the $\pi [651]3/2(r=-i)$ orbital. For both
bands the dynamic moments of inertia $J^{(2)}$ increase
slightly with the increase of the rotational frequency, a
feature not reproduced in our previous CRMF calculations
(see Figs. 10,12 of Ref. \cite{AKR.96}).  The increase of
$J^{(2)}$ is not well understood and one of the
possibilities is that pairing still persists even at high
frequencies.  If this is the case, then the similarity of
$J^{(2)}$ of these bands suggests that pairing  effects are
quite similar for both bands. Indeed, better agreement
between experiment and calculations is obtained in this
case, compared with the $^{150}$Gd(1)/$^{151}$Tb(1) and
$^{149}$Gd(1)/$^{150}$Tb(1) pairs (see Figs. 2b and
3a).

%%%%%%%%%%%%%%%%%%%%%%%%%%%%%%%%%%%%%%%%%%%%%%%%%%%%%%%%%%%%%%%%%%
\subsection{The $\pi[651]3/2(r=+i) + \pi[651]3/2(r=-i) $ orbitals:} 
%%%%%%%%%%%%%%%%%%%%%%%%%%%%%%%%%%%%%%%%%%%%%%%%%%%%%%%%%%%%%%%%%%

The comparison with experiment of the calculated total
effective alignments of the $\pi[651]3/2(r=+i)$ and
the $\pi[651]3/2(r=-i)$ orbitals is shown in Fig. 2c.
One finds rather good agreement with experiment for
the $^{149}$Gd(1)/$^{151}$Dy(1) pair at $\Omega_x>0.5$ MeV
and for the $^{150}$Gd(1)/$^{152}$Dy(1) pair at
$\Omega_x>0.6$ MeV. With decreasing rotational frequency,
the discrepancy between theory and experiment increases.
This could be partially connected with the absence of
pairing in our calculations. Indeed, in all cases, where
one of the $\pi[651]3/2(r=\pm i)$ orbitals or their
combinations were involved, the largest discrepancies
between experiment and calculations at low rotational
frequencies were observed when the comparison was made
between band 1 in $^{150}$Gd and another band. One should
note that this band shows one of the largest variations in
dynamic moment of inertia observed in the $A \sim 150$ mass
region.  This was explained in the CWS approach in terms of
consecutive neutron and proton alignments \cite{NWJ.89}. At
low frequencies a smaller discrepancy between theory and
experiment exists if the $^{149}$Gd(1) band is compared
with another band (see Fig. 2). Since a paired neutron
band crossing is blocked in the configuration assigned to
this band (conf. $\pi 6^2 \nu 7^1(-,+i)$, see Ref.
\cite{AKR.96}), the discrepancy most likely arises from
pairing correlations in the proton subsystem. This is
consistent with the interpretation given in Ref.
\cite{NWJ.89}.  Even smaller discrepancies at low
rotational frequencies are observed when yrast bands in
$^{150,151}$Tb are compared with yrast bands in
$^{151,152}$Dy. In these bands the paired proton band
crossing is blocked because the assigned configurations are
$\pi 6^3 \nu 7^2(+,-i)$ ($^{151}$Tb(1)) and $\pi 6^3 \nu
7^1(-,+1)$ ($^{150}$Tb(1)), (see Ref. \cite{AKR.96}). On
the other hand, they are different in the sense that paired
neutron band crossing is allowed in the $^{151}$Tb(1) band
and it is blocked in the $^{150}$Tb(1) band. The fact that
no large deviations from experiment have been calculated at
low rotational frequencies for the
$^{150}$Tb(1)/$^{151}$Dy(1) and $^{151}$Tb(1)/$^{152}$Dy(1)
pairs, compared with those involving $^{150}$Gd(1),
indicates that the large self-consistent deformation of the
$^{151}$Tb(1) band is crucial in quenching the neutron
pairing correlations.

One should note, however, that the discrepancy between
experiment and calculations might be partly connected
either with the correctness of the description of the
alignment properties of the $\nu [651]3/2(r=\pm i)$
orbitals or with the correctness of the self-consistent
deformations of the studied configurations. This is also
corroborated by the fact that there are still some
discrepancies between experiment and calculations at
the highest observed frequencies, where the role of
pairing correlations is expected to be rather small.

%%%%%%%%%%%%%%%%%%%%%%%%%%%%%%%%%%%%%%%%%%%%%%%%%
\subsection{The $\nu [402]5/2(r=\pm i)$ orbitals:}
%%%%%%%%%%%%%%%%%%%%%%%%%%%%%%%%%%%%%%%%%%%%%%%%%

The effective alignments of bands 2 and 3 in $^{153}$Dy
relative to the yrast band in $^{152}$Dy are close to zero.
The CRMF results for the configurations in which the 87th
neutron is placed either in the $\nu[402]5/2(r=-i)$ orbital
(band 3) or in the $\nu[402]5/2(r=+i)$ orbital (band 2) are
very close to experiment (Fig. 3c). Another example is the
pair of signature partner bands in $^{152}$Tb which are formed
by placing the 87th neutron either in the $\nu[402]5/2(r=-i)$
(band 1) or in the $\nu[402]5/2(r=+i)$ (band 2) orbital (see
Ref. \cite{AKR.96}). The agreement between calculations and
experiment is rather good (see Fig. 3b). It should be noted
that the interpretation of these bands is in agreement with
that given in Ref. \cite{Tb152a} based on the CN model. In
addition, the calculated values of the effective alignment
associated with the $\nu [402]5/2(r=\pm i)$ orbitals are
rather similar to those obtained in the CN model, (see Fig.
3 of \cite{Tb152a} and Fig. 4 of \cite{Rag93}).

Using an effective alignment approach within the CN model,
it was suggested \cite{Tb152a} that bands 4 and 5 in
$^{152}$Dy are based on the promotion of the last neutron
from the second $N=7$ orbital below the $N=86$ SD shell gap
into the $\nu [402]5/2(r=\pm i)$ orbitals. A similar
scenario was discussed within the CRMF approach
\cite{AKR.96} using the properties of the dynamic moment of
inertia $J^{(2)}$. The present calculations confirm this
interpretation because of the very good agreement between
calculations and experiment (Fig. 3d). According to the
present analysis, the $\nu [402]5/2(r=-i)$ orbital is
occupied in band 5, while the $\nu [402]5/2(r=+i)$ orbital
is occupied in band 4.

Let us consider band 1 in $^{154}$Dy as another example.
Based on the similarities of the observed dynamic moments
of inertia $J^{(2)}$ in $^{152}$Dy(1) and  $^{154}$Dy(1),
it was suggested \cite{Dy154a} that this band has the same
high-$N$ configuration as the yrast band in $^{152}$Dy,
namely a $\pi 6^4 \nu 7^2$ configuration. CRMF calculations
\cite{AKR.96} support this configuration assignment. The
similarity of the dynamic moments of inertia $J^{(2)}$ of
these two bands was reproduced rather well assuming that
two neutrons above the $N=86$ SD shell gap sit in the $\nu
[402]5/2$ orbital. Indeed, the effective alignment in the
$^{153}$Dy(1)/$^{154}$Dy(1) pair, (Fig. 5b), reveals that a
third $N=7$ neutron is not occupied in band 1 of
$^{154}$Dy. This is because the drop in the effective
alignment of the $^{153}$Dy(1)/$^{154}$Dy(1) pair, (Fig.
5b), approximately corresponds to an increase of the
effective alignment in the pair $^{152}$Dy(1)/$^{153}$Dy(1)
connected with the occupation of a third $N=7$ orbital,
(see subsection 4.6. below).  Moreover, the effective
alignment in the $^{152}$Dy(1)/$^{154}$Dy(1) pair is nearly
constant for frequencies $0.35-0.65$ MeV and equal to
$\approx -0.5\hbar$ for the spin assignment given in Table
2. The increase in $i_{eff}$ above 0.65 MeV is connected
with an increase of $J^{(2)}$ in $^{154}$Dy(1), (see Fig. 9
in Ref. \cite{AKR.96}). The last feature is not well understood.
In all, the analysis indicates that the $^{154}$Dy(1) band
has the same configuration as the $^{152}$Dy(1) band in
terms of the occupation of high-$N$ intruder orbitals.

The comparison of the experimental effective alignments of
the $^{152}$Dy(1)/$^{154}$Dy(1) and
$^{153}$Dy(1,2)/$^{154}$Dy(1) pairs with the calculated
ones shows a systematic discrepancy of about  $0.5\hbar$
(Figs. 5d and 5b).  One should mention that the calculated
effective alignments of the configurations with one or both
occupied $\nu [402]5/2(r=\pm i)$ orbitals follow the
pattern of alignment of these orbitals, seen in the
$^{151}$Tb(1)/$^{152}$Tb(1,2),
$^{151}$Dy(1)/$^{152}$Dy(4,5) and
$^{152}$Dy(1)/$^{153}$Dy(2,3) pairs (see Fig. 3).
However, the experimental effective alignment for the
$^{153}$Dy(2)/$^{154}$Dy(1) pair deviates significantly
from the pattern of the effective alignment expected for
the occupation of the $\nu[402]5/2(r=-i)$ orbital, (see
Fig. 1 in Ref. \cite{Tb152a} and Fig. 3 in the present
work). Similar situation holds also for the
$^{152}$Dy(1)/$^{154}$Dy(1) pair.

Let us now consider alternative possibilities for the
interpretation of band 1 in $^{154}$Dy.  It seems that the
options connected with non-pairwise occupation of different
orbitals lying above the $N=86$ SD shell gap can be ruled
out because in such a case it is reasonable to expect the
observation of several SD bands. The remaining
possibilities are associated with pairwise occupation of
the orbitals lying above the $N=86$ SD shell gap.  For
example, one of them would be the occupation of the
orbitals occupied by the 87th neutron in the $^{153}$Dy(4,5)
bands. In Fig. 5 it is seen that they are signature
degenerated with nearly constant effective alignment
$i_{eff}=(\sim 0.35\hbar)\pm 1\hbar {\rm n}$ (n is integer)
with respect to the $^{152}$Dy(1) band.  Such a scenario,
however, will give $i_{eff}=(\sim 0.7\hbar)\pm 2\hbar {\rm
n}$ if the additivity principle is fulfilled with high
accuracy.  This is again in contradiction to the
experimental effective alignment $i_{eff}=(\sim
-0.4\hbar)\pm 2\hbar {\rm n}$ observed in the
$^{152}$Dy(1)/$^{154}$Dy(1) pair (see Fig. 5).

One should note that similar to the bands 7 and 8 observed
in $^{151}$Tb and discussed in detail within the CRMF
theory in Ref. \cite{ALR.97}, we have not found a
consistent interpretation for the bands 4 and 5 observed in
$^{153}$Dy within a pure single-particle picture. Note that
we have restricted ourselves to several orbitals lowest in
energy located above the $N=86$ SD shell gap shown, for
example, in Fig. 4. The similarity of the
effective alignments of these bands relative to the yrast
bands in nuclei with one less neutron, (Fig. 5a), suggests
that the same neutron orbitals above the $N=86$ SD shell
gap are occupied in the $^{151}$Tb(7,8) and $^{153}$Dy(4,5)
bands. However, according to the CRMF calculations it is
not likely the $\nu [521]3/2(r=\pm i)$ orbitals, lying above
the $\nu [402]5/2(r=\pm i)$ orbitals, (see Fig. 4), to be
occupied in these bands. In the case of the $^{151}$Tb(7,8)
bands, a strong argument against such an interpretation
comes from the large signature splitting between the two
signatures of the $\nu [521]3/2$ orbital. The signature
splitting is rather similar both in the CRMF theory and in
the CN model and contradicts to the experimental data
\cite{ALR.97}.  This is also the case in the
$^{153}$Dy(4,5) bands. Here, however, the signature
splitting between the configurations with the $\nu
[521]3/2(r=\pm i)$ orbitals occupied is smaller.  This is
connected to a large admixture of the $N=7$ states in the
wave function of the $\nu [521]3/2(r=+ i)$ orbital. On the
other hand, the calculated slope of the effective alignment
of these configurations relative to the yrast band
configuration in $^{152}$Dy is in large disagreement with
the one observed in the $^{152}$Dy(1)/$^{153}$Dy(4,5)
pairs. One should note that the next orbitals above the
$N=86$ SD shell gap, namely $\nu [514]9/2(r=\pm i)$, are
signature degenerated and the configurations based on them
have nearly constant effective alignment (essentially close
to zero) relative to the $^{152}$Dy(1) band configuration.

We take also into account that the difficulty of a
consistent interpretation of the $^{154}$Dy(1) and
$^{153}$Dy(4,5) bands might be connected with the influence
of residual interactions, neglected in the present
approach.  Concerning the $^{154}$Dy(1) band, we still
consider pairwise occupation of the $\nu [402]5/2$ orbitals
as the most probable interpretation. The alternative
possibilities discussed above are less probable. The
preference to this assignment is also based on the
systematics of the configuration assignments for the bands
in the $A\sim 150$ mass region which have nearly constant
effective alignment $i_{eff}$ with respect to the reference
band. Observations of excited bands in $^{154}$Dy would
have been highly desirable for the resolution of the
existing inconsistency in the interpretation of this band.
     
In view of the problems with the interpretation of band 1
in $^{154}$Dy it is worth for better understanding to use
recent experimental data for band 1 in $^{155}$Dy
\cite{Dy155a}.  The experimental alignments in the
$^{153}$Dy(1)/$^{155}$Dy(1), $^{152}$Dy(1)/$^{155}$Dy(1)
and $^{154}$Dy(1)/$^{155}$Dy(1) pairs, (see Fig. 5),
and the similarity of the dynamic moments of inertia
$J^{(2)}$ of the $^{155}$Dy(1) and $^{153}$Dy(1) bands
\cite{Dy155a} strongly suggest the $\pi 6^4 \nu 7^3$
configuration for this band. The effective alignment in the
pair $^{153}$Dy(1)/$^{155}$Dy(1) is nearly constant with
$i_{eff} \approx -0.35 \hbar$. Using similar arguments, as
in the case of band 1 in $^{154}$Dy, one could  show that a
more or less uncontradictory picture would  emerge if only
two neutrons are located in the $\nu [402]5/2(r=\pm i)$
orbitals.  Although some discrepancy between calculations
and experiment still remains (see Fig. 5c), as in
the case of band 1 in $^{154}$Dy, other options for the
occupation of the orbitals above the $N=86$ SD shell gap
seem to be less probable.

Our analysis indicates that it is rather difficult to find
a consistent explanation for the deviations of the
experimental effective alignments in the $^{154}$Dy(1) and
$^{155}$Dy(1) bands from the systematics observed in
lighter nuclei.  A similar situation is also observed in
the $^{144}$Gd and $^{145}$Tb nuclei. The details are given
in section 5.  It is also noted that in this case the
orbitals with $i_{eff} \approx 0 \hbar$, namely $\pi
[404]9/2$, are most likely involved. Both cases suggest
that the additivity of effective alignments is broken at
experimental level.

%%%%%%%%%%%%%%%%%%%%%%%%%%%%%%%%%%%%%%%%%%%%%%%%%
\subsection{The $\nu [761]3/2(r=+i)$ orbital:}
%%%%%%%%%%%%%%%%%%%%%%%%%%%%%%%%%%%%%%%%%%%%%%%%%

The yrast band in $^{153}$Dy is built by the occupation of
the third $N=7$ neutron intruder orbital relative to the
yrast band in $^{152}$Dy \cite{NWJ.89,AKR.96,Dy153a}. This
leads to an effective alignment which increases by $\sim
1.5~\hbar$ in the rotational frequency range
$\Omega_x=0.35-0.75$ MeV. The CRMF calculations for this
configuration are in good agreement with experiment (see
Fig. 5a). It is also seen in Fig. 5c that the effective
alignment in the $^{154}$Dy(1)/$^{155}$Dy(1) pair,
connected with this orbital, is reproduced reasonably well.
This also confirms that the same non-intruder orbitals
above the $N=86$ SD shell gap are occupied in these two
bands. Moreover, in this specific case the uncertainty
related to the description of their alignment properties is
eliminated by such a comparison.

%%%%%%%%%%%%%%%%%%%%%%%%%%%%%%%%%%%%%%%%%
\subsection{The nucleus $^{148}$Gd} 
%%%%%%%%%%%%%%%%%%%%%%%%%%%%%%%%%%%%%%%%%

In our previous investigation \cite{AKR.96} we have
discussed in detail the difficulty to find a consistent
interpretation of the yrast and first excited bands in the
chain of $^{146-148}$Gd isotopes. This difficulty is
connected with the relative positions of the $\nu[642]5/2$
and $\nu[651]1/2$ orbitals at self-consistent deformations
of interest. The interpretation based on the CN model
\cite{Rag91,Gd46-50} associates the jumps in $J^{(2)}$
observed in the $^{148}$Gd(2), $^{147}$Gd(1) and
$^{146}$Gd(1) bands with the crossing of the $r=+i$
signatures of the $\nu [651]1/2$ and the $\nu [642]5/2$
orbitals. A situation like this could appear only if the
$\nu [651]1/2$ orbital were higher in energy than the $\nu
[642]5/2$ orbital at $\Omega_x=0.0$ MeV. In the CRMF
calculations, however, using the parameter set NL1 we find
that the relative  positions of these two orbitals are
opposite to those found in the CN calculations (see Fig. 18
of Ref. \cite{AKR.96}).
  
It should be noted, however, that this difficulty does not
forbid the identification of the underlying configuration
by means of the effective alignment approach since in this
case we are dealing with the alignment properties of the
single-particle orbitals and not with their energies in the
vicinity of the Fermi level.  For the sake of clarification
we discuss here, as an example, the two lowest SD
configurations calculated for $^{148}$Gd in Ref.
\cite{AKR.96}. The effective alignments of the yrast
configuration of $^{149}$Gd (conf. $\pi 6^2 \nu 7^1
(-,+i)$) relative to the $\pi 6^2 \nu 7^1 (-,\pm 1)$
configurations of $^{148}$Gd  reflect the alignments of the
$\nu [642]5/2(r=\pm i)$ orbitals.  The comparison between
experiment and calculations, (see Fig. 6a), suggests that
one can attribute the $\pi 6^2 \nu 7^1 (-,-1)$
configuration to band 5 and the $\pi 6^2 \nu 7^1 (-,+1)$
configuration to band 2 above the crossing. Such an
interpretation of band 5 is in accord with the analyses of
Refs. \cite{Gd147c,Gd148d}.  In the first case  we have an
excellent agreement with experiment, while in the second
only above the crossing reasonable agreement is observed.
This is in keeping with the interpretation of Ref.
\cite{Gd46-50} where the effective alignment in the
$^{148}$Gd(2)/$^{149}$Gd(1) pair is connected with the
alignment of the $\nu [651]1/2(r=+i)$ orbital before
crossing in the $^{148}$Gd(2) band and with the alignment
of the $\nu [642]5/2(r=+i)$ orbital after crossing. Since
there is no crossing between these orbitals in the CRMF
theory, only the last part is in agreement with experiment.
It is noted that for this configuration assignment, the
calculated values of the dynamic moment of inertia
$J^{(2)}$ are in rather good agreement with experiment for
bands 2 (above the crossing) and 5 of $^{148}$Gd (see Fig.
6b).

The above discussion confirms the crucial role of the
relative position of the $\nu [651]1/2$  and $\nu [642]5/2$
orbitals for a quantitative understanding of the properties
of some bands in the $^{146-148}$Gd isotopes.  Additional
confirmation comes from the fact that by considering only
two lowest SD solutions in $^{148}$Gd we have not been able
to find a relevant configuration for band 1 of $^{148}$Gd
(Fig. 6a).  The interpretation suggested by the CN model
\cite{Rag91,Gd46-50} connects this band with the hole in
the $\nu [651]1/2(r=-i)$ orbital. In the CRMF calculations
this orbital is below the $\nu [642]5/2(r=\pm i)$ orbitals,
(see, for example, bottom panel of Fig. 18 in Ref.
\cite{AKR.96}), i.e. one can find a theoretical counterpart
for band 1 only if more excited neutron configurations were
considered.  For this however, additional calculations are
necessary and a more detailed investigation of the SD bands
in the Gd region, which could be the subject of a future
study.  One should mention that this interpretation of
bands 1 and 2 has been questioned in Ref. \cite{Gd148d}.
The authors of this article, however, were not able to
provide any explanation for the jump in $J^{(2)}$ of band 2
and therefore we prefer the interpretation given above.

%%%%%%%%%%%%%%%%%%%%%%%%%%%%%%%%%%%%%%%%%%%%%%%%%%%%%%%%%%
\section{Effective alignments in nuclei around $^{143}$Eu}
%%%%%%%%%%%%%%%%%%%%%%%%%%%%%%%%%%%%%%%%%%%%%%%%%%%%%%%%%

Our starting point for nuclei in the neighborhood of
$^{143}$Eu is the yrast $\pi 6^1 \nu 6^4 7^0 (+,-i)$
configuration in $^{143}$Eu.  This is the lowest SD
configuration in the CWS model \cite{NWJ.89} and in CRMF
theory \cite{AKR.96}. A few experiments have been focused
on the linking of the observed SD band in $^{143}$Eu to the
low-spin level scheme but so far no definite success has
been achieved. A spin value $\frac{35}{2}\hbar$ and as an
alternative possibility the value $\frac{37}{2}\hbar$ have
been suggested for the lowest observed SD state in one
experiment \cite{Eu143a}.  The analysis of another
experiment \cite{Eu143c} indicates that possible spin
values for the lowest observed SD state are in the range
from $\frac{33}{2}\hbar$ up to $\frac{37}{2}\hbar$.  Based
on the comparison between calculated and experimental
kinematic moments of inertia (see Fig. 22 in Ref.
\cite{AKR.96}) the spin value of $\frac{37}{2}^+$ is most
favored for the lowest state observed in the yrast SD band
of $^{143}$Eu. However, one should note that the CWS
calculations with pairing correlations \cite{Gd144new}
favor the spin value $\frac{33}{2}^+$ for this state. This
reminds the situation seen in the case of $^{152}$Dy where
the inclusion of pairing lowers the spins within the yrast
band by  $2\hbar$ (see section 4). The analysis of the
effective alignments in the $^{143}$Eu(1)/$^{149}$Gd(1)
pair indeed favors this spin assignment (see Fig. 7).
Larger than typical discrepancy between experimental and
calculated effective alignments could be expected
considering that these nuclei differ by 1 proton and 5
neutrons. As a result, we adopt the spin value of
$\frac{33}{2}^+$ for the lowest state in the yrast SD band
of $^{143}$Eu (see Table 2.) One should note that the
available experimental estimates of the initial spin values
for the bands discussed below are not conclusive enough to
exclude one of the above mentioned ($\frac{33}{2}^+$ or
$\frac{37}{2}^+$) possibilities which agree with the
signature of the configuration assigned to the
$^{143}$Eu(1) band.

%%%%%%%%%%%%%%%%%%%%%%%%%%%%%%%%%%%%%%%%
\subsection{\bf The nucleus $^{144}$Gd } 
%%%%%%%%%%%%%%%%%%%%%%%%%%%%%%%%%%%%%%%%

Our previous CRMF calculations \cite{AKR.96} suggested that
the yrast band in the $^{144}$Gd nucleus could be
associated with the $\pi 6^2 \nu 7^0 (+,+1)$ configuration,
where the 64th proton is located into the $\pi
[660]1/2(r=+i)$ orbital.  The analysis of the effective
alignment in the $^{143}$Eu(1)/$^{144}$Gd(yrast) pair (Fig.
8) supports the validity of this configuration assignment.
Although the calculations overestimate the effective
alignment by $\approx 0.4\hbar$ at $\Omega_x\sim 0.55$ MeV
and by $\approx 0.2\hbar$ at the highest observed
frequency, it is impossible to find an alternative proton
orbital above the $Z=63$ SD shell gap with similar
alignment properties. Band 1 in $^{144}$Gd undergoes
the paired proton band crossing and this leads to a large
discrepancy between theory and experiment at $\Omega_x<0.5$
MeV. As the pairing correlations are neglected, we are not
able to reproduce the large drop in the effective alignment
observed in these frequencies.

In our analysis of Ref. \cite{AKR.96}, using the relative
properties of the dynamic moments of inertia $J^{(2)}$ of
the excited SD bands with respect to the yrast SD band, it
was found that the excited bands 1 and 2 as well as the
excited bands 3 and 4 (both pairs being signature partners)
could be based on configurations in which the 64th proton
sits either in $\pi [404]9/2(r=\pm i)$ or in $\pi
[413]5/2(r=\pm i)$. We should also note that it was not
possible to attribute  definite configurations to the
observed excited bands using only the properties of their
dynamic moments of inertia if the 64th proton were
considered in one of these orbitals. This is due to the
rather large similarities in $J^{(2)}$'s of the excited
bands (see Fig. 21 of \cite{AKR.96}). The present analysis,
based on the effective alignment approach, indicates that
the excited band 3 in $^{144}$Gd most likely has the 64th
proton in $\pi [413]5/2(r=-i)$, (see Fig. 8).  It
seems reasonable to associate the excited band 4 in
$^{144}$Gd with a configuration in which the 64th proton is
placed into $\pi [413]5/2(r=+i)$ because the calculated
signature splitting of the $\pi [413]5/2(r=\pm i)$ orbitals
is close to zero (Fig. 14 in Ref. \cite{AKR.96}). For such a
configuration assignment,  the effective alignments of
these two bands relative to the yrast band in $^{143}$Eu
are perfectly reproduced (see Fig. 8).

The effective alignments of the excited bands 1 and 2
relative to yrast band in $^{143}$Eu are close to zero
(see Fig. 8).  Moreover, the experimental effective
alignments of these bands indicate no signature splitting.
The only orbital with such properties, located above the
$Z=63$ SD shell gap, is $\pi [404]9/2$, (see Fig. 14 in
Ref. \cite{AKR.96}).  We have not made calculations for
configurations of $^{144}$Gd in which the 64th proton is in
$\pi [404]9/2(r=+ i)$ or in $\pi [404]9/2(r=-i)$ because in
self-consistent calculations it is sometimes rather
difficult to distinguish orbitals belonging to the same
shell. However, the results for $^{145}$Tb (see subsection
5.2) indicate that the total effective alignment of the
$\pi [404]9/2(r=+i)$ and $\pi [404]9/2(r=-i)$ orbitals is
close to zero. These orbitals are signature degenerated,
therefore each of them will also have an effective
alignment close to zero. As a result, one could associate
the excited band 2 with a configuration in which the 64th
proton is located in the $\pi [404]9/2(r=-i)$ orbital,
while in the excited band 1 it is located in the $\pi
[404]9/2(r=+i)$ orbital.

%%%%%%%%%%%%%%%%%%%%%%%%%%%%%%%%%%%%%%%%%%%%%
\subsection{\bf The nucleus $^{145}$Tb} 
%%%%%%%%%%%%%%%%%%%%%%%%%%%%%%%%%%%%%%%%%%%%%

The configuration $\pi 6^1 \nu 7^0 (+,-i)$, with two
protons above the Z=63 SD shell gap sitting in the $\pi
[404]9/2(r=\pm i)$ orbitals, has been considered as the
most likely candidate for the observed yrast SD band of
$^{145}$Tb in our previous CRMF calculations \cite{AKR.96}.
Indeed, the second $N=6$ proton is not occupied in the
yrast SD band because the experimental effective alignment
in the pair $^{144}$Gd(yrast)/$^{145}$Tb(1) is strongly
down-sloping as a function of the rotational frequency
(Fig. 8).  The calculated effective alignment between the
corresponding configurations is in rather good agreement
with experiment at $\Omega_x\geq 0.5$ MeV. At lower
frequencies, the large discrepancy is connected with the
paired proton band crossing observed in the yrast band of
$^{144}$Gd. The effective alignment of the $\pi 6^1 \nu 7^0
(+,-i)$ configuration with respect to the configuration
assigned to the yrast band of $^{143}$Eu originates from
the alignment of two protons in $\pi [404]9/2$ and it is
close to zero. The experimental effective alignment in the
$^{143}$Eu(1)/$^{145}$Tb(1) pair is nearly constant
($i_{eff}\approx -0.2\hbar$) at $\Omega_x \geq 0.45$ MeV
and close to the calculated one.

One should note that the effective alignments in the
$^{144}$Gd(Exc.band 1)/$^{145}$Tb(1) and
$^{144}$Gd(Exc.band 2)/$^{145}$Tb(1) pairs (see Fig. 9)
show a stronger deviation from the expected $i_{eff}
\approx 0~\hbar$ compared with the
$^{143}$Eu(1)/$^{145}$Tb(1) pair (see Fig. 8). The
experimental effective alignment $i_{eff} \approx
+0.2~\hbar$ extracted from the
$^{143}$Eu(1)/$^{144}$Gd(Exc.  bands 1 or 2) pairs reflects
the alignment of the $\pi [404]9/2$ orbitals (see
subsection 5.1). Assuming no change in the pairing field
and using the additivity of effective alignments, the
expected value of effective alignment of the yrast band in
$^{145}$Tb relative to the yrast band in $^{143}$Eu is
found to be $i_{eff}\approx +0.4\hbar$.  This value
deviates significantly from the observed one.  Since
alternative explanations for the yrast band in $^{145}$Tb
are less probable (see the discussion below) it might be
that the breaking of the additivity of the experimental
effective alignments is connected with the influence of
residual interactions. A similar situation has been also
found in $^{154}$Dy (see subsection 4.5 for details).

An additional challenge is provided by the effective
alignment of the $^{144}$Gd(Exc.band 3)/$^{145}$Tb(1) and
$^{144}$Gd(Exc.band 4)/$^{145}$Tb(1) pairs, which is nearly
constant (apart from the last two points) and close to zero
if the spin in band 1 of $^{145}$Tb were increased by
$1\hbar$ (Fig. 9).  This suggests the existence of some
ambiguity in our interpretation of the observed band.  For
this we have also considered alternative possibilities. One
of them is to place the last two protons into the $\pi
[413]5/2 (r=\pm i)$ orbitals, because the analysis of the
SD bands observed in the $^{144}$Gd nucleus indicates the
closeness of these orbitals to the $Z=63$ SD shell gap.
Keeping in mind the approximate additivity of the
calculated effective alignments and the results for the
$\pi [413]5/2 (r=\pm i)$ orbitals in $^{144}$Gd (Fig. 8),
the expected effective alignment of this configuration,
relative to the yrast SD band configuration in $^{143}$Eu,
will be increasing from $i_{eff} \approx 1.0\hbar$ at
$\Omega_x \sim 0.3$ MeV up to $i_{eff} \approx 1.6\hbar$ at
$\Omega_x \sim 0.7$ MeV. The general slope of the effective
alignment of these two orbitals as function of the
rotational frequency is now in better agreement with
experiment.  However, even in the best case in which the
spin of the initial state of the observed band in
$^{145}$Tb is increased by $2\hbar$ relative to the value
given in Table 2, the discrepancy in absolute values will
be $\approx 0.4\hbar$. We consider this interpretation as
less probable because it leads to larger deviations from
experiment, though one cannot fully exclude it.

Comparing the effective alignment of the $\pi 6^1 \nu 7^0
(+,+i)$ configuration of $^{145}$Tb (the last two protons
being in $\pi[404]9/2(r=-i)$ and  $\pi[413]5/2(r=-i)$)
relative to the $\pi 6^1 \nu 7^0 (+,-1)$ configuration of
$^{144}$Gd (the 64th proton is in $\pi[413]5/2(r=-i)$),
which is close to zero at the rotational frequency range of
interest, with the effective alignment in the
$^{144}$Gd(Exc. band 3)/$^{145}$Tb(1, $I_0$=21.5$^+$) pair
(according to interpretation given in subsection 5.1.
$^{144}$Gd(Exc. band 3) has the 64th proton in
$\pi[413]5/2(r=-i)$), see Fig. 9, one gets the impression
that the $\pi 6^1 \nu 7^0 (+,+i)$ configuration could be
considered as a candidate for the yrast SD band observed in
$^{145}$Tb.  Despite the reasonable agreement between
theory and experiment, this configuration cannot be
assigned to the observed band. This is because (see Ref.
\cite{AKR.96} for a detailed discussion) it appears more
reasonable to expect the observation of four signature
partner bands closely lying in energy.

These arguments suggest that the $\pi 6^1 \nu 7^0 (+,-i)$
configuration, with two protons above the Z=63 SD shell gap
sitting in the $\pi [404]9/2(r=\pm i)$ orbitals, could be
considered as the most probable candidate for the yrast SD
band observed in $^{145}$Tb.  An additional confirmation of
the interpretation of the bands 1 and 2 in $^{144}$Gd and
the yrast band in $^{145}$Tb (as connected with the
occupation of the $\pi [404]9/2 (r=\pm i)$ orbitals) could
come from accurate measurements of the charge quadrupole
moments $Q_0$ of these bands relative to the yrast band in
$^{143}$Eu in a way similar to that of Refs.
\cite{Q-dy5152,GdDy-Q} for the Gd/Dy region. This is
because the occupation of the $\pi [404]9/2$ orbitals leads
to a considerable decrease in $Q_0$. For example, the
configuration $\pi 6^1 \nu 7^0 (+,-i)$ of $^{145}$Tb with
two occupied $\pi [404]9/2$ orbitals has $Q_0=12.15$ $e$b
at spin $I\sim 30\hbar$, which is smaller than the one
($Q_0=12.9$ $e$b) for the $\pi 6^1 \nu 6^4 7^0 (+,-i)$
configuration of the yrast band in $^{143}$Eu (see Table 2
in Ref. \cite{AKR.96}).  This indicates that the occupation
of the two $\pi [404]9/2$ orbitals decreases the
self-consistent deformation $\beta_2$ by $\approx 10\%$. On
the other hand, the occupation of the $\pi [413]5/2$
orbitals has a rather small effect on $Q_0$. Finally, it is
also noted that the role of the proton holes in the $\pi
[404]9/2$ orbitals on the stabilization of superdeformation
has been studied in detail within the CN model for nuclei
in the $A\sim 135$ mass region
\cite{AR.96}.

%%%%%%%%%%%%%%%%%%%%%%%%%%%%%%%%%%%%%%%%%%%%%%%%%%%%%%%%%%%%%%%%%%%%%
%
\section{Dependence of collective and single-particle properties on RMF 
parameterizations}
%
%%%%%%%%%%%%%%%%%%%%%%%%%%%%%%%%%%%%%%%%%%%%%%%%%%%%%%%%%%%%%%%%%%%%%

In the present paper the Lagrangian parameterization NL1
has been used for the analysis of the SD rotational bands
in the $A \sim 140-150$ mass region. The same
parameterization was also employed in our previous work
\cite{AKR.96} where our analysis was based on the
properties of the dynamic moments of inertia.  This set
behaves very well for stable nuclei in the vicinity of the
valley of beta-stability \cite{R.96} as it was tailored for
this area. However, on going away from the beta-stability
line the agreement with the experiment becomes less
satisfactory due to the isospin properties of this set,
which predicts a too large value for the asymmetry parameter
$J$ in nuclear matter.  In general, the NL1 set provides a
rather good agreement with the experimental data for the SD
bands in the $A \sim 140-150$ mass region.  However, the
description of the yrast and the first excited SD bands in
the $^{146-148}$Gd isotopes is not very good. This is
related to the single-particle energies of the $\nu
[642]5/2$ and $\nu [651]1/2$ orbitals and indicates that
the use of another parameter set could perhaps lead to
better results.

There are several parameter sets for the Lagrangian of RMF
theory.  Among them the set NLSH \cite{NLSH} and the
recently proposed set NL3 \cite{NL3} give very good results
for the ground state properties of finite nuclei at and
away from stability.  In this section, our aim is to
examine the dependence of the collective and
single-particle properties in the superdeformed minimum on
the various parameterizations of the RMF theory.  Thus,
additional calculations have been performed using the
parameter sets NLSH and NL3.  However, due to the very
large computing time needed for a systematic investigation
we have limited ourselves to the study of several nuclei,
namely $^{142}$Sm, $^{143}$Eu, $^{147}$Gd, $^{151}$Tb and
$^{151,152}$Dy. These nuclei are representative of the
different parts of the SD periodic chart (see Fig. 27 in
Ref. \cite{AKR.96}) in the $A\sim 140-150$ mass region and
their detailed investigation is expected to provide the
general features which are connected with the use of
different parameterizations in RMF theory.  Their selection
is optimal because, as it will be shown below, it allows us
to study the dependence of the CRMF results on the
parameterization at the two ends (nuclei around $^{152}$Dy
and $^{143}$Eu) and in the middle (nuclei around
$^{147}$Gd) of the SD periodic chart.  The key question is:
how much the {\bf collective properties}, such as the
kinematic $J^{(1)}$ and dynamic $J^{(2)}$ moments of
inertia and the charge quadrupole moments $Q_0$, and the
{\bf single-particle properties}, such as single-particle
ordering at the self-consistent deformation in the SD minimum,
alignment properties of the single-particle orbitals,
single-particle contributions to the kinematic and dynamic
moments of inertia and to the charge quadrupole moments are
affected by the various RMF parameterizations and how
good is the agreement with experiment in each case. One
should keep in mind that all Lagrangian parameterizations
used in this study have been determined from ground
state properties of a few spherical nuclei and no
adjustment to the properties of the SD bands has been made.
 
%%%%%%%%%%%%%%%%%%%%%%%%%%%%%%%%%%%%%%%%%%%%%%%%%%%%%%%
\subsection{The alignment properties
of the single-particle orbitals }
%%%%%%%%%%%%%%%%%%%%%%%%%%%%%%%%%%%%%%%%%%%%%%%%%%%%%%%

In Fig. 10  the effective alignments of the
$^{151}$Tb(1-4)/$^{152}$Dy(1) and
$^{151}$Dy(1)/$^{152}$Dy(1) pairs are compared with the
results obtained using the sets NL1, NL3 and NLSH. It turns
out that the bands 2 -- 4 in $^{151}$Tb are based on the
proton holes in the $^{152}$Dy doubly magic core connected
with the $\pi [301]1/2(r=+i)$, $\pi [651]3/2(r=-i)$ and
$\pi [301]1/2(r=-i)$ orbitals, respectively.  The
interpretation of these bands is in accord with earlier
findings within  the CN model \cite{Rag93} and the CHF
approach with Skyrme forces \cite{DD.95,BFH.96,EDD.97}.
The structure of the $^{151}$Dy(1) and $^{151}$Tb(1) bands
has been discussed in detail in sections 4.1 and 4.2.  Note
that in all three parameterizations, the $\pi
[651]3/2(r=\pm i)$, $\pi [301]1/2(r=\pm i)$ and $\nu
[770]1/2(r=-i)$ orbitals are calculated just below the
$Z=66$ and $N=86$ SD shell gaps, respectively (see Figs. 4,
11 and 12). Comparing the results obtained with different
RMF parameterizations, one can conclude that the effective
alignments calculated with NL1 and NL3 are rather similar.
On the other hand, the results obtained with the NLSH set
deviate  more strongly (especially, in the cases of the
$\pi [651]3/2(r=+i)$ and $\nu [770]1/2(r=-i)$ orbitals)
from those obtained with NL1.

The calculated effective alignments for the bands 1 -- 4 in
$^{151}$Tb are in reasonable agreement with experiment for
all three sets.  However, for the
$^{151}$Dy(1)/$^{152}$Dy(1) pair, large disagreement is
systematically observed for all parameterizations. The
possible origin of this discrepancy has been discussed in
section 4.1.  One should note that the establishment of the
spin of band 1 in $^{151}$Dy relative to yrast band in
$^{152}$Dy would be highly ambiguous in the case of NL3 and
especially for NLSH.

The kinematic moment of inertia $J^{(1)}$ is defined as 
\begin{eqnarray}
J^{(1)}=\frac{I}{\Omega_x}
\label{j1}
\end{eqnarray}
thus, the single-particle contributions to the kinematic
moments of inertia $\Delta J^{(1)}$ are related to the
effective alignments of the single-particle orbitals via 
\begin{eqnarray}
\Delta J^{(1)}_{B,A}=J^{(1)}_B-J^{(1)}_A=\frac{I_B-I_A}{\Omega_x}=
\frac{i^{eff}_{B,A}}{\Omega_x}
\label{dj1}
\end{eqnarray}
This shows that the conclusions made above, concerning the
dependence of the alignment properties of the
single-particle orbitals on the parameterization of the RMF
theory, are also valid for the single-particle
contributions to the kinematic moment of inertia $\Delta
J^{(1)}_{B,A}$.

%%%%%%%%%%%%%%%%%%%%%%%%%%%%%%%%%%%%%%%%%%%%%%%%%%%%%%%%%%%%%%%%
\subsection{The dynamic and kinematic moments
of inertia }
%%%%%%%%%%%%%%%%%%%%%%%%%%%%%%%%%%%%%%%%%%%%%%%%%%%%%%%%%%%%%%%%%

In Fig. 13 the dynamic moments of inertia $J^{(2)}$
calculated with NL1, NL3 and NLSH are compared with
experiment. The configuration assignment of the previous
subsection has been used for the bands 1 -- 4 in $^{151}$Tb.
It is seen that the results of the calculations for the
bands 1 -- 3 are in rather good agreement with the
experimental data for all sets.  In addition, the relative
properties of their dynamic moments of inertia are also
reproduced. A somewhat larger disagreement is observed in
band 4. The experimental $J^{(2)}$ is a slightly increasing
function of the rotational frequency. This feature is not
reproduced in calculations. The influence of pairing
correlations which are neglected in the present work might
be the reason for this discrepancy. On the other side, all
sets provide absolute values of $J^{(2)}$ close to
experiment for $\Omega_x>0.5$ MeV.  The values of $J^{(2)}$
calculated with NL3 and NLSH are lower than those of the
NL1 set, with the NL3 results always being closer to the
NL1 ones. The results for the lowest SD solution in
$^{143}$Eu (conf. $\pi 6^1 \nu 6^4 7^0 (+,-i)$), (Fig. 14)
are rather similar for the three parameterizations
manifesting the same dependence of the $J^{(2)}$ values as
in the cases of bands 1-4 in $^{151}$Tb.
 
The occupation of different orbitals has different impact
on the dynamic moment of inertia $J^{(2)}$ and this is
exemplified in Fig. 15. According to the calculations, the
occupation of the $\pi [651]3/2(r=+i)$ orbital has
significant impact on the dynamic moment of inertia, while
the occupation of the $\pi [651]3/2(r=-i)$ and $\pi
[301]1/2(r=\pm i)$ orbitals are not expected to lead to
considerable changes in $J^{(2)}$.  Considering that
$J^{(2)}$ is a very sensitive quantity because it is
calculated from the second derivative of the total energy
as a function of spin, one can conclude that the results
(see Fig. 15) are in reasonable agreement with experiment
when bands 1 -- 3 in $^{151}$Tb are compared with band 1 in
$^{152}$Dy.  The comparison of the $^{151}$Tb(3) and
$^{152}$Dy(1) bands shows significant deviations from the
trend of the calculations only for experimental points
above $\Omega_x>0.7$ MeV.  We cannot exclude that this
might be connected with the less accurate measurements of
the $\gamma$-transition energies on the top of the SD bands
and for the excited bands compared with the yrast ones.
The large discrepancy between calculations and experiment
for $\Delta J^{(2)}=J^{(2)}(^{152}{\rm Dy}(1))-
J^{(2)}(^{151}{\rm Tb}(4))$ (see Fig. 15) could 
be attributed to the upslope of the experimental $J^{(2)}$ 
of the $^{151}$Tb(4) band, the feature not reproduced
in calculations. It is also seen in Fig. 15 that the
results for the single-particle contributions to the
dynamic moments of inertia $\Delta J^{(2)}$ are rather
similar in all RMF parameterizations.  The largest
difference between the parameterizations appears when the
single-particle contribution to the dynamic moment of
inertia $\Delta J^{(2)}$ is coming from the occupation of
the $\pi [651]3/2(r=+i)$ orbital.

The difference  $\Delta J^{(1)}=J^{(1)}({\rm
NL3\,\,or\,\,NLSH}) -J^{(1)}({\rm NL1})$ in the kinematic
moments of inertia $J^{(1)}$ is presented in Fig. 16 for
four SD configurations assigned  to the bands 1 -- 4 in
$^{151}$Tb  and in the top panel of Fig. 14
for SD configuration assigned to the yrast band of
$^{143}$Eu. Our results indicate that changes of a few
percent in the kinematic moment of inertia are expected
when the non-linear parameter sets NL3 and NLSH are used
instead of NL1.  The kinematic moments of inertia $J^{(1)}$
obtained with NL3 are typically lower than those calculated
with NL1. A more complicated relation exists between the
$J^{(1)}$ values calculated with NL1 and NLSH.

%%%%%%%%%%%%%%%%%%%%%%%%%%%%%%%%%%%%%%%%%%%%%%%%%%%%%%%%%%%%
\subsection{The charge quadrupole $Q_0$ and mass 
hexadecupole $Q_{40}$ moments }
%%%%%%%%%%%%%%%%%%%%%%%%%%%%%%%%%%%%%%%%%%%%%%%%%%%%%%%%%%%%

The charge quadrupole $Q_0$ and the mass hexadecupole
$Q_{40}$ moments calculated with the sets NL1, NL3, and
NLSH are listed in Table 3 for a number of nuclei using
some configurations at different rotational frequencies.
It is seen that the $Q_0$ values calculated with NL3 and
NLSH are systematically smaller than those of NL1. It
should be noted that similarly to $i_{eff}$, $J^{(1)}$ and
$J^{(2)}$, the $Q_0$ and $Q_{40}$ values calculated with
NL1 and NL3 are rather similar. Only for the set NLSH, the
$Q_0$ values are typically smaller by $ \approx 2-3\%$
compared with the ones obtained with NL1. The difference in
the calculated mass hexadecupole moments $Q_{40}$ is rather
small (typically $\sim 2\%$) and it is less important.

The results of Table 3 also suggest that the occupation of
the same single-particle orbital has a rather similar
impact on the $Q_0$ and $Q_{40}$ values in different
parameterizations. Furthermore, it is seen from Table 4
that our results for the relative charge quadrupole moments
$\Delta Q_0$ are in good agreement with the precise
measurements of this quantity. They are also close to the
values obtained within the CHF approach with Skyrme forces
\cite{Satula}.

The comparison of the results for the $\pi[651]3/2(r=+i)$
and $\nu[770]1/2(r=-i)$ orbitals given in Tables 3 and 4
shows that the $\Delta Q_0$ values extracted at fixed
rotational frequency (Table 3) are rather close to the ones
obtained from the $Q_0^{th}$ values averaged over the
observed spin range (Table 4). This suggests that
it is essentially enough to perform the calculations at
fixed rotational frequency to compare theoretical and
experimental values of $\Delta Q_0$. The comparison of
the values in columns 4 and 5 of Table 4 indicates that the
additivity of the charge quadrupole moments discussed in
Ref. \cite{Satula} is also fulfilled in the CRMF theory.

%%%%%%%%%%%%%%%%%%%%%%%%%%%%%%%%%%%%%%%%%%%%%%%%%%%%%%%%%%%
%
\subsection{Single-particle energies in the SD minimum }
%
%%%%%%%%%%%%%%%%%%%%%%%%%%%%%%%%%%%%%%%%%%%%%%%%%%%%%%%%%%%

The $\pi 6^0 \nu 6^4 7^0(+,+1)$ configuration of $^{142}$Sm
has been selected to illustrate the dependence  of the
single-particle energies at superdeformation on the
parameterization.  This choice was motivated by the fact
that the change of parameterization leads to smaller
variations in the calculated $Q_0$ and $Q_{40}$ values
compared with the other cases given in Table 3.  In Fig. 17
we show the single-particle states around the SD shell gaps
at $Z=62$ and $N=80$ calculated at the corresponding
self-consistent deformations for $\Omega_x=0.0$ MeV using
the three parameterizations. One can see that large
similarities exist for the single-particle states appearing
in the vicinity of these SD shell gaps.

In all parameterizations, the $\pi [413]5/2$, $\pi
[404]9/2$, $\pi 1/2[660]$ and $\pi [523]7/2$ states are
above the $Z=62$ SD shell gap. Our analysis for the
isotonic ($N=80$) nuclei $^{143}$Eu, $^{144}$Gd and
$^{145}$Tb nuclei (see section 5), has shown that the
observed SD band structures are most likely connected with
the first three orbitals. The $\pi [411]3/2$ orbital is
also reasonably close to those orbitals in the NL1 results.
Large similarities are also observed for the proton states
below the $Z=62$ SD shell gap.  The $\pi [541]1/2$, $\pi
[301]1/2$ and $\pi [532]5/2$ states are located below this
gap in all parameterizations.  For the neutron states below
and above the $N=80$ shell gap (see bottom panel of Fig.
17) all sets give also similar results.

Regarding the absolute energies of the single-particle
states, it is clear that they depend strongly on the
parameterization.  It turns out that the energies of the
single-particle states obtained with NL3 are closer to the
ones calculated with NL1 than those of NLSH.  For some
states (for example, $\pi [411]3/2$) the difference in the
single-particle energies  calculated with the NL1 and NLSH
is around 2 MeV. The small differences in the
self-consistent equilibrium deformations cannot
explain the differences in the single-particle energies.
This can be clearly seen from the study of the proton and
neutron $[660]1/2$ states.  The self-consistent
deformations are decreasing on going from NL1 to NL3 to
NLSH. This should have lead to an increase of the energy of
this state because the $[660]1/2$ orbital is strongly
down-sloping as a function of deformation.  The
$\pi[660]1/2$ orbital manifests such a feature. However,
the changes in the single-particle energies are not
proportional to those of the equilibrium deformations.  On
the other hand, the single-particle energies of the
$\nu[660]1/2$ orbital show opposite trends (see Fig. 17).
This example clearly indicates that the difference in the
single-particle energies at superdeformation in the various
parameter sets is not strongly connected with the small
differences of the calculated self-consistent equilibrium
deformations.

In Fig. 17 is shown the variation of the single-particle
energies of various states as a function of the
parameterization, while in Figs. 18 and 19 similar results
are shown for the energies of the spherical subshells. The
comparison gives important correlations.  For example, the
positions of the $\nu [402]5/2$ and $\nu [404]7/2$ states
are inversed on going from NL1 to NL3 to NLSH. A similar
inversion takes also place in $^{151}$Tb, (see Figs. 4, 11
and 12).  The origin of this feature could be traced to the
increase of the energy difference between the $\nu
2d_{5/2}$ and $\nu 1g_{7/2}$ spherical subshells, from
which these states originate, when the parameterization is
changed from NL1 to NL3  to NLSH. Another example is the
$\pi [301]1/2$ and $\pi [301]3/2$ states originating from
the $\pi 2p_{1/2}$ and $\pi 2p_{3/2}$ subshells,
respectively. The energy difference between these states is
nearly constant for all parameterizations, while the
absolute energies are lower in the case of NL1 by $\sim 1$
MeV and $\sim 1.3$ MeV compared with NL3 and NLSH,
respectively (see Fig. 17).  Similar is also the case for
the $\pi 2p_{1/2}$ and $\pi 2p_{3/2}$ subshells (Fig. 18)
and in general a careful comparison of the Figs. 17, 18 and
19 will provide us with additional examples. Hence, one can
conclude that the differences in the single-particle
energies at superdeformation when different RMF
parameterizations are used is to a great extend connected
with their energy differences at spherical shape.

A few comments are in place.  The energy splitting at
superdeformation between the Nilsson states originating
from the high-$j$ subshells ($1g_{9/2}$, $1h_{11/2}$ and
$1i_{13/2}$) is practically independent on the
parameterization. Indeed, the energy splitting between the
$\pi [404]9/2$ and $\pi [413]7/2$ states originating from
$\pi 1g_{9/2}$, between the $\pi [660]1/2$, $\pi [651]3/2$
($\nu [660]1/2$, $\nu [651]3/2$ and $\nu [642]5/2$) states
originating from $\pi 1i_{13/2}$ ($\nu 1i_{13/2}$) and
between the $\pi [532]5/2$ and $\pi [523]7/2$ ($\nu
[532]5/2$ and $\nu [523]7/2$ states) originating from the
$\pi 1h_{11/2}$ ($\nu 1h_{11/2}$) subshell is very similar
(with an accuracy of $\leq 130$ keV) in the various
parameterizations, (see Fig. 17).  This suggests that the
energy dependence of these orbitals from the deformation
should be similar in all parameter sets. One can draw
similar conclusions for the Nilsson states originating from
the low-$j$ subshells ($\pi 2d_{5/2}$, $\pi 1h_{9/2}$,
$\nu 1g_{7/2}$ and $\nu 2d_{5/2}$) with somewhat smaller
accuracy (typically with an accuracy of $\leq 350$ keV).
The smaller accuracy is due to stronger fragmentation of
the wave functions of the states originating from the
low-$j$ subshells at the considered deformations.  We have
also analyzed the $^{152}$Dy nucleus. The analysis has lead
to conclusions similar to those reported above.
 
As it was discussed in Ref. \cite{AKR.96} and in the
subsection 4.7 of the present work, it is impossible to
reproduce the properties of some yrast and some excited
bands in the $^{146-148}$Gd isotopes using the parameter
set NL1. This difficulty is connected with the relative
energies of the $\nu[642]5/2$ and $\nu[651]1/2$ orbitals at
the self-consistent deformations. In order to study how the
results are changed when the parameter sets NL3 and NLSH
are used, the 'central' nucleus $^{147}$Gd and more
specifically, the configuration $\pi 6^2 \nu 7^1 (-,+i)$
have been investigated in detail.  According to the CN
results \cite{Gd46-50}, this configuration is assigned to
the band 1 of this nucleus. The dynamic moment of inertia
$J^{(2)}$ of band 1 shows a large peak at $\Omega_x \sim
0.65$ MeV, (see Fig. 19 in Ref. \cite{AKR.96}).  The
interpretation provided in Ref. \cite{Gd46-50} suggests
that it originates from the crossing of the $r=+i$
signatures of the $\nu[651]1/2$ and $\nu[642]5/2$ orbitals.
To have such a situation the $\nu[651]1/2$ state should be
higher in energy than the $\nu[642]5/2$ state at
$\Omega_x=0.0$ MeV. However, in NL1 the $\nu[651]1/2(r=+i)$
orbital is below the $\nu[642]5/2(r=+i)$ orbital at the
self-consistent deformation of this configuration (Fig. 18
of Ref. \cite{AKR.96}).  Rather similar results are
obtained with the parameter sets NLSH and NL3.  Based on
the relevant neutron single-routhian diagrams one can
conclude that in all parameterizations the position of the
$\nu[651]1/2$ orbital should be shifted higher in energy by
$\approx 1$ MeV relative to $\nu[642]5/2$. This is
necessary in order to get the crossing of two orbitals at
the correct frequency.

Keeping in mind that the $\nu [651]1/2$ orbital is more
strongly down-sloping with the increase of the quadrupole
deformation (charge or mass quadrupole moments) compared
with the $\nu [642]5/2$ orbital (see for example Fig. 1.6.
in Ref. \cite{Sol}), a slight decrease of the
self-consistent deformation could bring the results closer
to experiment. For example, the inclusion of the pairing
correlations might produce such a decrease of the
self-consistent deformation. However, taking into account
that the jumps in $J^{(2)}$ in the SD bands of the Gd
isotopes take place at high rotational frequency ($\Omega_x
\sim 0.6-0.7$ MeV), it seems that the inclusion of pairing
will not significantly improve the results.  An alternative
possibility, which perhaps more likely holds, is that the
relative energies of the spherical subshells from which
these states originate are not optimal in these RMF
parameterizations.

Our analysis shows that all parameter sets, considered in
this work, have some drawbacks which are related to the
description of the energies of the single-particle states
at superdeformation. However, this should have been
expected as no single-particle properties have been used
for the determination of the various RMF parameterizations.
One should also keep in mind the rather small ``effective
mass'' of the RMF theory which has an effect on the
single-particle energies.  Thus, the role of the effective
alignment approach is important for the theoretical
interpretation of the structure of SD bands.  To answer the
question which parameter set is more suitable for the
description of the energies of single-particle states at
superdeformation, a systematic investigation of the SD
bands employing different parameterizations is needed.
This, however, is an extremely time consuming procedure and
beyond the scope of the present article.

Of course, the study of the single-particle spectra of the
spherical doubly magic nuclei, obtained in various RMF
parameterizations, could  give some insight to this
problem.  We have chosen the $^{208}$Pb nucleus, for which
rich experimental information is available.  In Fig. 20 are
shown the calculated energies of the single-particle states
around the Fermi surface together with the empirical
values.  It is seen that the calculated levels have a close
correspondence with experiment. The theoretical spectra are
less dense compared with experiment. The same is known from
the non-relativistic density dependent Hartree-Fock
calculations \cite{BFNQ.75,QF.78,BQBGH.82}.  Such an
effect, which appears mainly in medium heavy and heavy
nuclei, is well explained in terms of the interplay between
single-particle motion and low-lying collective vibrations
(see e.g. the discussion in Ref. \cite{QF.78}).  The
neutron $N=126$ spherical gap is more or less reproduced in
all three parameterizations.  On the other hand, the
non-linear set NL1 shows a gap at $Z=82$ closer to
experiment.  In the NLSH parameterization, the $Z=84$
spherical gap is more pronounced than the $Z=82$ spherical
gap in contradiction with experiment. Intermediate
situation is observed for the NL3 parameterization.  In
this case, the $Z=84$ spherical gap is similar to the
$Z=82$ spherical gap.  All that might be considered as an
indication that the non-linear set NL1 is more optimal for
the description of the single-particle energies for the
nuclei in the valley of beta-stability compared with NL3
and NLSH. However, more detailed investigation of this
question is needed since the coupling between
single-particle motion and low-lying collective vibrations
is neglected in the present calculations.

%%%%%%%%%%%%%%%%%%%%%%%%%%%%%%%%%%%%%%%%%%%%%%%%%%%%%%%%%
\section{Conclusions}
%%%%%%%%%%%%%%%%%%%%%%%%%%%%%%%%%%%%%%%%%%%%%%%%%%%%%%%%%

The single-particle properties at superdeformation and the
dependence of the results for various observables on the
Lagrangian parameterization have been investigated in the
framework of the cranked relativistic mean field theory.
For our study the SD bands observed in the $A\sim 140-150$
mass region have been used. In the following we summarize
the main results of the present work.

The alignment properties of the single-particle orbitals at
superdeformation are reproduced adequately in the CRMF
theory in almost all cases studied with exception of yrast
SD bands observed in $^{154,155}$Dy and bands 4 and 5 in
$^{153}$Dy.  This allows us to apply the effective
alignment (or similar) approach for the identification of
the configurations of the SD bands.  The interpretation of
the SD bands observed around $^{152}$Dy, first given by the
CN model, have been confirmed. However, the effective
alignments of the single-particle orbitals, calculated in
the CRMF theory, are typically in better agreement with
experiment compared with the CN model.

In the case of yrast bands in $^{154,155}$Dy the
experimental effective alignments related to non-intruder
orbitals deviate significantly from the pattern expected
from the experimental systematics seen in lighter nuclei.
It was not possible to find a consistent interpretation for
these bands in terms of non-intruder orbitals in a pure
single-particle picture. This might be due to the influence
of residual interactions neglected in the present
investigation.

Though in our calculations pairing correlations have been
neglected good agreement with experiment has been obtained
in most of the cases.  This is an indication that many
features of the observed superdeformed bands, in this mass
region, can be well understood in terms of an almost
undisturbed single-particle motion.

The stability of the results obtained within the CRMF
theory with respect to parameterization have been
investigated employing the non-linear parameter sets NL1,
NL3 and NLSH for the SD bands observed in $^{151}$Tb and
$^{151,152}$Dy nuclei.  The calculated kinematic $J^{(1)}$
and dynamic $J^{(2)}$ moments of inertia, the charge
quadrupole $Q_0$ and mass hexadecupole $Q_{40}$ moments are
rather similar in all parameterizations we have used. They
are also close to experimental data. It turns out that the
single-particle contributions to the charge quadrupole
$Q_0$ and mass hexadecupole $Q_{40}$ moments are rather
independent from the RMF parameterization.  The effective
alignments $i_{eff}$ of the high-$N$ intruder orbitals and
the single-particle contributions to the dynamic moments of
inertia, arising from the occupation of the high-$N$
intruder orbitals show, however, some dependence from the
parameterization.

Although in all parameterizations a similar group of
single-particle states appear in the vicinity of the SD
gaps, the relative energies of different single-particle
states are strongly depended on the RMF parameterization.
To define the parameter set which provides the better
description of the energies of the single-particle states
at superdeformation, a more systematic investigation of the
SD bands within the various RMF parameterizations is
needed.  The fact, however, that based only on the lowest
and the slightly excited SD configurations, calculated in
Ref. \cite{AKR.96} with NL1, it was possible to reproduce
the effective alignments of the observed SD bands shows
that this parameter set provides a reasonable description
of the single-particle energies at superdeformation.
Undeniably a parameterization more oriented for
superdeformation would provide improved results. Keeping in
mind however, that the limited number of the parameters of
the RMF theory have been determined by means of a few
ground state bulk properties of some spherical nuclei, it
is rather amazing that in the limit of superdeformation the
RMF theory describes the energies and the ordering of the
single-particle states in the vicinity of the SD shell gaps
with reasonable accuracy.

\section{Acknowledgments}

One of the authors (A.\ V.\ A.) acknowledges financial
support from the {\it Konferenz der Deutschen Akademien der
Wissenschaften through the Volkswagenstiftung}, the NORDITA
(Nordisk Institute for Theoretisk Physik: NORDITA Baltic
Fellowship 1996) and from the Crafoord Foundation (Lund,
Sweden). The authors are grateful to I.\ Ragnarsson for
careful reading of manuscript and valuable discussions.
This work has been also supported by the Bundesministerium
f\"ur Bildung und Forschung under contract 06 TM 875.

%%%%%%%%%%%%%%%%%%%%%%%%%%%%%%%%%%%%%%%%%%%%%%%%%%%%%%%%%%%

%%%%%%%%%%%%%%%%%%%%%%%%%%%%%%%%%%%%%%%%%%%%%%%%%%%%%%%%%%%%%%%%
\newpage

\section{Figure captions}

Fig. 1.
The calculated differences $\Delta i_{eff}$ (in units
$\hbar$), see eq. (\protect\ref{addit2}) in the text for
definition.  The 'independent' effective alignments
$i_{eff}$ for the $\nu[402]5/2(r=\pm i)$, $\pi
[651]3/2(r=\pm i)$ and $\nu[770](r=-i)$ orbitals have been
extracted by comparing the relevant configurations in
$^{153}$Dy, $^{151}$Tb and $^{151}$Dy with lowest SD
configuration in $^{152}$Dy, respectively. The
$i_{eff}^{AB}$ values have been extracted from the relevant
configurations in the pairs $^{149}$Gd/$^{152}$Dy
(differing in the occupation of the
$(\pi[651]3/2)^2+\nu[770](r=-i)$ orbitals),
$^{150}$Gd/$^{152}$Dy (differing in the occupation of the
$(\pi[651]3/2)^2$ orbitals) and $^{152}$Dy/$^{154}$Dy
(differing in the occupation of the $(\nu[402]5/2)^2$
orbitals).
\\
\\ 
Fig. 2.
Effective alignments, $i_{eff}$ (in units $\hbar$),
extracted from experiment (unlinked large symbols) are
compared with those extracted from the corresponding
calculated configurations (linked small symbols of the same
type) for orbitals active in the SD bands when Z increases
from 64 to 66 and N from 85 to 86. The experimental
effective alignment between SD bands A and B is indicated
as ``A/B''.  The band A in the lighter nucleus is taken as
a reference, so the effective alignment measures the effect
of the additional particle. The spin-parity assignment for
the experimental bands is given in Table 2. The compared
configurations differ in the occupation of the orbitals
shown in the panels.  The long-dashed lines are used in
panel (a) in order to show the effective alignments of the
compared bands under the spin assignments obtained with the
CN calculations (see Table 2 and Ref.
\protect\cite{Rag93}).  The figure has been designed in the
same fashion as Fig. 3 of Ref. \protect\cite{Rag93} for the
sake of comparison of the CRMF and CN (see Ref.
\protect\cite{Rag93}) results.
\\
\\
Fig. 3.
The same as in Fig. 2, but for the
effective alignments of the $\pi [402]5/2(r=\pm i)$ and
$\pi [651]3/2(r=-i)$ (panel (a)) orbitals.
\\
\\
Fig. 4.
Proton (top) and neutron (bottom) single-particle
energies (routhians) in the self-consistent rotating
potential as a function of the rotational frequency
${\sl\Omega}_x$. They are given along the deformation path
of the lowest superdeformed configuration in $^{151}$Tb and
they have been calculated using the non-linear parameter
set NL1. Solid, short-dashed, dot-dashed and dotted lines
indicate $(\pi=+,\,\,r=-i)$, $(\pi=+,\,\,r=+i)$,
$(\pi=-,\,\,r=+i)$ and $(\pi=-,\,\,r=-i )$ orbitals,
respectively.  At ${\sl\Omega}_x=0.0$ MeV, the
single-particle orbitals are labeled by means of the
asymptotic quantum numbers $[Nn_z\Lambda]\Omega$ (Nilsson
quantum numbers) of the dominant component of the wave
function.  The proton hole in doubly magic $^{152}$Dy core
is indicated by an open square.
\\
\\
Fig. 5.
The same as in Fig. 2, but for the
bands observed in $^{153,154,155}$Dy. The effective
alignments of the $^{150}$Tb(1)/$^{151}$Tb(7,8) pairs are
shown in the inset of the panel (a). The spin values
$I_0=32.5\hbar$ and $I_0=33.5\hbar$, corresponding to the
lowest observed transitions with energies 758.0 and 785.0
keV within the bands 7 and 8 of $^{151}$Tb, are used for
the lowest observed states of the bands 7 and 8,
respectively.  Note, that they differ by $-1\hbar$ from the
spins assumed for these bands in Ref.
\protect\cite{ALR.97}.
\\
\\
Fig. 6. (a) The same as in Fig. 2, but for
the effective alignments of the $\nu [642]5/2(r=-i)$ (long
dashed line) and $\nu [642]5/2(r=+i)$ (dot-dashed line)
orbitals. The long-dashed line should be compared with open
triangles, while the dot-dashed line with solid triangles.
(b) The dynamic moment of inertia $J^{(2)}$ of the band 2
(open circles) and band 5 (solid squares) in $^{148}$Gd
versus the calculated values for the assigned
configurations. Based on the analysis of the effective
alignments (see panel (a) and text for details) the
following correspondence is established: band $2 \equiv \pi
6^2 \nu 7^1 (-,+1)$; band $5 \equiv \pi 6^2 \nu 7^1
(-,-1)$.
\\
\\
Fig. 7. Experimental effective alignments $i_{eff}$ in the
$^{143}$Eu(1)/$^{149}$Gd(1) pair under different
assumptions for the spin $I_0$ of the lowest state of the
$^{143}$Eu(1) band compared with calculated one (solid
line). The notation of the symbols is given in the figure.
\\
\\
Fig. 8. The same as in Fig. 2, but for the
calculated effective alignments of the $\pi
[413]5/2(r=-i)$, $(\pi [404]9/2)^2$ and $\pi
[660]1/2(r=+i)$ orbitals which are compared with the
available experimental effective alignments in the
$^{143}$Eu(1)/$^{144}$Gd(N) (N is the band label),
$^{143}$Eu(1)/$^{145}$Tb(1) and
$^{144}$Gd(yrast)/$^{145}$Tb(1) pairs.
\\
\\
Fig. 9. Experimental effective alignments $i_{eff}$ in the
$^{144}$Gd(N)/$^{145}$Tb(1) pairs. In cases where the spin
values are different from those of Table 2, the used values
are shown in the figure.
\\
\\
Fig. 10. The same as in Fig. 2, but for the
effective alignments of the $^{151}$Tb(N)/$^{152}$Dy(1) (N
is the band label) and $^{151}$Dy(1)/$^{152}$Dy(1) pairs.
The calculations have been carried out using the non-linear
parameter sets NL1, NL3 and NLSH.
\\
\\
Fig. 11. The same as in Fig. 4, but for
calculations performed with the non-linear parameter set
NL3.
\\
\\
Fig. 12. The same as in Fig. 4, but for
calculations performed with the non-linear parameter set
NLSH.
\\
\\
Fig. 13.
Dynamic moments of inertia $J^{(2)}$ of the bands
1-4 in $^{151}$Tb (unlinked solid circles) versus those of
the assigned calculated configurations (lines without
symbols), obtained with the NL1, NL3 and NLSH forces. 
\\
\\
Fig. 14.
Upper panel: The difference $\Delta
J^{(1)}=J^{(1)}({\rm NL3\,\,or\,\,NLSH})-J^{(1)}({\rm
NL1})$ of the kinematic moments of inertia $J^{(1)}$
calculated with NL3 or NLSH from those obtained with NL1
for the lowest SD configuration in $^{143}$Eu.  Lower
panel: The same as in Fig. 13, but for
the lowest SD configuration in $^{143}$Eu.
\\
\\
Fig. 15. The contributions to the dynamic moments of
inertia, defined as $\Delta J^{(2)}=J^{(2)}(^{152}{\rm
Dy}(1))- J^{(2)}(^{151}{\rm Tb}(N))$ (N is the number of
the band in $^{151}$Tb), from the occupation of different
proton orbitals shown in the figure, calculated using the
NL1, NL3 and NLSH forces.  The experimental values (solid
unlinked circles), have been extracted via a quadratic
least-squares fit to the $J^{(2)}$ of the observed bands.
The corresponding bands in $^{151}$Tb are also shown in the
figure.  The experimental values of $\Delta
J^{(2)}=J^{(2)}(^{152}{\rm Dy}(1))- J^{(2)}(^{151}{\rm
Tb}(4))$ vary from $\Delta J^{(2)}\sim 3.4$ MeV$^{-1}$ at
$\Omega_x\sim 0.4$ MeV to $\Delta J^{(2)}\sim -2.5$
MeV$^{-1}$ at $\Omega_x\sim 0.73$ MeV. Only part of this
curve is shown in the figure. 
\\
\\
Fig. 16. The difference $\Delta J^{(1)}=J^{(1)}({\rm
NL3\,\,or\,\,NLSH})-J^{(1)}({\rm NL1})$ of the kinematic
moments of inertia $J^{(1)}$ calculated with NL3 or NLSH
from those obtained with NL1 for four SD configurations
assigned  to the bands 1-4 in $^{151}$Tb.
\\
\\
Fig. 17.
The single-particle states around the $Z=62$ and
$N=80$ SD shell gaps calculated with NL1, NL3 and NLSH at
the corresponding equilibrium deformations of the $\pi 6^0
\nu 6^4 7^0(+,+1)$ configuration in $^{142}$Sm at
$\Omega_x=0.0$ MeV.  The single-particle states are labeled
by means of the asymptotic quantum numbers
$[Nn_z\Lambda]\Omega$ (Nilsson quantum numbers) of the
dominant component of their wave function. In addition, the
spherical subshells from which these states originate, if
the diabatic continuation from deformed to spherical shapes
is built, are shown in parentheses. Because of the change
of the relative positions of the $\nu 1h_{9/2}$ and $\nu 2
f_{7/2}$ spherical subshells in NL3 and NLSH  compared
with NL1, see Fig. 19 below, two labels for the spherical subshells separated
by '/' are given in  parentheses for neutron $N=5$ Nilsson
states.  The first label corresponds to NL1, while the
second to NL3 and NLSH. The calculated charge quadrupole
$Q_0$ and mass hexadecupole $Q_{40}$ moments are
$Q_0=12.40\,\,e$b, $Q_{40}=7.39 \cdot 10^3\,fm^4$ (NL1),
$Q_0=12.25\,\,e$b, $Q_{40}=7.25 \cdot 10^3\,fm^4$ (NL3) and
$Q_0=12.04\,\,e$b, $Q_{40}=7.05 \cdot 10^3\,fm^4$ (NLSH).
To minimize the numerical uncertainties in the
single-particle energies originating from the truncation of
the basis, the calculations have been performed in the
extended basis space i.e. all fermion states below
$15.5\hbar \omega_0^F$ and all boson states below
$16.5\hbar \omega_0^B$ have been taken into account. 
\\
\\
Fig. 18. Single-proton states at spherical shape in
$^{142}$Sm in the energy range from $-25$ MeV up to 10 MeV.
On the left side are shown only the proton spherical
subshells from which the single-particle states, located
in the vicinity of the $Z=62$ SD shell gap (see top panel
of Fig. 17), originate.
\\
\\
Fig. 19. The same as in Fig. 18, but for
neutrons. On the left side are shown mainly the neutron
spherical subshells from which the single-particle states,
located in the vicinity of the $N=80$ SD shell gap (see
bottom panel of Fig. 17), originate.
\\
\\
Fig. 20. The experimental neutron and proton
single-particle spectra of $^{208}$Pb are compared with the
predictions of the RMF forces NL1, NL3 and NLSH. The
experimental data is taken from \protect\cite{VB.72}.

%%%%%%%%%%%%%%%%%%%%%%%%%%%%%%%%%%%%%%%%%%%%%%%%%%%%%%%%%%%%
\newpage

\begin{table}
\caption{ \sf The non-linear parameter sets NL1, NLSH and
NL3. The masses are given in MeV, the parameter $g_2$ in
$fm^{-1}$, while the rest of the parameters are
dimensionless. The nuclear matter properties, predicted
with these effective forces, namely, the baryon density
$\rho_0$ (in units $fm^{-3}$), the binding energy per
particle $E/A$ (in MeV), the incompressibility $K$ (in MeV), 
the effective mass $m^*/m$ and the asymmetry parameter $J$ 
(in MeV) are also shown.}
\vspace{0.5cm}   
\begin{center}
\begin{tabular}{|c|c|c|c|} \hline
Parameter       &   NL1      &  NLSH     & NL3  \\ \hline
\multicolumn{4}{|c|}{\bf Masses} \\ \hline
$m_N$           & 938.0      & 939.0     & 939.0    \\
$m_{\sigma}$    & 492.25     & 526.059   & 508.194  \\
$m_{\omega}$    & 795.36    & 783.0     & 782.501  \\
$m_{\rho}$      & 763.0      & 763.0     & 763.0    \\ \hline
\multicolumn{4}{|c|}{\bf Coupling constants}        \\ \hline
$g_{\sigma}$     & 10.138     & 10.4444    & 10.217   \\
$g_2$           & $-12.172$    & $-6.9099$   & $-10.431$  \\
$g_3$           & $-36.265$    & $-15.8337$  & $-28.885$  \\
$g_{\omega}$     & 13.285     & 12.945    & 12.868   \\
$g_{\rho}$       & 4.976     & 4.383     & 4.474    \\ \hline
\multicolumn{4}{|c|}{\bf Nuclear matter properties } \\ \hline
$\rho_0$        & 0.153      & 0.146     & 0.148   \\
$E/A$           & $-16.488$    & $-16.346$   & $-16.299$   \\
$K$             & 211.29      & 355.36     & 271.76   \\
$m^*/m$         & 0.57      & 0.60     & 0.60     \\
$J$             & 43.7       & 36.1      & 37.4     \\ \hline
\end{tabular}
\end{center}
\end{table}

%%%%%%%%%%%%%%%%%%%%%%
% Table 2
% Spin assignments
%%%%%%%%%%%%%%%%%%%%%%
\newpage
\begin{table}
\caption{\sf The spin assignment for the lowest state
$I_0^{CRMF}$ of the SD bands. The corresponding lowest
transition energies $E_{\gamma}(I_0+2 \rightarrow I_0)$ are
also shown.  In the present formalism only the relative
spins are "determined" therefore shifts of all bands in
steps of $\pm 2\hbar$ could not be excluded. In the last
two columns, the spin assignments for the lowest state of
the SD bands in the Dy/Tb/Gd region based on the CN model
are given according to "Alternative 1" ($I_0^{Nils}$ (Alt.
1)) and "Alternative 2" ($I_0^{Nils}$ (Alt. 2)) of Ref.
\protect\cite{Rag93}, respectively. When the spin values
assigned to the lowest states are the same in CRMF and CN
results, the CN values are marked with an asterisk (*).
Note, that the spin values for bands 4 and 5 in $^{153}$Dy
used in Fig. 5 are not based on the
comparison between calculations and experiment (see text
for details).  }
\vspace{0.5cm}
\begin{tabular}{|c|c|c|c|c|c|c|} \hline
 Nucleus &  Band & Ref. & $E_{\gamma}(I_0+2 \rightarrow I_0)$
& $I_0^{CRMF}$ & $I_0^{Nils}$ (Alt. 1) & $I_0^{Nils}$ (Alt. 2) \\ \hline
$^{155}$Dy & band 1 & \cite{Dy155a} & 909.6  & 37.5$^-$   &  & \\
$^{154}$Dy & band 1 & \cite{Dy154a} & 701.7  & 28$^+$   &  & \\
$^{153}$Dy & band 1 & \cite{Dy153a} & 721.4  & 29.5$^-$ & 29.5$^-$(*) 
& 31.5$^-$  \\ 
           & band 2 & \cite{Dy153a} & 678.6  & 27.5$^+$ & 27.5(*) & 29.5  \\
           & band 3 & \cite{Dy153a} & 702.0  & 28.5$^+$ & 28.5(*) & 30.5   \\
           & band 4 & \cite{Dy153a} & 723.4  & 28.5 &         &   \\
           & band 5 & \cite{Dy153a} & 743.2  & 29.5 &         &   \\
$^{152}$Dy & band 1 & \cite{Dy152a} & 602.4  & 24$^+$ & 24$^+$(*) & 26$^+$ \\  
           & band 4 & \cite{Dy152a} & 669.6  & 27$^-$   &         &     \\
           & band 5 & \cite{Dy152a} & 642.1  & 26$^-$   &         &     \\
$^{151}$Dy & band 1 & \cite{Dy151a} & 523.7  & 21.5$^-$ & 21.5$^-$(*) 
& 23.5$^-$  \\
$^{152}$Tb & band 1 & \cite{Tb152a} & 823.0  & 35$^+$ &         &   \\
           & band 2 & \cite{Tb152a} & 801.0  & 34$^+$ &         &   \\
$^{151}$Tb & band 1 & \cite{Tb151b} & 726.5  & 30.5$^+$ & 28.5$^{+}$ & 
30.5$^{+}$(*) \\
           & band 2 & \cite{Tb151b} & 602.1  & 24.5$^-$ & 24.5$^{-}$(*) &  
26.5$^{-}$ \\
           & band 3 & \cite{Tb151b} & 681.5  & 27.5$^+$ &         &   \\
           & band 4 & \cite{Tb151b} & 768.6  & 31.5$^-$ &         &   \\
$^{150}$Tb & band 1 & \cite{Tb150b} & 596.8  & 26$^-$ & 24$^{-}$ & 
26$^{-}$(*) \\
           & band 2 & \cite{Tb150b} & 662.5  & 27$^-$ &       &       \\
$^{150}$Gd & band 1 & \cite{Gd150Tb151} & 815.2  & 32$^+$ & 30$^+$ & 
32$^+$(*)  \\
$^{149}$Gd & band 1 & \cite{Gd149b} & 617.8  & 25.5$^-$ &  23.5$^-$ & 
25.5$^-$(*)   \\ 

$^{148}$Gd & band 2 & \cite{Gd148d} & 790.2  & 32$^-$ &  30$^-$ & 
32$^-$(*) \\ 
           & band 5 & \cite{Gd148d} & 891.1  & 37$^-$ &   & \\ 
\hline \hline
$^{143}$Eu & band 1 & \cite{BFS.96} & 483.7  & 16.5$^+$ &    &   \\ 
$^{144}$Gd & yrast band  & \cite{Gd144new} & 802.8 & 22$^+$ &  &    \\ 
           & exc. band 1 & \cite{Gd144new} & 774.5 & 26$^+$ & &  \\ 
           & exc. band 2 & \cite{Gd144new} & 743.6 & 25$^+$ & &  \\ 
           & exc. band 3 & \cite{Gd144new} & 852.9 & 29$^+$ & &   \\ 
           & exc. band 4 & \cite{Gd144new} & 936.8 & 32$^+$ & &   \\ 
$^{145}$Tb & band 1      & \cite{Tb145a}         & 627.1 & 20.5$^+$ & &   \\ 
&        & &       &          &         &            \\ \hline   
\end{tabular}
\end{table}

%%%%%%%%%%%%%%%%%%%%%%%%%%%%%%%%%%%%%%%%%%%%%%%%%%%%%%%%%%%%%%%%%%%%%%%%%%
\newpage
\begin{table}
\caption{\sf {\bf Upper part:} The absolute values of the
charge quadrupole $Q_0$ (in units $e$b) and mass
hexadecupole $Q_{40}$ (in units $10^3$ fm$^4$) moments
calculated with NL1 and given in the form
$\frac{Q_0}{Q_{40}}$ (second column), are shown for the
self-consistent deformations of the corresponding
configurations obtained at $\Omega_x=0.0$ MeV for
$^{152}$Dy and $^{142}$Sm and at $\Omega_x=0.55$ MeV for
$^{147}$Gd.  For the NL3 and NLSH (third and fourth
columns), the changes of the $Q_0$ and $Q_{40}$ values
relative to the values obtained with NL1 are given in the
form $\frac{\Delta Q_0}{\Delta Q_{40}}$.  To minimize the
numerical uncertainties in the absolute values of $Q_0$ and
$Q_{40}$ due to the use of the truncated basis, the
calculations have been performed in the extended basis
space i.e. all fermion states below  $15.5\hbar \omega_0^F$
and all boson states below  $16.5\hbar \omega_0^B$ have
been taken into account. {\bf Middle part:} The differences
$\Delta Q_0=Q_0(^{151}{\rm Tb})-Q_0(^{152}{\rm Dy})$ and
$\Delta Q_{40}=Q_{40}(^{151}{\rm Tb})-Q_{40}(^{152}{\rm
Dy})$, associated with different proton holes in the doubly
magic $^{152}$Dy core, indicated in the first column, are
given in the form $\frac{\Delta Q_0}{\Delta Q_{40}}$ for
different parameterizations. They are calculated at
$\Omega_x=0.5$ MeV.  In the calculations, all fermion
states below $11.5\hbar \omega_0^F$ and all boson states
below  $16.5\hbar \omega_0^B$ have been considered.
Compared with the truncated scheme used in the upper part, the
numerical uncertainties in $\Delta Q_0$ and $\Delta Q_{40}$
due to the truncation of the basis are very small.
{\bf Lower part:} Similar as in the middle part, but for
the differences between the $^{151}$Dy and $^{152}$Dy yrast
bands associated with the neutron hole $(\nu
[770]1/2(r=-i))^{-1}$ in the doubly magic $^{152}$Dy core.}
\vspace{0.5cm}
\begin{tabular}{|c|c|c|c|} \hline
       & NL1 & NL3 & NLSH \\ \hline
 & & & \\
$^{152}$Dy (conf. $\pi 6^4 \nu 7^2 (+,+1)$)
& $\frac{18.46}{19.67}$ & $\frac{-0.20}{-0.31}$ & 
             $\frac{-0.47}{-0.76}$ \\
 & & & \\
$^{147}$Gd (conf. $\pi 6^2 \nu 7^1 (-,+i)$) & $\frac{14.99}{12.77}$  &
$\frac{-0.18}{-0.30}$ & $\frac{-0.49}{-0.78}$  \\
 & & & \\
$^{142}$Sm (conf. $\pi 6^0 \nu 6^4 7^0 (+,+1)$)
& $\frac{12.40}{7.39}$ & $\frac{-0.15}{-0.14}$ & 
$\frac{-0.36}{-0.34}$ \\
 & & & \\ \hline \hline
 & & & \\
$(\pi [301]1/2(r=-i))^{-1}$ & $\frac{+0.24}{+0.12}$
                            & $\frac{+0.24}{+0.13}$
                            & $\frac{+0.24}{+0.22}$\\
 & & & \\
$(\pi [301]1/2(r=+i))^{-1}$ & $\frac{+0.20}{+0.02}$
                            & $\frac{+0.21}{-0.13}$
                            & $\frac{+0.21}{-0.15}$\\
 & & & \\
$(\pi [651]3/2(r=-i))^{-1}$ & $\frac{-0.90}{-1.05}$
                            & $\frac{-0.90}{-1.05}$
                            & $\frac{-0.89}{-0.98}$ \\
 & & & \\
$(\pi [651]3/2(r=+i))^{-1}$ & $\frac{-0.99}{-1.32}$
                            & $\frac{-1.00}{-1.40}$
                            & $\frac{-1.00}{-1.43}$ \\
 & & & \\ \hline \hline
%%%
 & & & \\
$(\nu [770]1/2(r=-i))^{-1}$ &  $\frac{-0.55}{-1.69}$
                            &  $\frac{-0.53}{-1.62}$
                            &  $\frac{-0.51}{-1.47}$ \\ 
 & & & \\ \hline
\end{tabular}
\end{table}

%%%%%%%%%%%%%%%%%%%%%%%%%%%%%%%%%%%%%%%%%%%%%%%%%%%%%%%%%%%%%%%%%%%%%%%%%%
\newpage
\begin{table}
\caption{\sf Experimental and calculated relative charge
quadrupole moments $\Delta Q_0=Q_0({\rm
Band})-Q_0(^{152}\rm Dy(1))$ of the $^{149}$Gd(1),
$^{151}$Tb(1) and $^{151}$Dy(1) bands. The experimental
data are taken from Refs. \protect\cite{Q-dy5152,GdDy-Q}.
The calculations have  been carried out with NL1. The
detailed structure of the configurations of these bands is
given relative to the doubly magic $^{152}$Dy core in
column 2. Since the experimental values $Q_0^{exp}$ (column
3) are averaged over the observed spin range, the
theoretical values $Q_0^{th}$ (column 4) have been also
averaged over the same spin range using the spin assignment
given in Table 2.  In column 5 are shown the values $\sum_i
\Delta Q_0^i$ , where $\Delta Q_0^i$ is the 'independent'
contribution of i-th particle to the charge quadrupole
moment calculated at $\Omega_x=0.50$ MeV and
given in Table 3.}
\vspace{0.5cm}
\begin{tabular}{|c|c|c|c|c|} \hline
Band           & Configuration & $\Delta Q_0^{exp}$ ($e$b) & 
               $\Delta Q_0^{th}$ ($e$b) & $\sum_i \Delta Q_0^i$   \\ \hline
  1 & 2 & 3 & 4 & 5 \\ \hline
$^{149}$Gd(1)  & $\nu [770]1/2(r=-i)^{-1} (\pi [651]3/2)^{-2}$
                            & $-2.5(0.3)$          & $-2.41$  &  $-2.44$  \\
$^{151}$Tb(1)  & $\pi [651]3/2(r=+i)^{-1}$  & $-0.7(0.7)$ & $-1.01$ 
&  $-0.99$ \\
$^{151}$Dy(1)  & $\nu [770]1/2(r=-i)^{-1}$  & $-0.6(0.4)$ & $-0.53$ 
&  $-0.55$ \\ \hline 
\end{tabular}
\end{table}

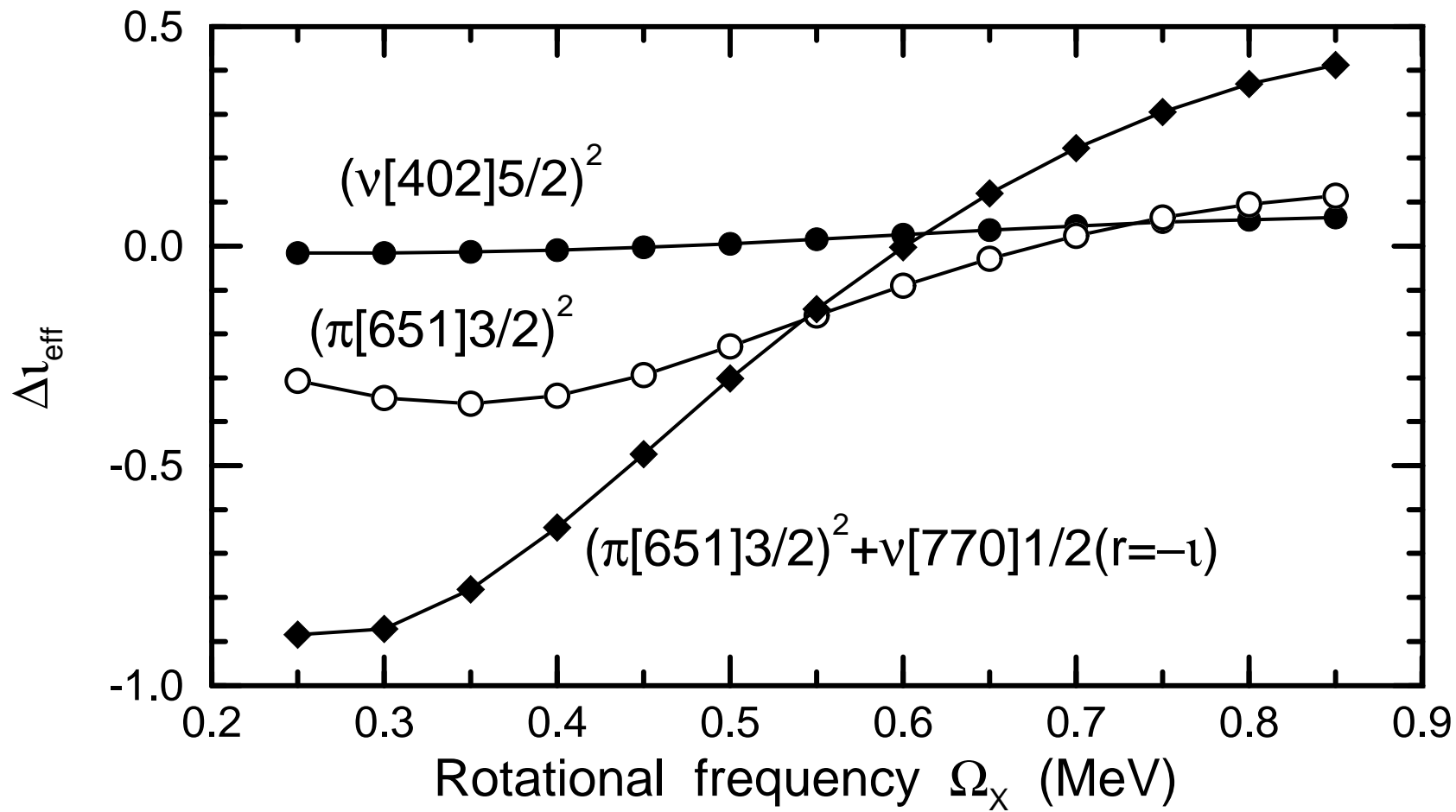
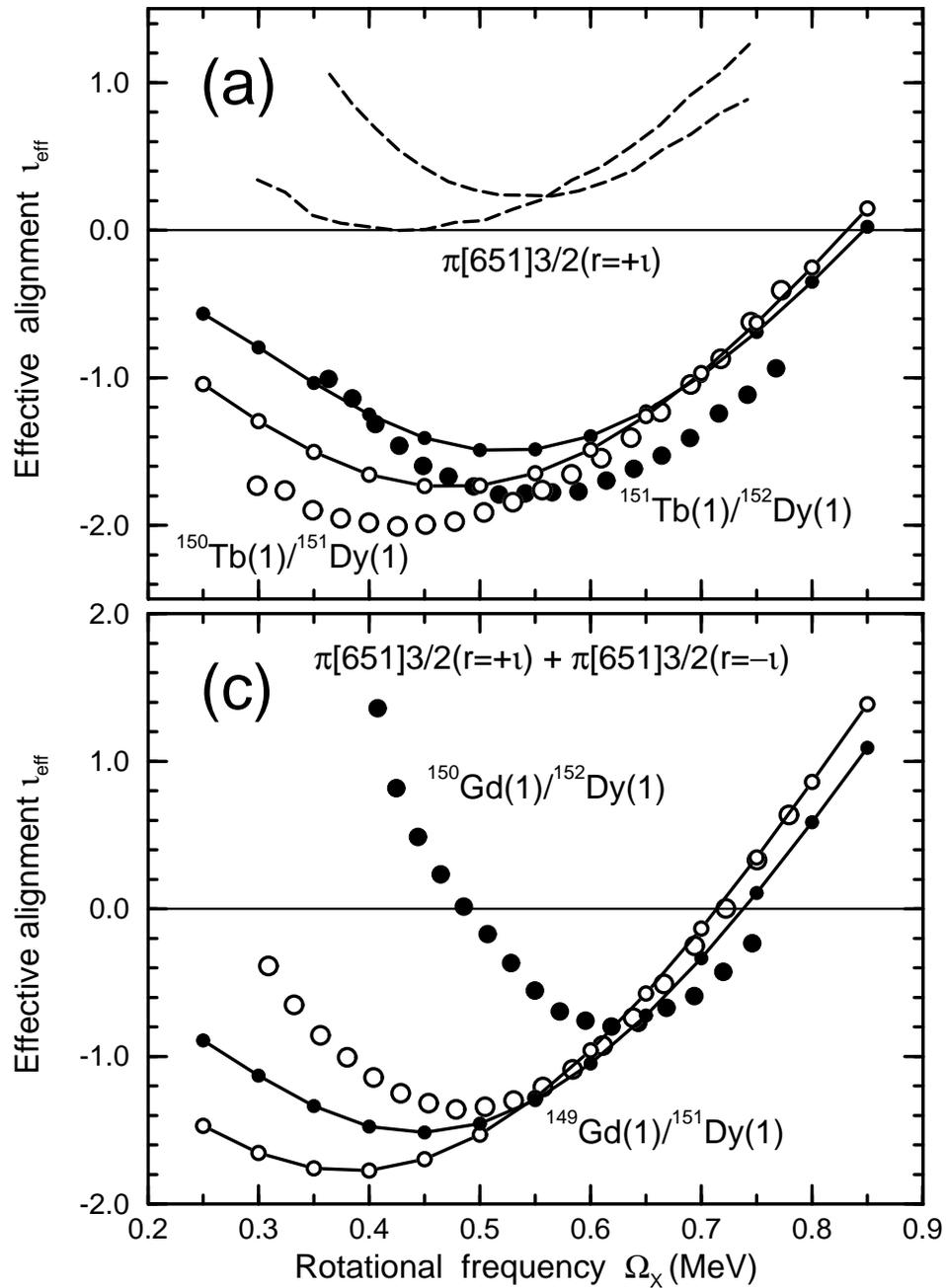
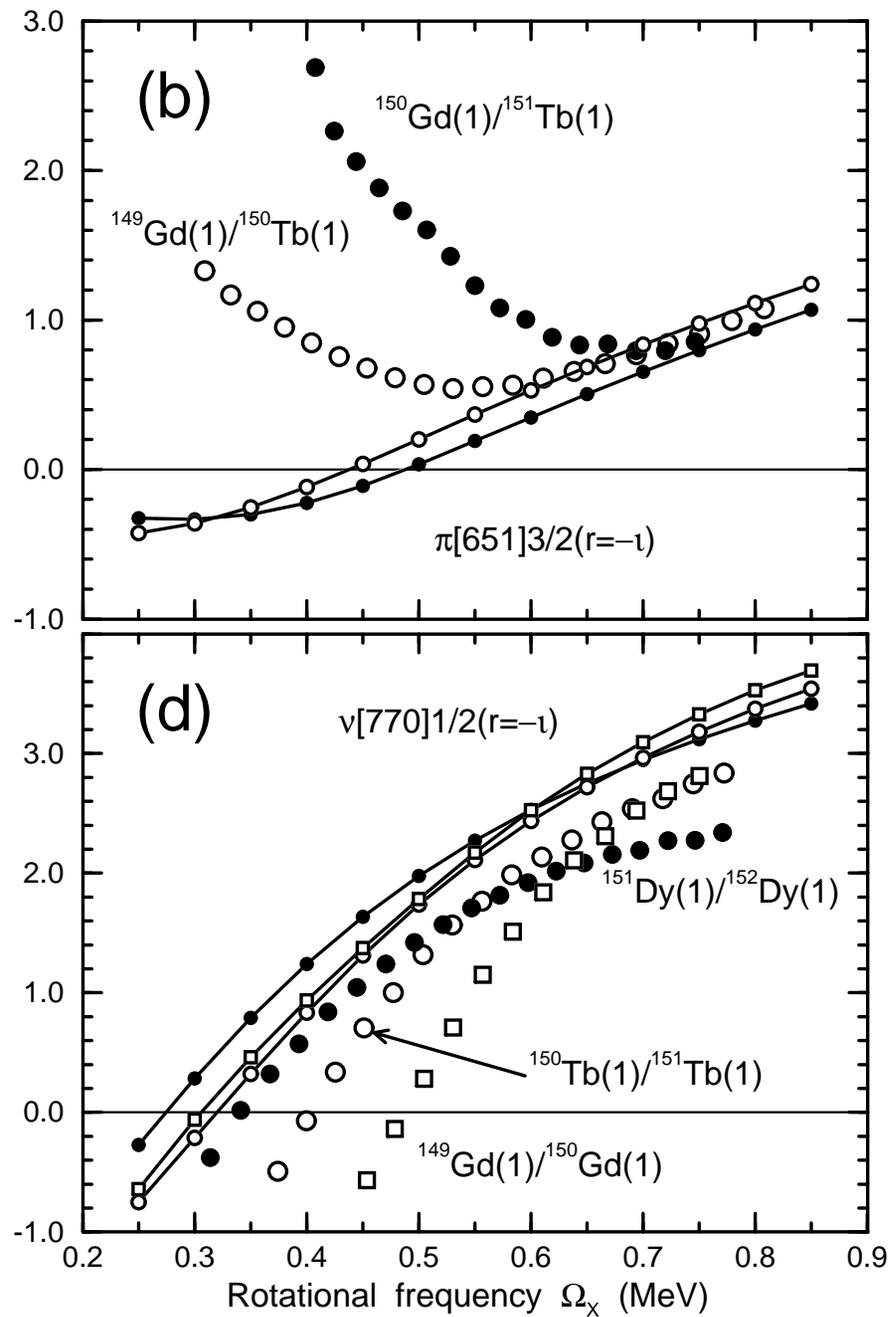
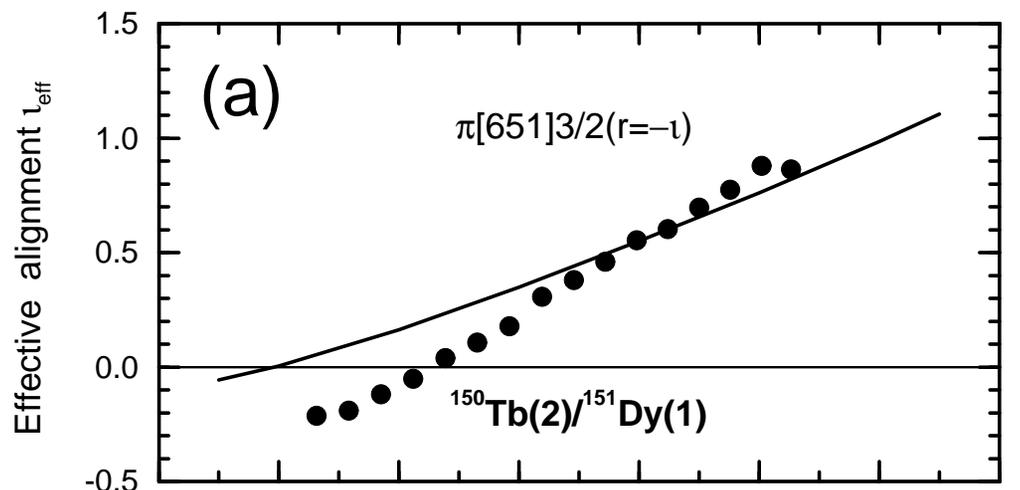
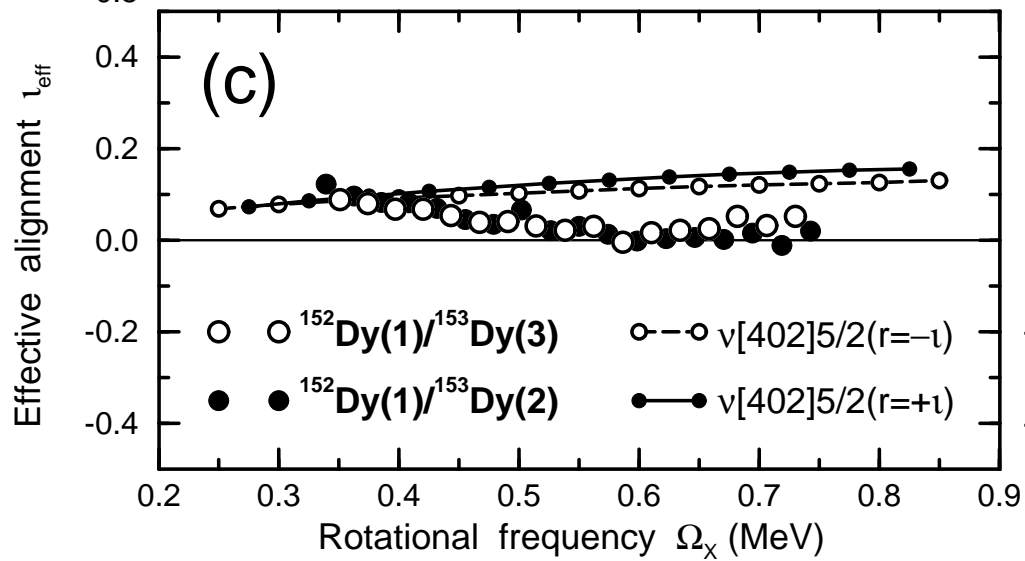
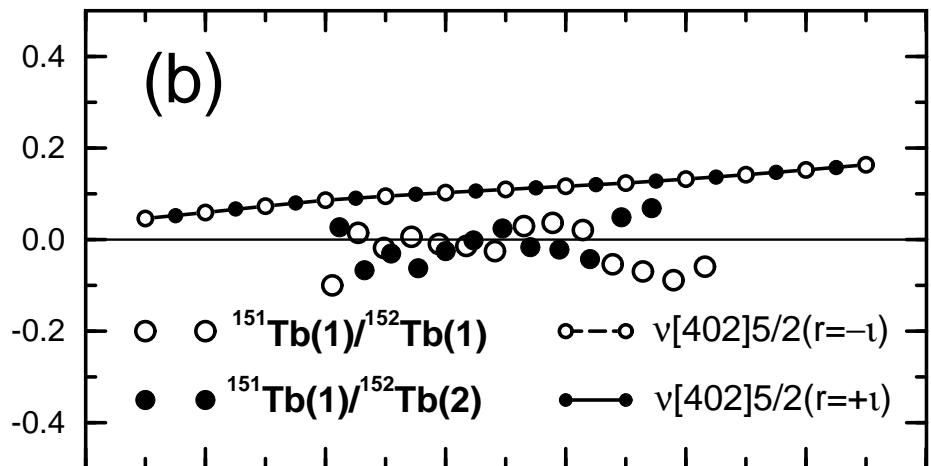
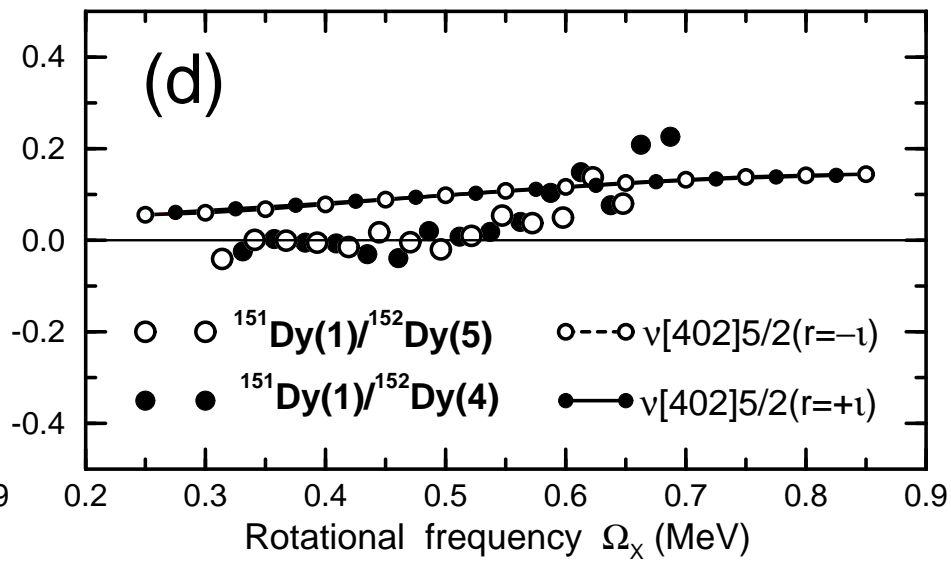
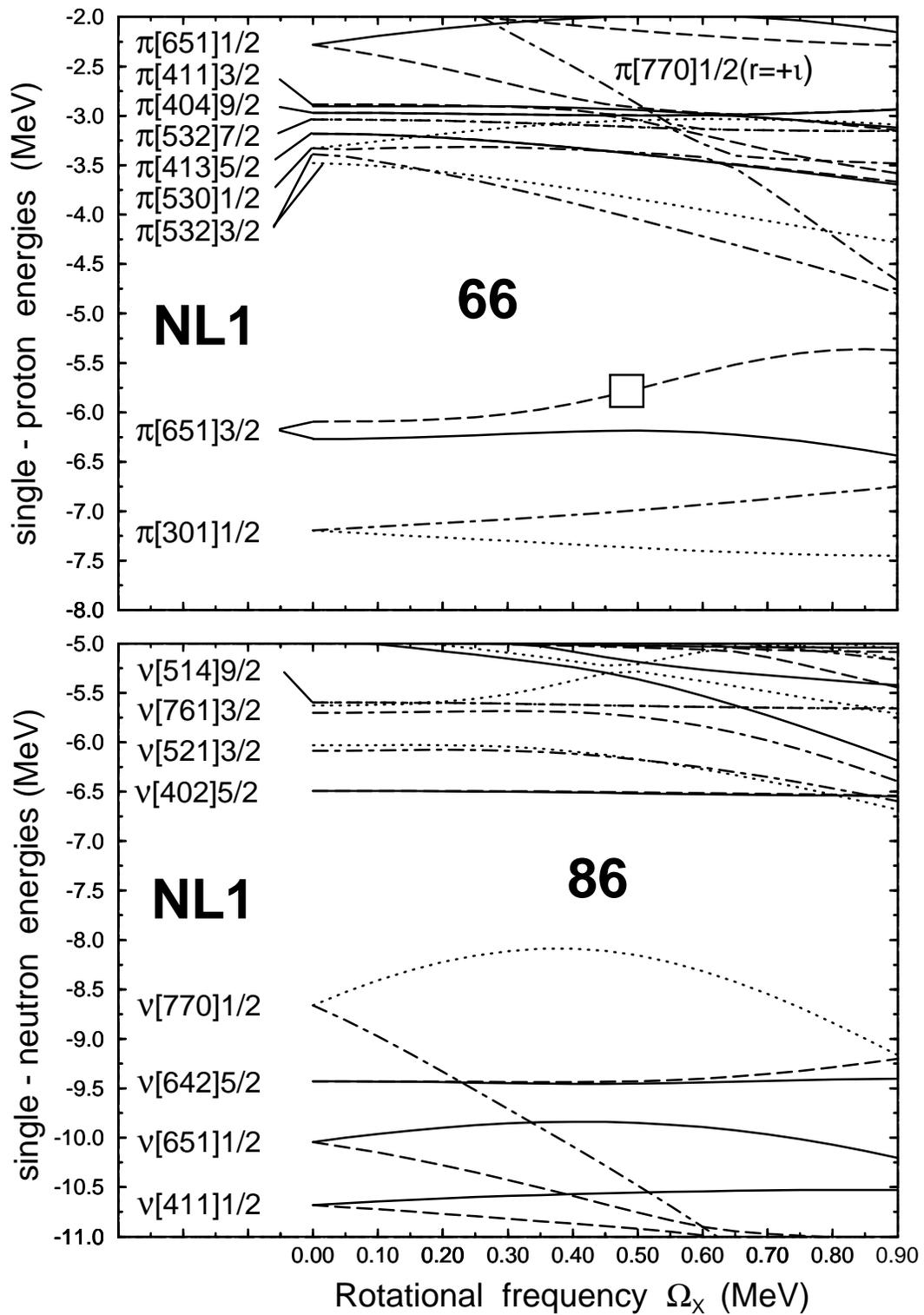
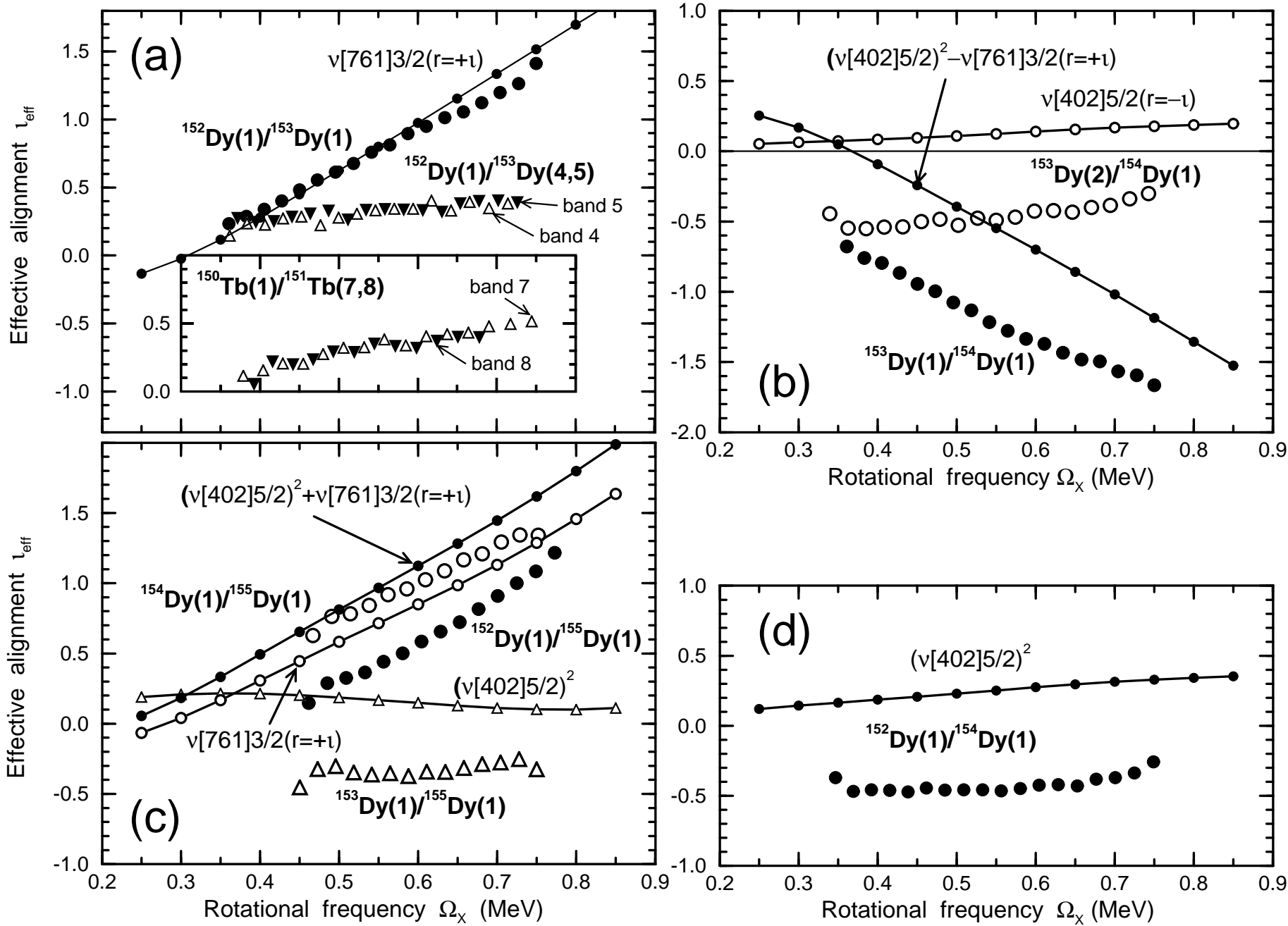
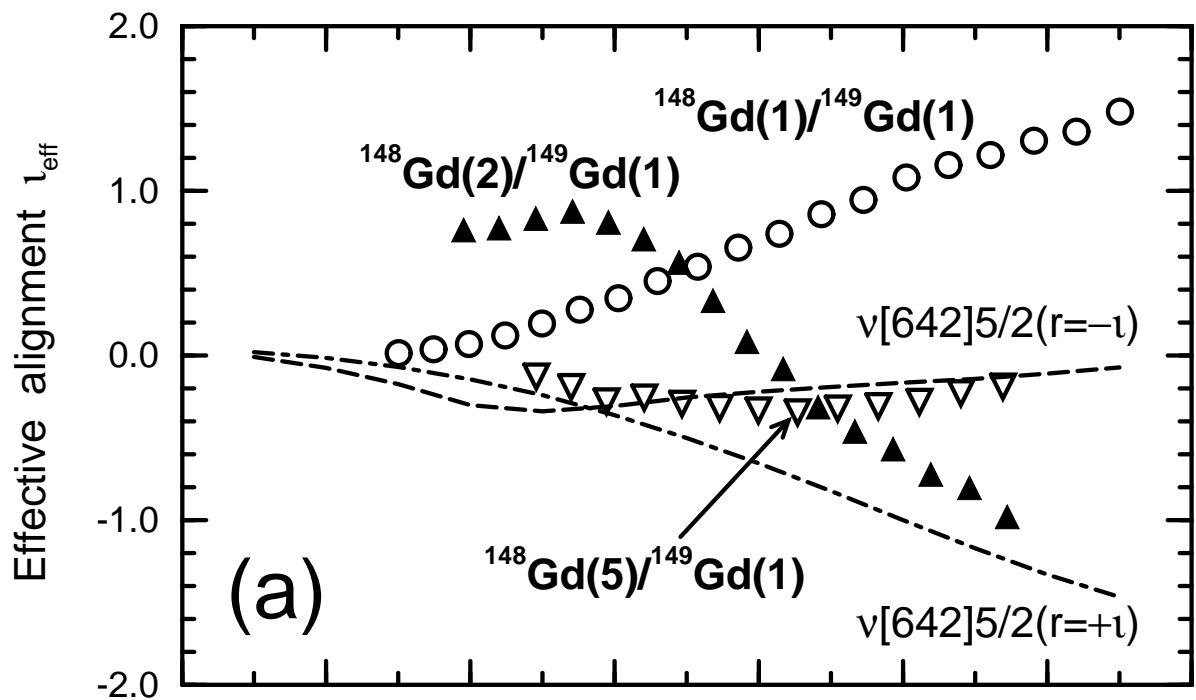
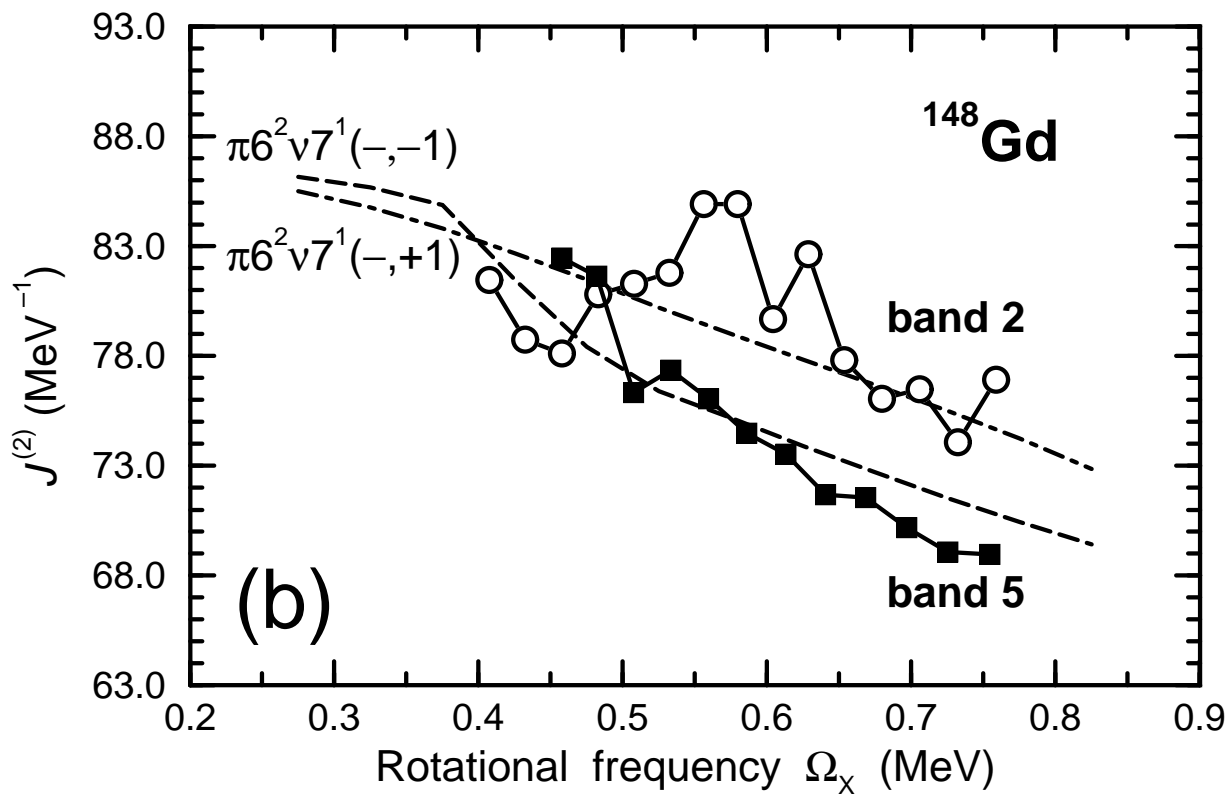
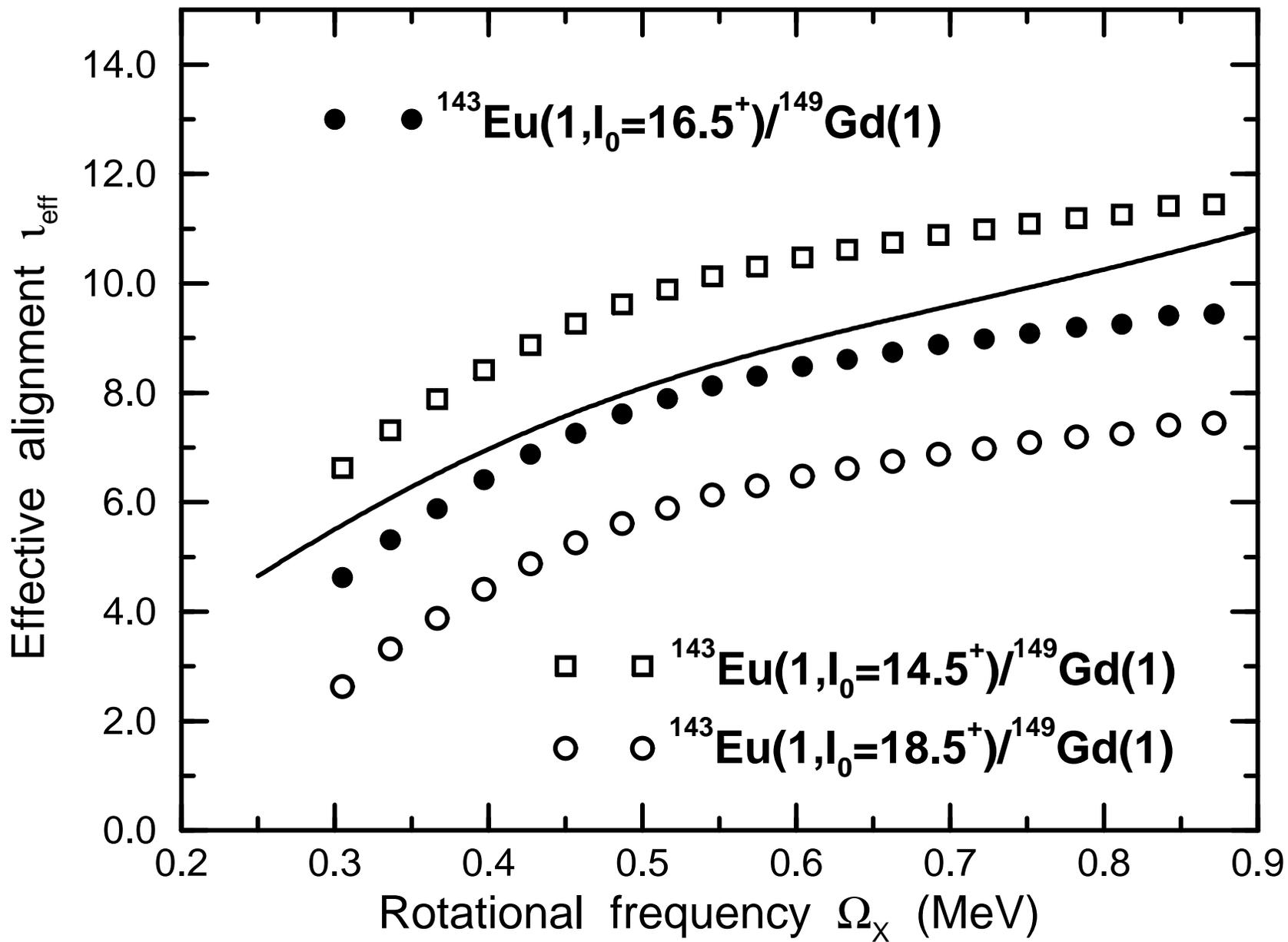
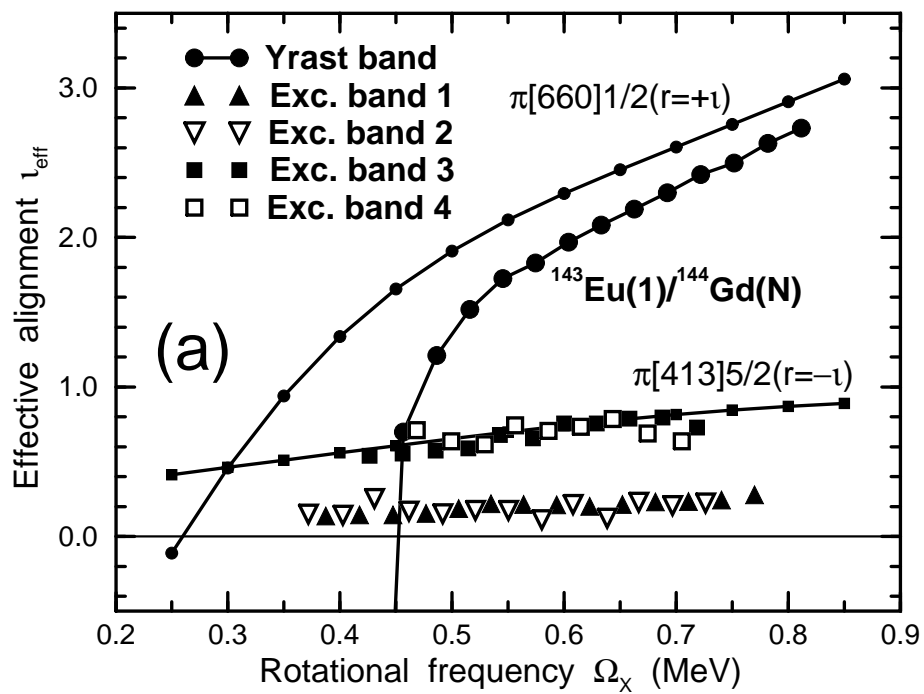
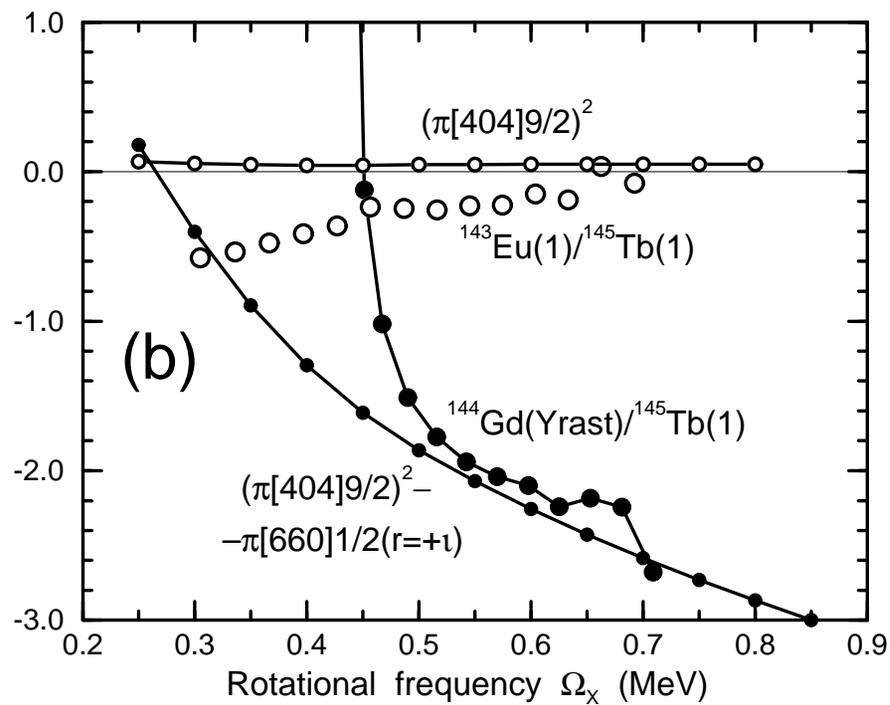
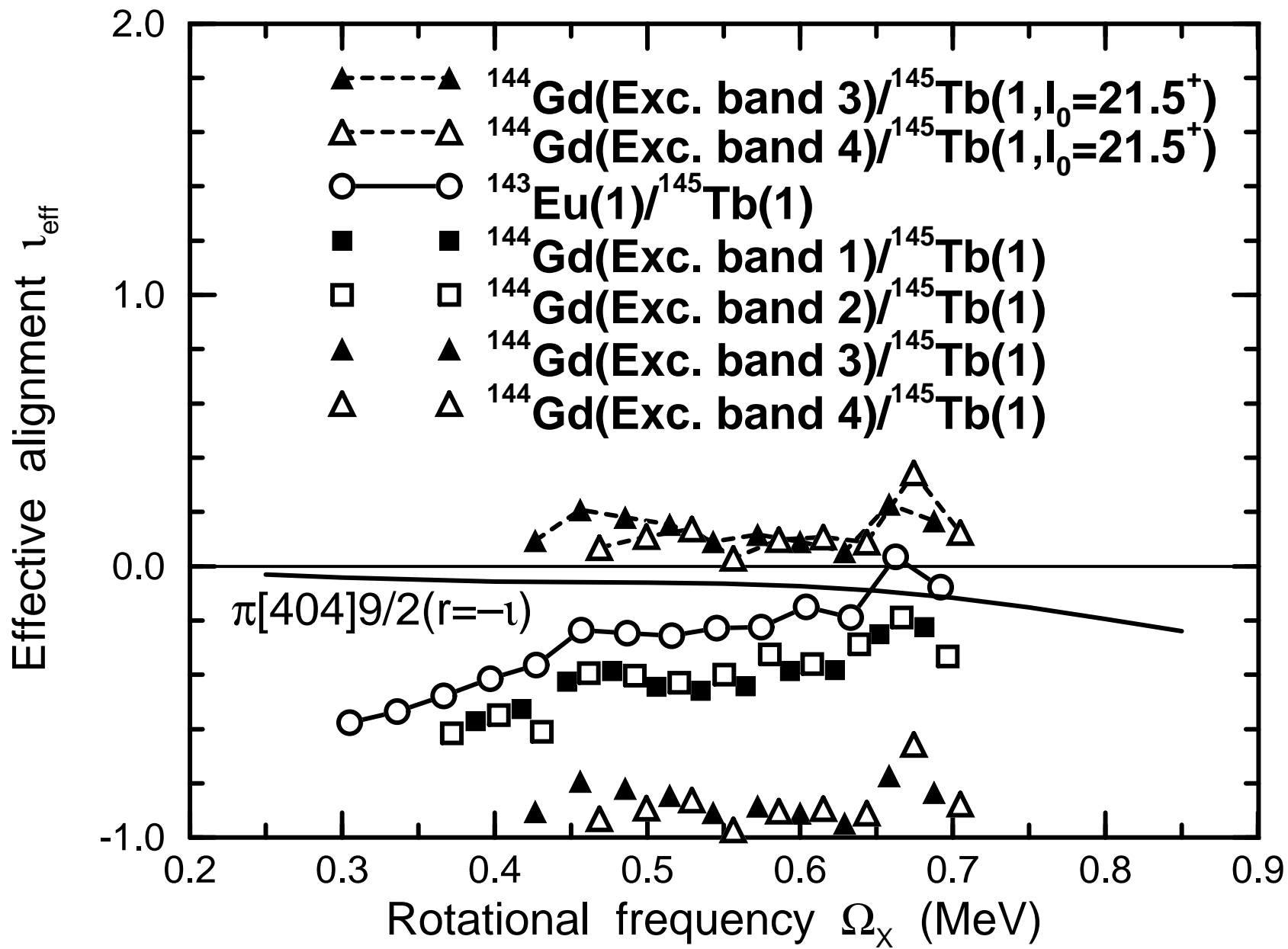
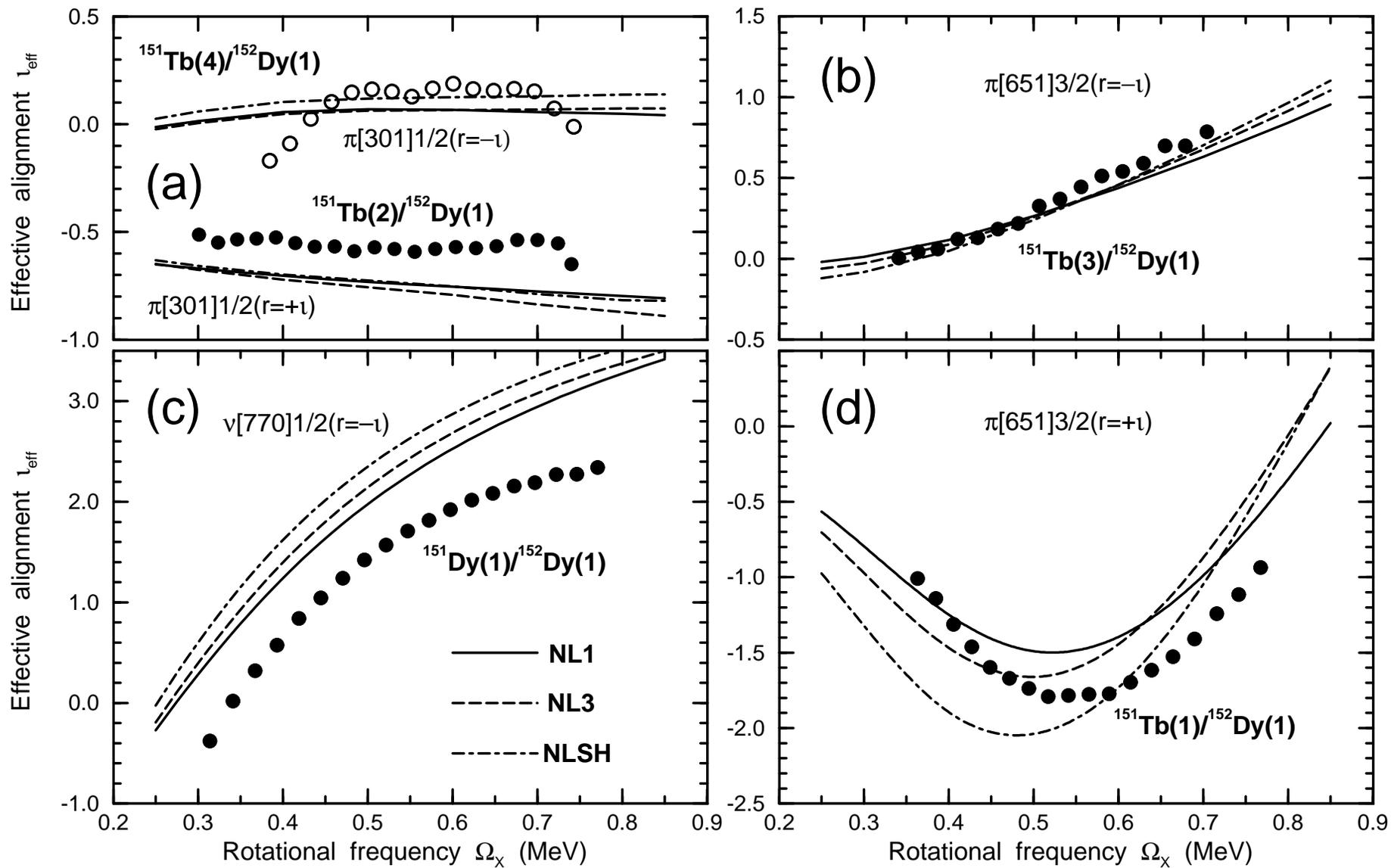
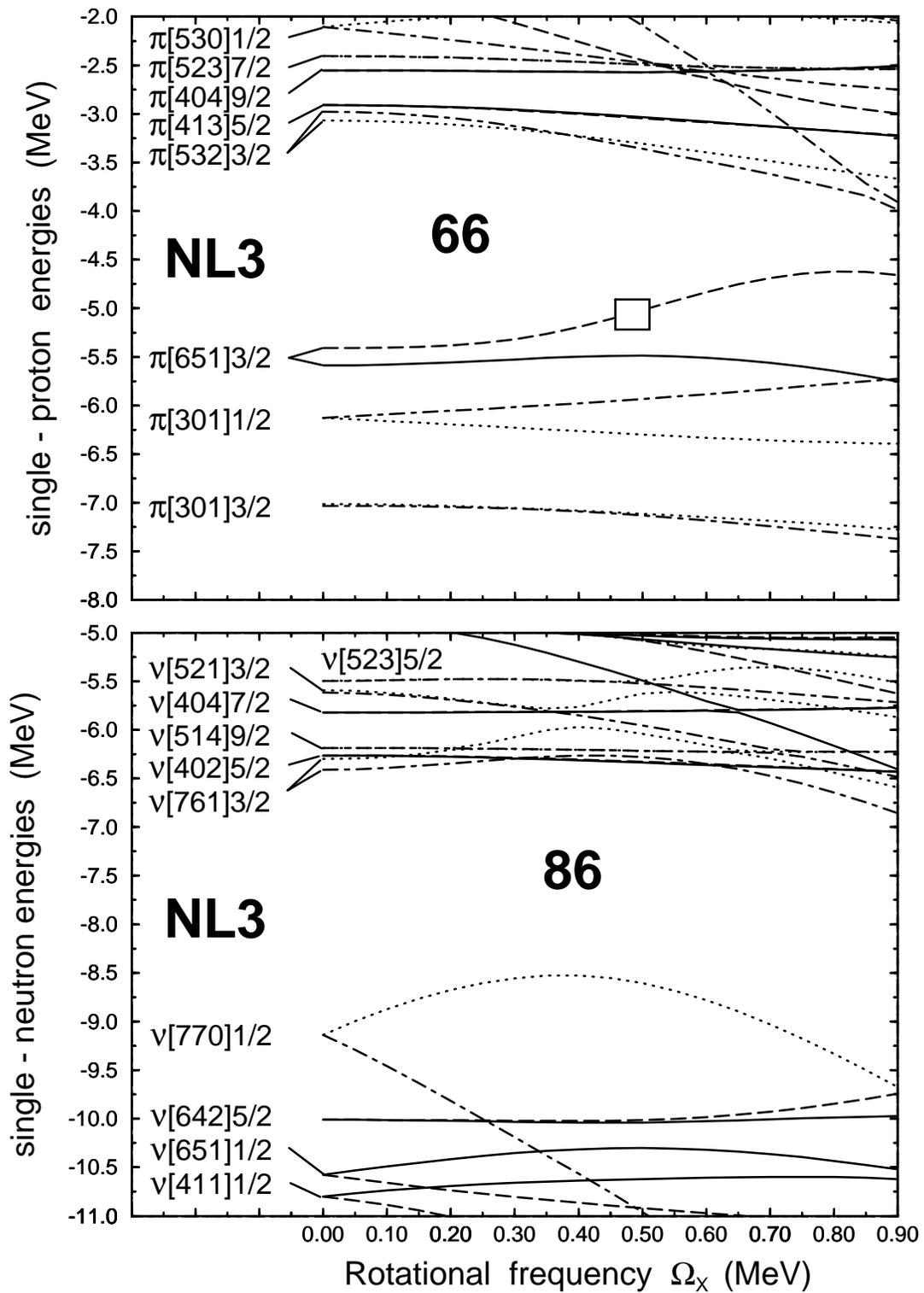
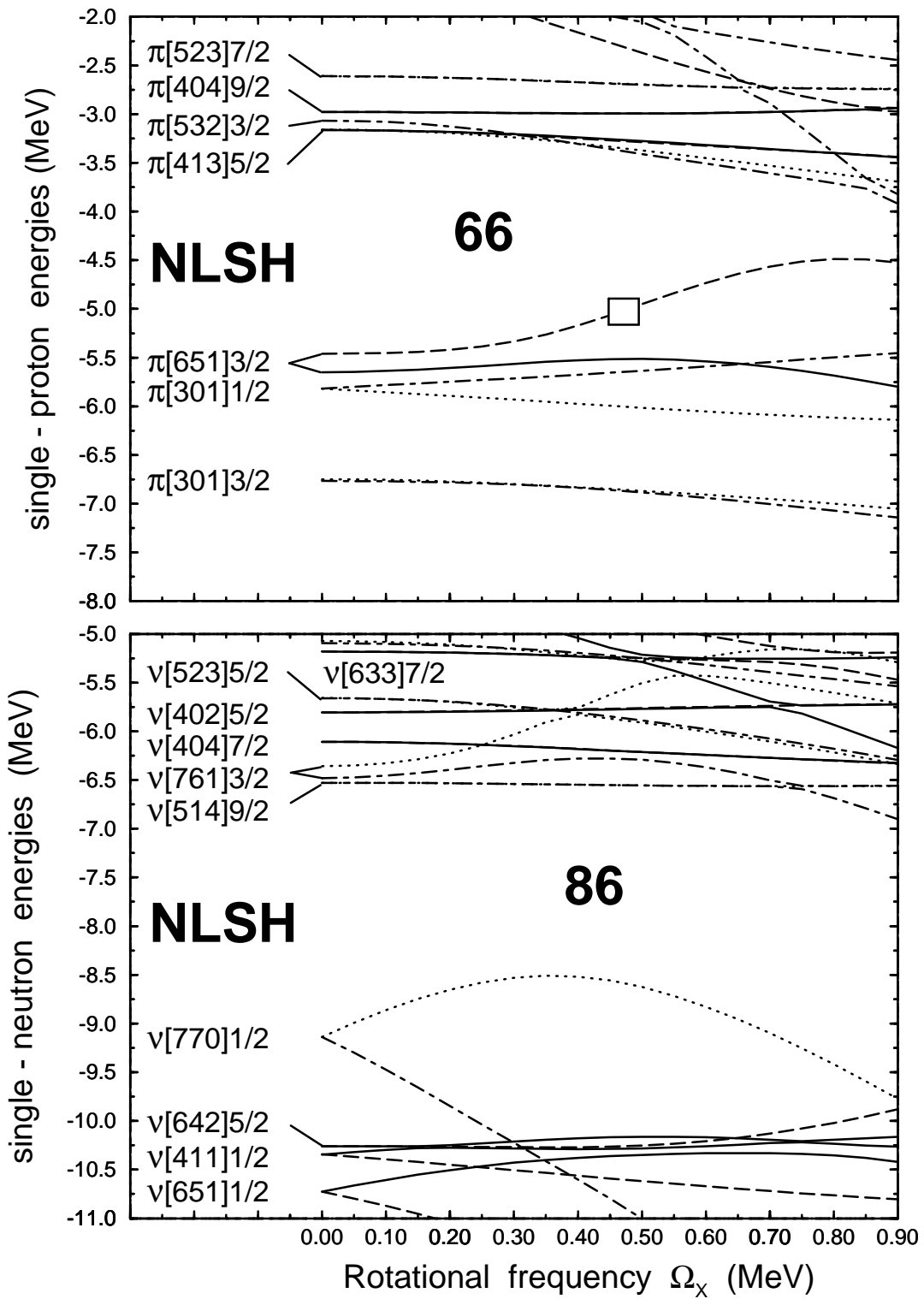
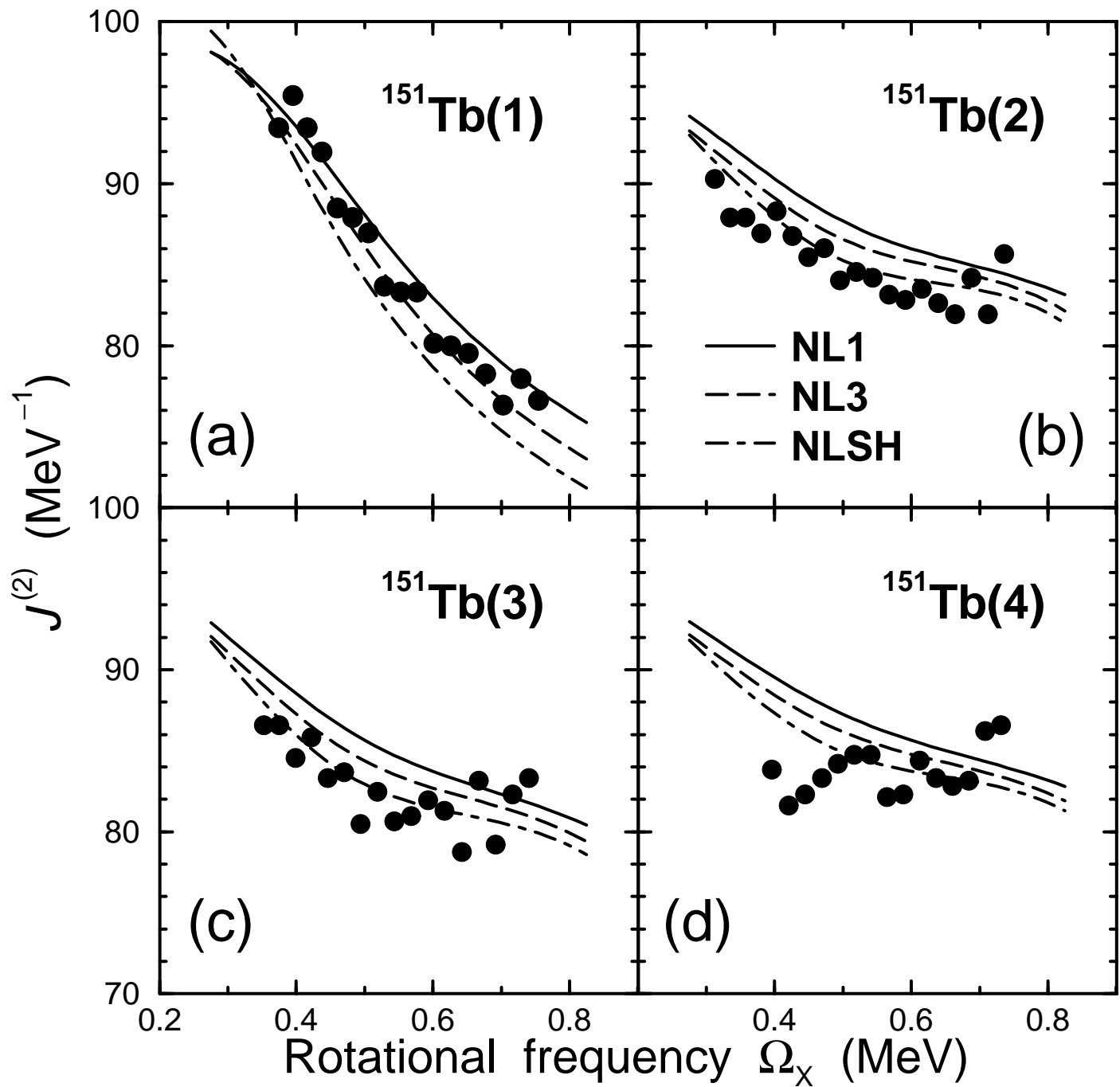
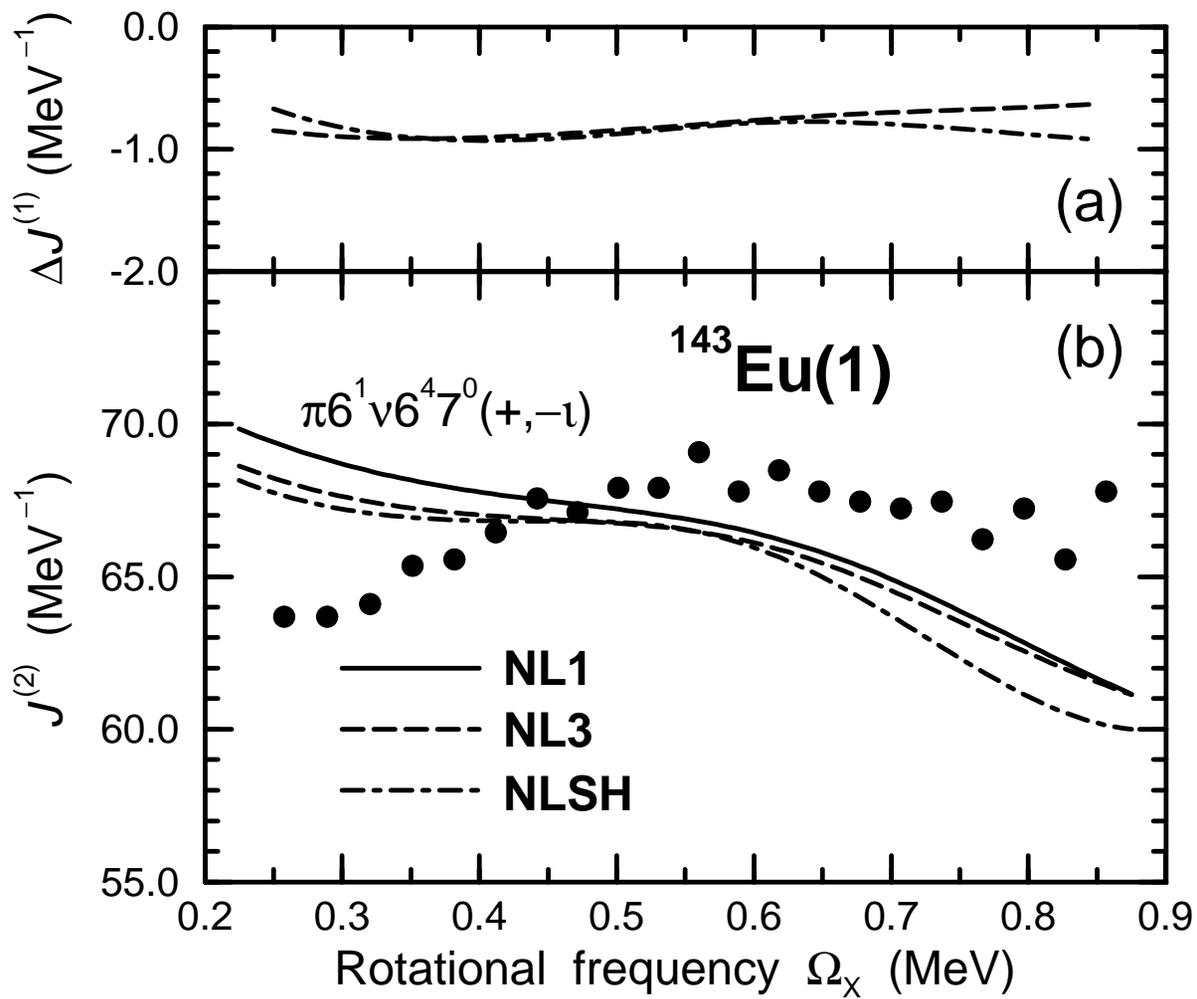
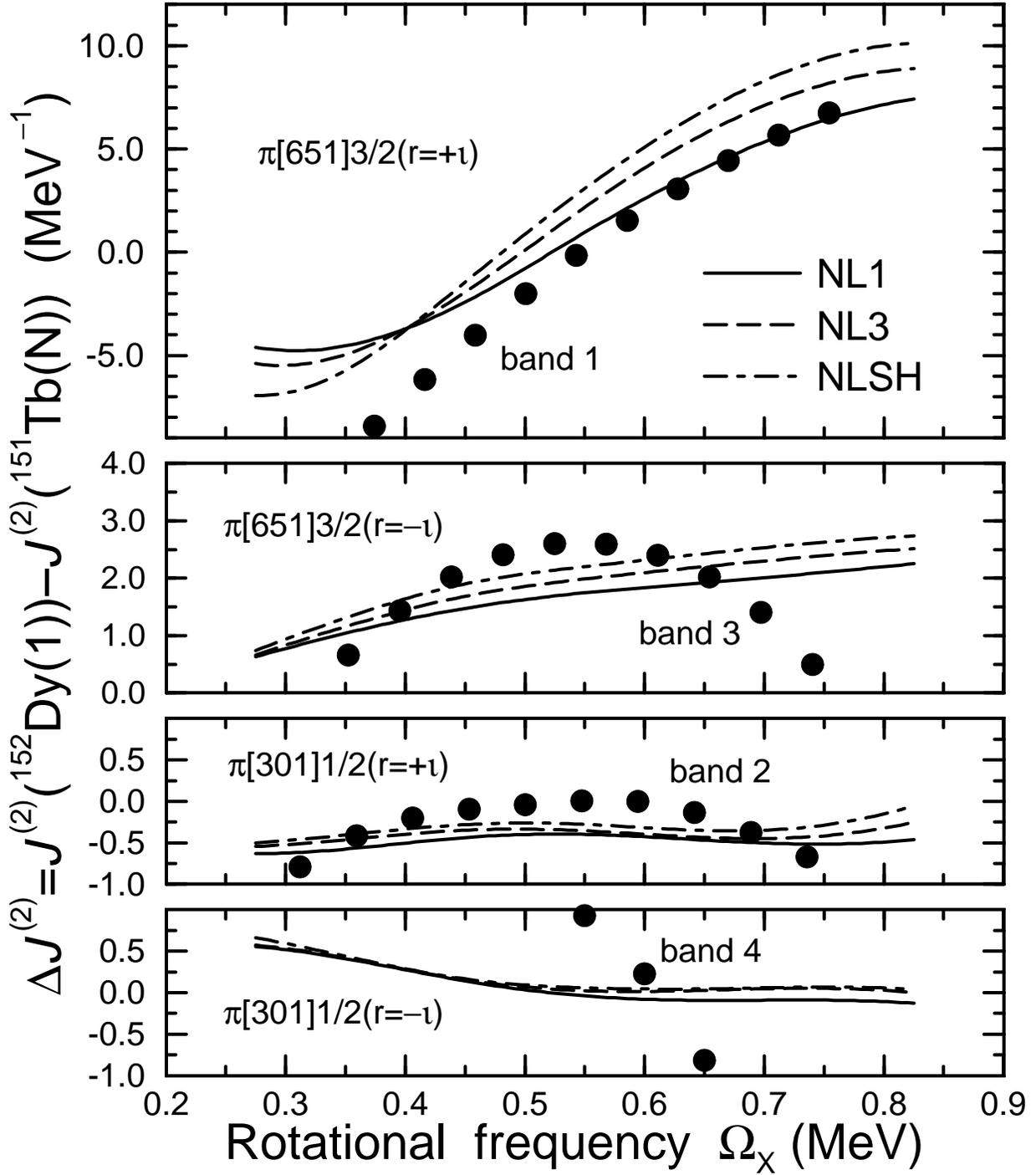
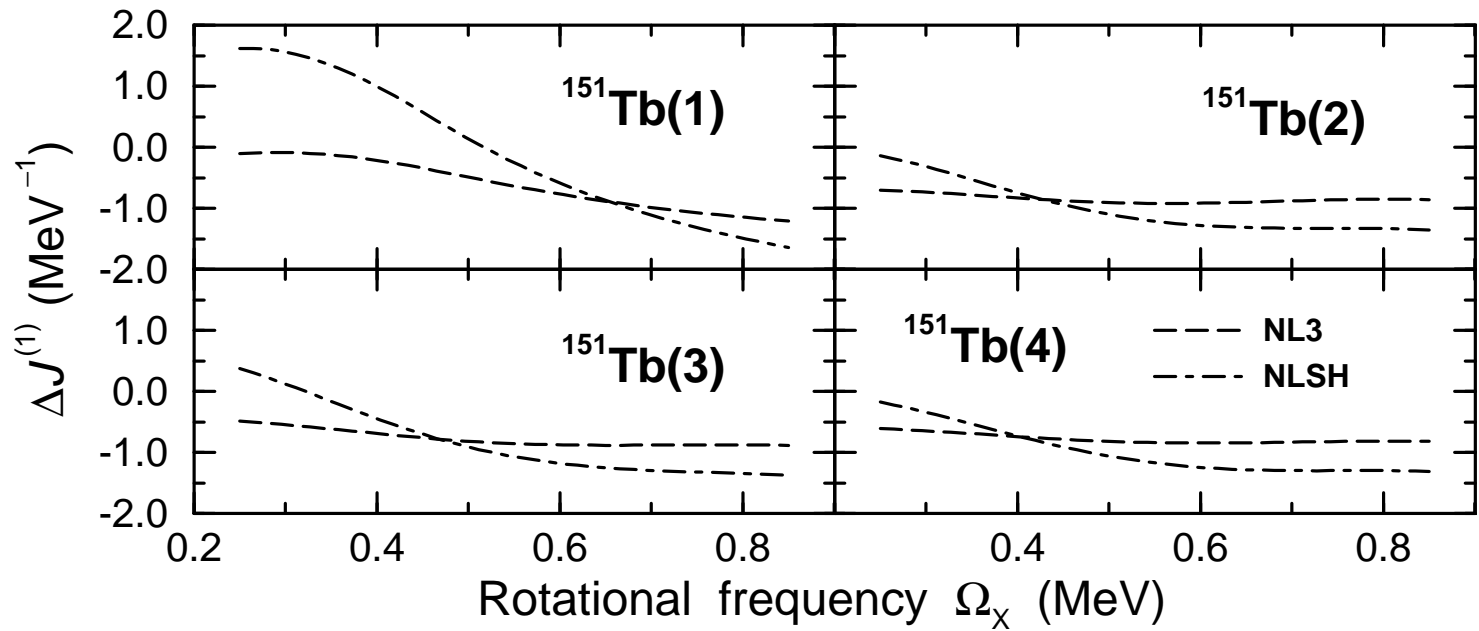
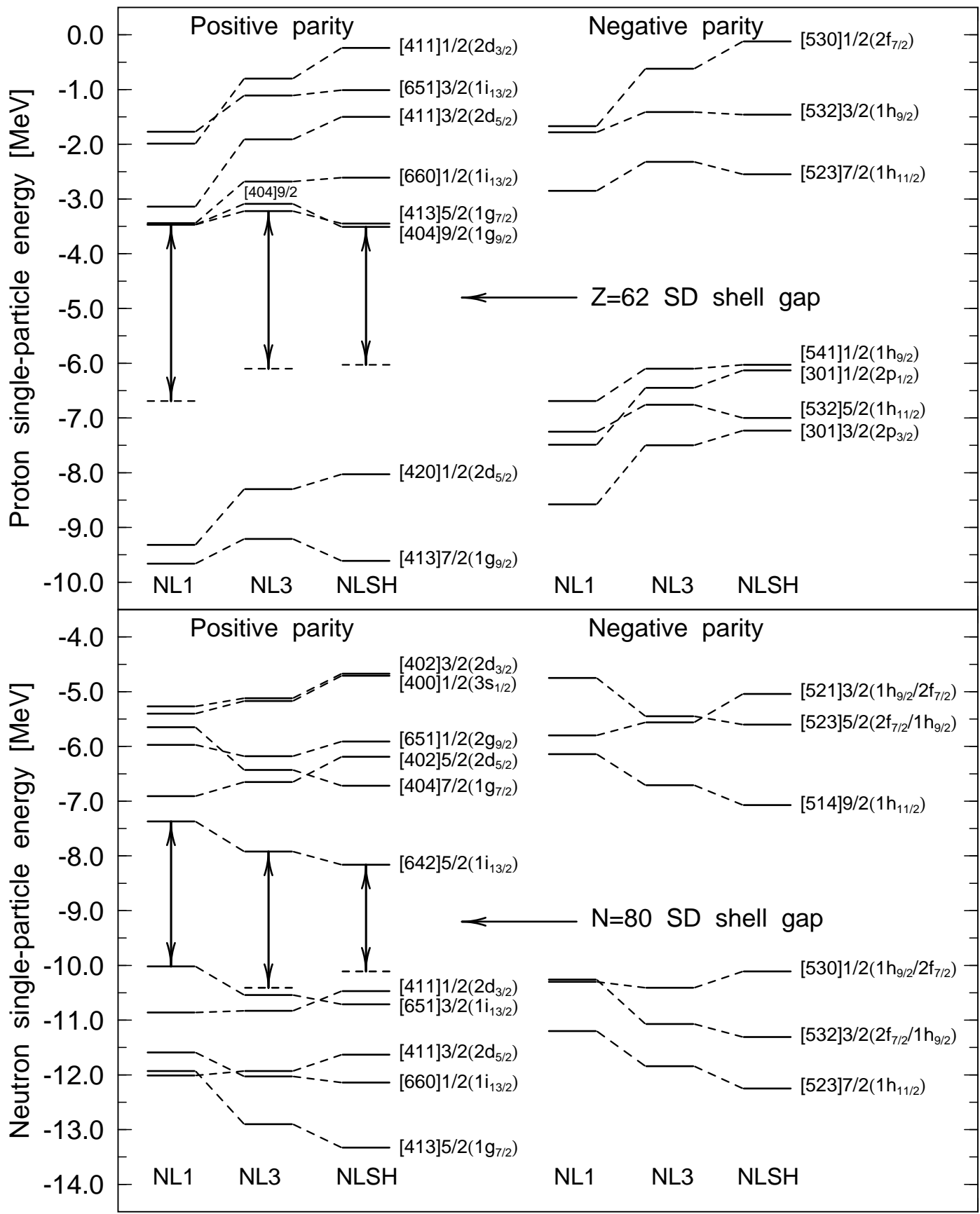
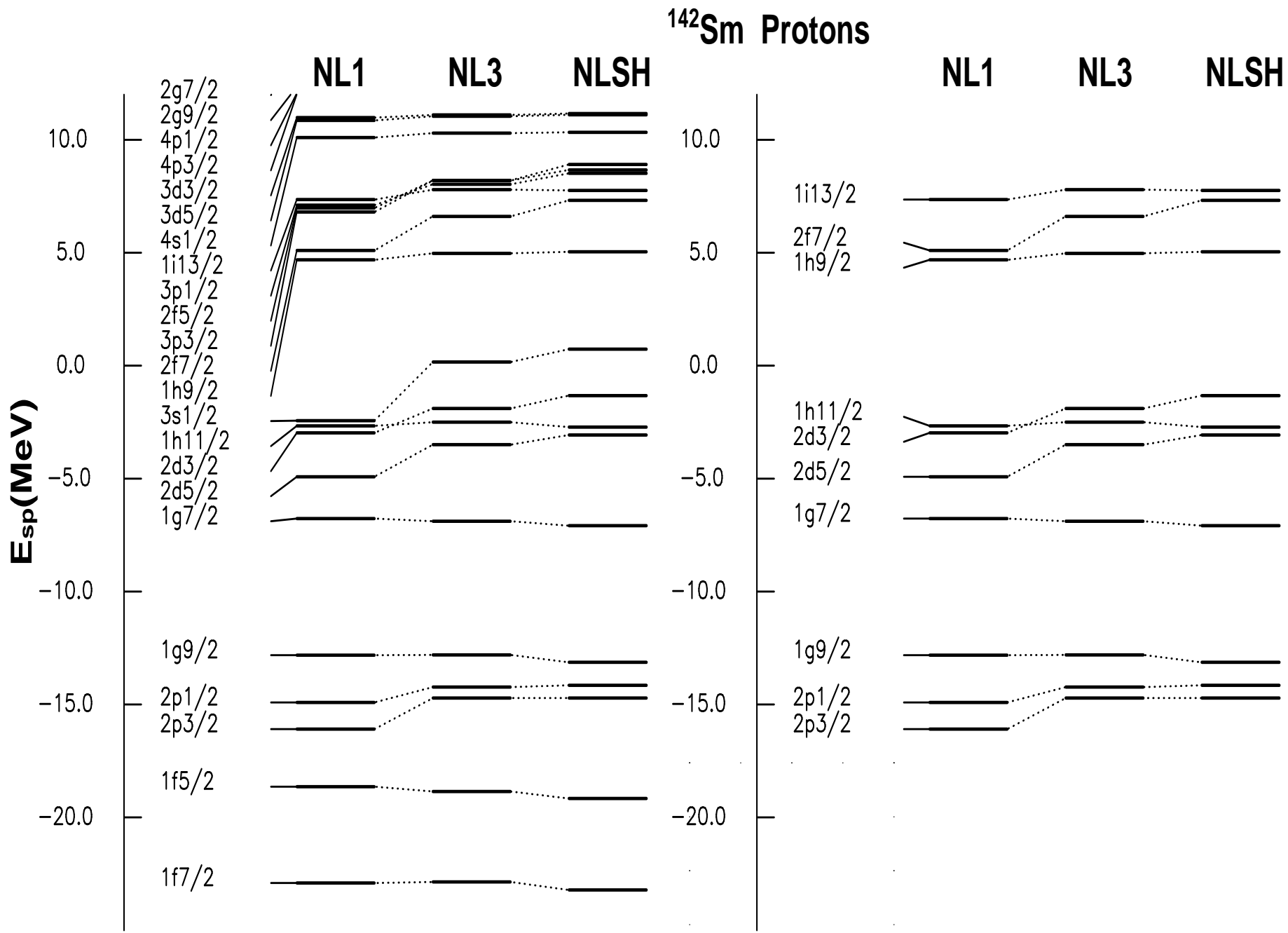
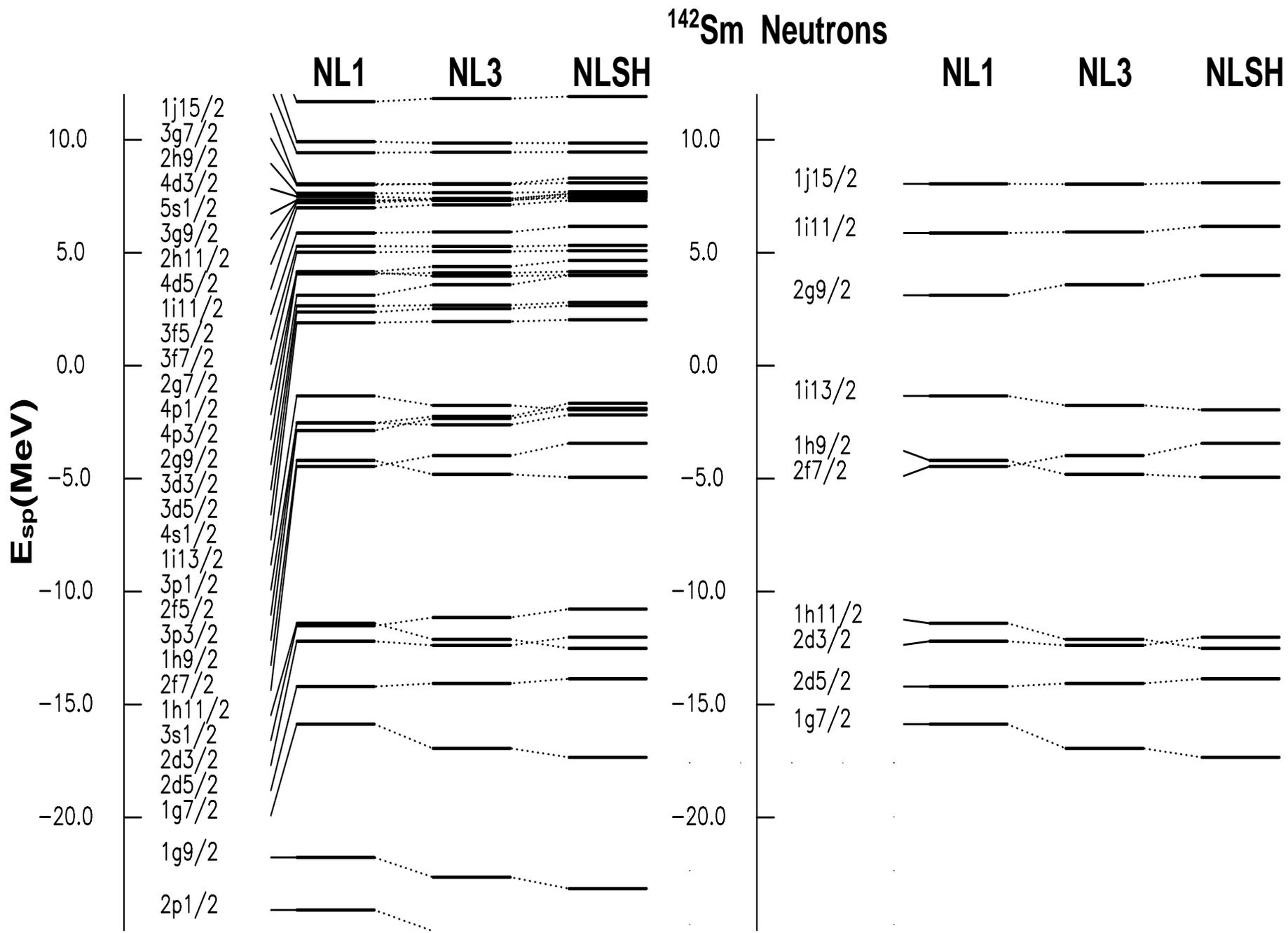
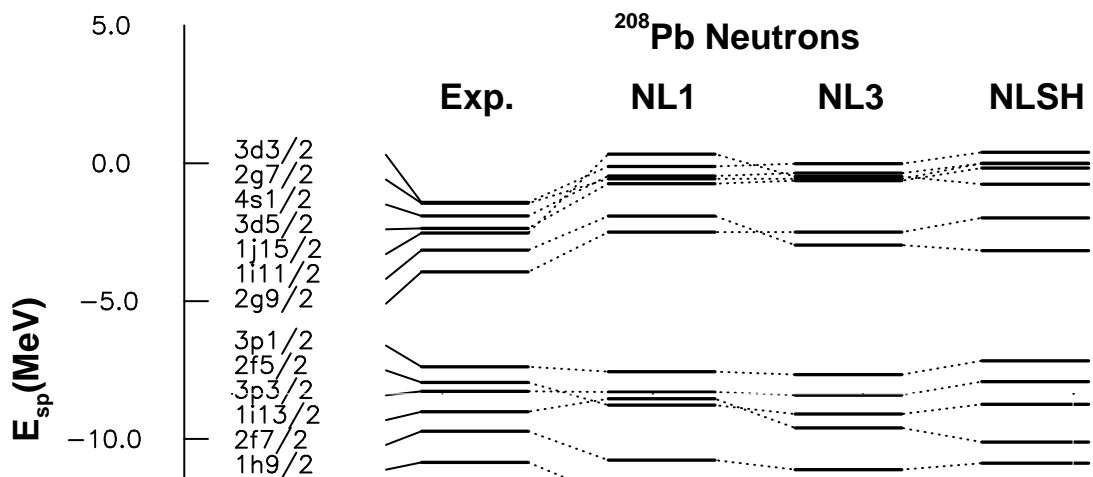
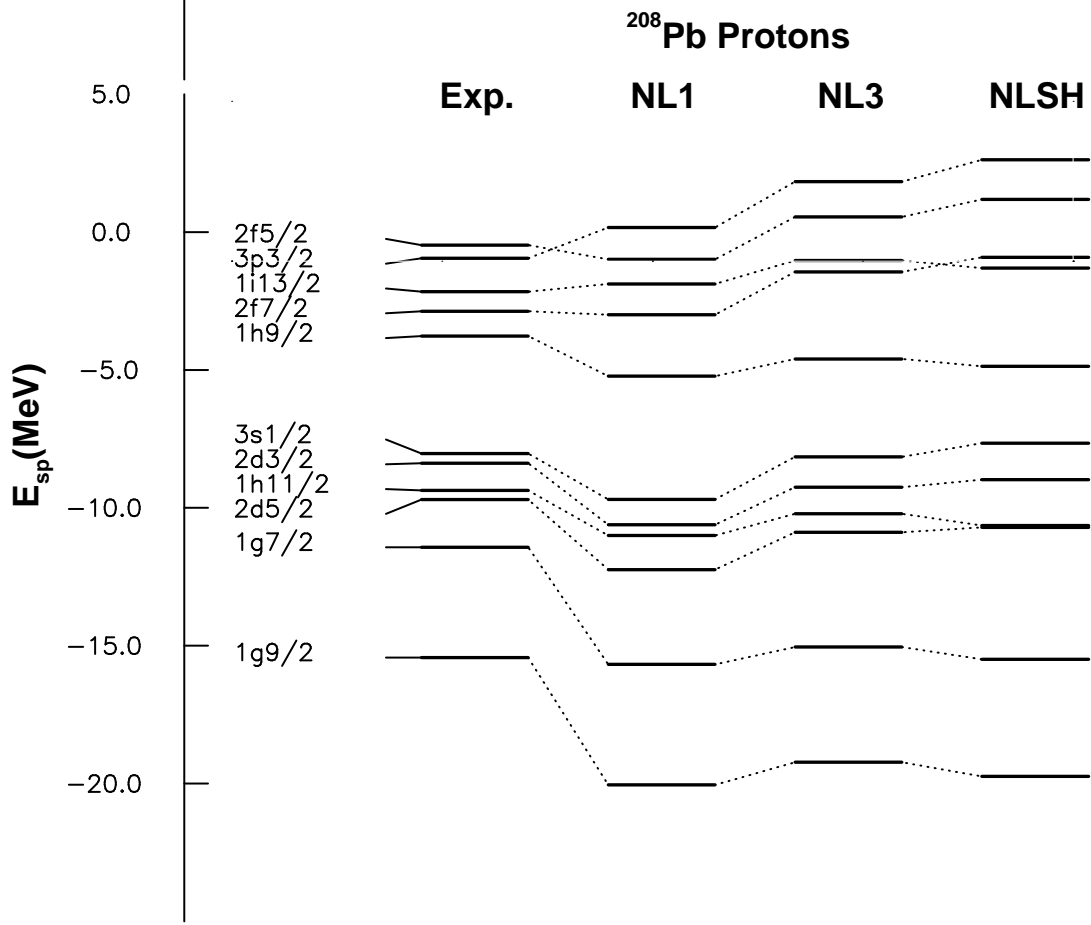
\end{document}